\documentclass[onecolumn]{aa}  
\usepackage{keyval}
\usepackage{graphicx}
\usepackage{subcaption}
\usepackage{caption}
\usepackage{multirow}
\usepackage{txfonts}
\usepackage{hyperref}
\usepackage{lipsum}

\begin{document}

   \title{Anomaly detection in Fink}

   \subtitle{I. Discovery, follow-up, and classification of unusual sources}

   \author{M.~V.~Pruzhinskaya
          \inst{1,2}
          \and
          M.~V.~Kornilov
          \inst{2,3}
          \and
          A.~V.~Dodin
          \inst{2}
          \and
          A.~Baluta
          \inst{4}
          \and
          T.~A.~Pshenichniy
          \inst{5}
          \and 
          A.~M.~Zubareva
          \inst{6,2}
          \and
          E.~E.~O.~Ishida
          \inst{1}
          \and
          J.~Peloton
          \inst{7}
          \and
          I.~Beschastnov
          \inst{8}
          \and
          I.~Ippolitov
          \inst{9}
          \and
          A.~A.~Belinski
          \inst{2}
          \and
          P. Golysheva
          \inst{2}
          \and
          N.~P.~Ikonnikova
          \inst{2}
          \and
          V.~A.~Kiryukhina
          \inst{2}
          \and
          V.~V.~Krushinsky
          \inst{10}
          \and
          A.~M.~Tatarnikov
          \inst{2}
          \and
          S.~G.~Zheltoukhov 
          \inst{2}
          \and
          D.~A.~Buckley
          \inst{11,12,13}
          \and
          A.~Kniazev
          \inst{11,14,2}
          \and
          S.~V.~Karpov
          \inst{15} 
          \and
          A.~Möller
          \inst{16}
          \and
          Yusuke Tampo
          \inst{11,12}
}

        \institute{Université Clermont Auvergne, CNRS, LPCA, Clermont-Ferrand, F-63000, France
             \email{pruzhinskaya@gmail.com}
         \and
            Lomonosov Moscow State University, Sternberg Astronomical Institute, Universitetsky 13, Moscow, 119234, Russia
        \and
            National Research University Higher School of Economics, 21/4 Staraya Basmannaya Ulitsa, Moscow, 105066, Russia
         \and
            Center for Astrophysics and Cosmology, University of Nova Gorica, Vipavska 11c, 5270 Ajdovščina, Slovenia
         \and
            Lomonosov Moscow State University, Faculty of Space Research, Leninsky Gori 1 bld. 52, Moscow, 119234, Russia
         \and
            Institute of Astronomy (Russian Academy of Sciences), Pyatnitskaya Str. 48, Moscow, 119017, Russia 
         \and
             Université Paris-Saclay, CNRS/IN2P3, IJCLab, Orsay, France
         \and 
             Institute of Mechanics of Lomonosov Moscow State University, Michurinskiy Prospekt 1, Moscow, 119192, Russia
         \and
             Independent researcher
         \and
             Laboratory of Astrochemical Research, Ural Federal University, ul. Mira d. 19, Yekaterinburg, 620062, Russia
         \and
            South African Astronomical Observatory, P.O Box 9, Observatory, 7935 Cape Town, South Africa
         \and
            Department of Astronomy, University of Cape Town, Private Bag X3, Rondebosch 7701, South Africa
         \and
            Department of Physics, University of the Free State, PO Box 339, Bloemfontein 9300, South Africa
         \and
            Southern African Large Telescope Foundation, PO Box 9, 7935 Observatory, Cape Town, South Africa
         \and
             Institute of Physics of the Czech Academy of Sciences, Na Slovance 1999/2, 182 00 Prague 8, Czech Republic
         \and
             Centre for Astrophysics and Supercomputing, Swinburne University of Technology, Hawthorn, VIC 3122, Australia
            }

   \date{Received September 15, 1996; accepted March 16, 1997}

 
  \abstract
   {Modern wide-field time-domain surveys produce alert streams whose scientific potential is often concentrated in rare and unusual events. Efficient discovery therefore requires automated pipelines  to be combined with rapid expert validation and follow-up.}
   {We present the first-year performance of the anomaly-detection (AD) pipeline operating within the Fink broker on the Zwicky Transient Facility alert stream, and assess its ability to identify scientifically valid outliers and enable discovery of rare phenomena.}
   {The pipeline transforms ZTF light curves into a compact set of features and ranks alerts using an Isolation Forest model trained on archival ZTF data. Each night, the 10 most anomalous candidates are distributed to experts via Slack/Telegram and exposed through an API. We also implement an expert-feedback loop using a public Telegram bot and retrain the model using the Active Anomaly Discovery algorithm.}
   {During the first year of operations (starting from 25 January 2023), the AD pipeline identified multiple high-interest sources and triggered dedicated photometric and spectroscopic follow-up. We report the discovery and multi-instrument (11-m SALT telescope, 2.5-m CMO telescope, 0.6-m ASA RC600, 0.25-m FRAM-ORM) follow-up of the rare AM~CVn system Fink J062452.88+020818.3 of the WZ Sge type, UX Ori-type star Fink J222324.32+744222.0 and the unusual transient with precursor SN~2023mtp. In addition, the module triggered 33 supernovae, including 30 previously unreported ones, with candidates for superluminous and hostless events. Furthermore, nine new dwarf novae were discovered.}
   {These results show that broker-level anomaly detection, coupled with rapid dissemination, expert assessment, and follow-up observations, provide an effective bridge between large-scale survey streams and domain expertise, turning anomaly scores into astrophysical insights and concrete discoveries.}


   \maketitle
%

\section{Introduction}

The identification of unusual events or objects has always been the trigger of astronomical discovery \citep{Dick_2013}. However, with the advent of large scale sky surveys that continuously monitor the sky, this exercise was transformed into a data mining task. Following a predetermined observation strategy, surveys like the Zwicky Transient Facility\footnote{\url{https://www.ztf.caltech.edu/}} \citep[ZTF,][]{bellm2019} construct large and complex data sets which hold the entire scientific potential of the experiment.

In this context, the use of machine learning tools are  unavoidable. Automatic search strategies play the crucial role of  identifying subsets of the data which will be subject to more detailed analysis. These sets can then be redirected to another pipeline, for example in the case of population studies \citep[e.g.][]{crawford2025} or photometric supernova cosmology \citep[e.g.][]{2022MNRAS.514.5159M,2024MNRAS.533.2073M,mitra2025}. However, in many situations, specifically in the presence of edge cases or in the study of individual anomalous sources, visual inspection by an expert is mandatory for adding value and placing the newly acquired data within the current state of the art \citep{pruzhinskaya2025}. In the specific case of analyzing astronomical transients, the problem becomes even more complex since, for many sources,  there is a window of opportunity for identification and follow-up. 

The alert stream infrastructure, currently operating with ZTF \citep{patterson2019}, distributes every detection resulting from difference imaging to community brokers whose task is to add value, filter and redistribute smaller data volumes to the astronomical community. It has been used as a precursor for broker teams\footnote{AMPEL \citep{ampel}, ANTARES \citep{matheson2021}, ALeRCE \citep{alerce}, Babamul \citep{babamul}, Fink \citep{fink}, Lasair \citep{lasair} and Pitt-Google \citep{pitt-google}.} who were chosen to receive the raw alert stream from the Vera C. Rubin Observatory\footnote{\url{https://rubinobservatory.org/}} Legacy Survey of Space and Time \cite[][]{lsst2009}. 

Beyond large efforts in providing photometric classification as a way to filter the stream \citep[see][and references therein]{sanchez2021,  desoto2024, fraga2024, sheng2024, nordin2025}, a significant fraction of resources have been devoted to the automatic identification of anomalous sources. The latter task is as promising as it is challenging. Considering that ZTF follows a pre-determined observation strategy which returns to the same place in the sky repeatedly, there is  a high probability of detecting rare instances of highly scrutinized astronomical classes \citep[e.g.][]{bag2025}, as well as truly unexpected events \citep{Norris_2017}. Despite this certainty, the conversion of this potential into scientific development is conditional to our capability of identifying, and interpreting, such objects among hundreds of thousands of newly detected candidates every night. 

Foreseeing the complexity of the challenge ahead, broker teams have been experimenting with different strategies for anomaly detection (AD). In \citet{carrasco2023}, the ALeRCE broker \citep{alerce} presents a multi-strategy AD technique which benefits from their own machine learning (ML) light curve classifier and taxonomy \citep{sanchez2021}. Six AD algorithms were trained in a controlled data environment, each one considering as anomaly a particular sub-set  of 3 general classes (transient, stochastic and periodic). Once optimized configurations were achieved, the models were applied to the ZTF stream and a group of 12 experts scrutinized top 10 anomalies from each of the 15 subclasses. Among 150 visually inspected candidates, 7 were considered scientifically interesting sources (including a microlensing and a confirmed tidal disruption event --- TDE). 

The anomaly detection pipeline implemented in the ANTARES broker \citep{aleo2024} employs statistical features built from light curves \citep{2021MNRAS.502.5147M}, as well as environmental information from the host galaxy \citep{gagliano2021}. These are input to a random forest classifier aimed to identify three different anomaly types (spectroscopical, behavioral and contextual). The pipeline includes a similarity search, where identified anomalies are used to retrieve more sources with similar characteristics. The authors report the discovery of 325 transients, observed between 2018 and 2021, which were absent from previous catalogs. 

The Lasair broker \citep{lasair} presented their AD approach in \citet{iskandarli2026}. They employed three  autoencoder architectures, targeting different data products available or derived from ZTF alerts: statistical features calculated from light curves, image cutouts and light curves augmented with Gaussian processes. All standard supernova (SN) classes were considered  nominal (2987 objects), while active galactic nuclei, TDEs, superluminous SNe (SLSNe), catalclysmic variables (CVs), and nuclear transients or supernova-labeled objects with unusual properties were tagged as anomalous (156 objects). There was no overlap between the anomalies identify by each of the models during training, showcasing the diverse potential of multiple data products. The final models were applied to 25 days of the ZTF stream which identified 206 anomalies. Taking advantage of contextual information \citep{sherlock} and visual inspection, they provide lists of anomalies ranked by their potential scientific application. 

The Fink broker \citep{fink} anomaly detection pipeline is presented in this work. Our goal is to showcase the machine learning, delivery channels and, specially, the domain expert engagement necessary to transform interesting anomaly candidates into new scientific discoveries, with their place established within the current state of the art. The structure of the paper is as follows. 
In Section~\ref{sec:ztf_fink} we describe the ZTF alert stream and the Fink broker infrastructure. 
Section~\ref{sec:ad_pipeline} presents the anomaly detection pipeline, including the feature extraction, Isolation Forest (IF) model, and the anomaly notification system. 
The results obtained during the first year of operation are presented in Section~\ref{Results}, where we discuss several representative anomalies and the main classes of objects detected by the module. 
In Section~\ref{sec:discussion} we discuss limitations of this approach, contaminants in the anomaly stream, and the role of expert feedback through the citizen science model. 
Finally, the main conclusions are summarized in Section~\ref{sec:conclusions}.

\section{ZTF alerts and the Fink broker}
\label{sec:ztf_fink}

As part of the ZTF nightly operations, after each exposure a difference is computed, by the telescope alert processing pipeline \citep{bellm2019,patterson2019}, between the current image and a co-added template of the same position. An alert is generated for each point in the difference image showing a signal stronger than 5$\sigma$ above the background level\footnote{\url{https://irsa.ipac.caltech.edu/data/ZTF/docs/ztf_explanatory_supplement.pdf}}.  The alert package contains photometric information about the current detection, metadata enabling the characterization of the observation conditions, 3 image stamps (science, template and difference) and a 30 day photometric history of that location. Originally, the public ZTF alert stream contained only $zg$ and $zr$-bands, with the $zi$-band only recently being introduced. Alerts are delivered to brokers systems, less than 20 minutes after being produced \citep{patterson2019}. 

Fink \citep{fink} is a community-driven astronomy broker which has been processing the ZTF alert stream since November 2019. 
It has a modular structure, designed to enable the smooth incorporation of filters and science modules, developed by domain experts. Thus allowing users to impose their own constraints and definitions while selecting a sub-stream. Data can be accessed through the webportal\footnote{\url{https://ztf.fink-portal.org/}}, data transfer\footnote{\url{https://ztf.fink-portal.org/download}} and cross-match services\footnote{\url{https://ztf.fink-portal.org/xmatch}}, as well as by real time Kafka streams\footnote{\url{https://doc.ztf.fink-broker.org/en/latest/services/livestream/}} and instant messaging bots\footnote{\url{https://doc.ztf.fink-broker.org/en/latest/broker/filters/}}. Domain experts have total autonomy to define their scientific goals, data specifications, and ways in which to interact with the broker. 

Since the start of ZTF operations, Fink has processed more than 230 million ZTF alerts, with more than half of these being associated to a classification, either via machine learning models or cross-match with known catalogs. 

\section{Anomaly detection pipeline}
\label{sec:ad_pipeline}
In this section we describe the Fink anomaly detection pipeline which includes the feature extraction, the Isolation Forest model used to identify anomalies, and the notification system for reporting high-interest candidates.

\subsection{AD features module}
\label{sec:ad_features}

The \texttt{AD features module}\footnote{\url{https://github.com/astrolabsoftware/fink-science/tree/master/fink_science/ztf/ad_features}} forms the first step of the Fink anomaly detection pipeline. Its purpose is to transform the photometric time series of alerts into a set of features suitable for machine-learning models. These features quantitatively characterize the variability and shape of the light curves, providing a compact representation of their temporal behavior.

The light curve data are processed using the \texttt{light-curve}\footnote{\url{https://github.com/light-curve/light-curve-python}} package, a time-series feature extraction tool developed as part of the SNAD pipeline \citep{2021MNRAS.502.5147M,lcpaper}. Features are extracted from $zg$ and $zr$-bands.
The extracted feature set consists of 22 feature extractors producing 26 numerical features in total, which have demonstrated good performance in the analysis of transients and variable stars (e.g.,~\citealt{2023A&A...672A.111P, 2024MNRAS.533.4309V,2024arXiv241021077M}). These include features such as amplitude, mean, standard deviation, Stetson K coefficient etc. The complete list and mathematical definitions of all features are provided in Appendix~\ref{sec:features}.

Feature extraction is performed either from the difference magnitude or from the apparent (DC) magnitude, depending on the presence of a nearby source in the reference image PSF-catalog within 30\arcsec. 
If no nearby source is detected, the difference magnitude is used directly. Otherwise, the apparent magnitude is reconstructed by adding the reference flux to the difference flux. 
This approach unifies the representation of observed light curves since,   
for variable stars, the difference magnitude is not astrophysically meaningful, while for transients, the difference magnitude usually corresponds to the apparent magnitude since the host galaxy is faint or absent.

The resulting feature vectors are passed to the anomaly detection model described in Section~\ref{sec:anomaly_detection}.

\subsection{AD module}
\label{sec:anomaly_detection}

We use the Isolation Forest technique~\citep{Liu2008} as the core anomaly detection method\footnote{\url{https://github.com/astrolabsoftware/fink-science/tree/master/fink_science/ztf/anomaly_detection}} (see also Section~\ref{CSM}). Since the features used by the module are not multicolor and are computed independently for each photometric passband ($zg$ and $zr$), we employ two separate IF models --- one per passband. This design allows us to handle cases in which only a limited number of observations, or none at all, are available in a given passband.

In our implementation, lower anomaly scores correspond to more anomalous behavior. When observations are available in both passbands, we define the final anomaly score of an object as the minimum of the two individual scores.

To build the initial IF model (\texttt{Base model}), we use archival ZTF alert data. Over the lifetime of the module, the \texttt{Base model} has been updated several times to account for the development phase of the pipeline, changes in the ZTF survey cadence between Phase~I and Phase~II, and seasonal variations in the data stream.

The latest version of the model was deployed on 11~December 2025. It was trained on 841\,200 alerts received by the Fink broker between 1~April and 31~July 2025, each containing at least four photometric measurements in each of the $zg$ and $zr$-bands.

\subsection{Anomaly notification module}
\label{sec:anomaly_notification}

The \texttt{Anomaly notification module}\footnote{\url{https://github.com/astrolabsoftware/fink-filters/tree/master/fink_filters/ztf/filter_anomaly_notification}} provides users with fast access to information about newly detected candidates in anomalies. After receiving anomaly scores for all alerts from the processed night, it generates data packages that are transmitted to the internal message-distribution modules for Slack and Telegram, and simultaneously exposes the same information through the Fink REST API~\footnote{\url{https://ztf.api.fink-portal.org/}}.

In order to run the Slack and Telegram notification services, we use Yandex Cloud\footnote{\url{https://yandex.cloud}} infrastructure. The notification services function as HTTP endpoints and are hosted on two virtual servers located in different availability zones.

\begin{figure}
    \centering
    \includegraphics[width=0.8\linewidth]{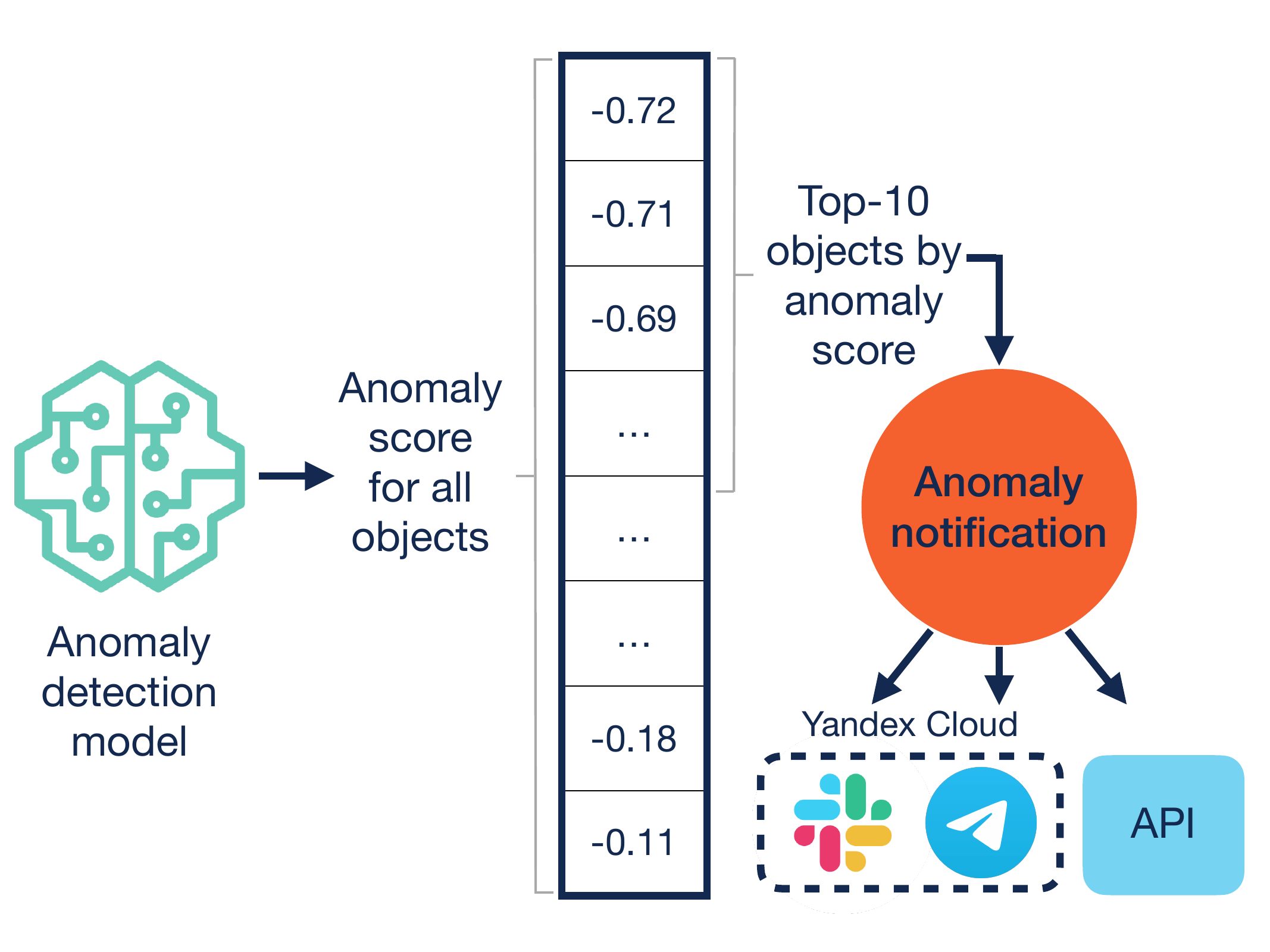}
    \caption{Schematic illustration of the anomaly notification process. Each night, the pipeline orders all alerts by their anomaly score. The \texttt{Anomaly notification} module selects the top-10 most anomalous objects and delivers them to the expert through Slack and Telegram messengers, as well as via the Fink API.}
    \label{fig:notification_scheme}
\end{figure}

The entire dataframe processing within the \texttt{Anomaly notification} module is implemented using \texttt{Apache Spark}\footnote{\url{https://spark.apache.org/docs/latest/}}.
The structure of the module is shown in Fig.~\ref{fig:notification_scheme}.

The Slack and Telegram messengers broadcast top-10 anomaly candidates, allowing convenient accumulation and browsing of notifications. Each notification contains the following information:

\begin{itemize}
    \item ZTF alert identifier (ID);
    \item the nearest ZTF data release object within $1^{\prime\prime}$ (DR OID);
    \item galactic coordinates in degrees (GAL);
    \item equatorial coordinates in degrees (EQU);
    \item alert time (UTC);
    \item real-bogus quality score from ZTF (range: 0-1, where values closer to 1 indicate higher reliability);
    \item anomaly score from the AD model;
    \item detection frequency --- how often the object appeared among the top-10 anomalies over the last 90~days.
\end{itemize}

\begin{figure*}
\centering
\includegraphics[width=1\linewidth]{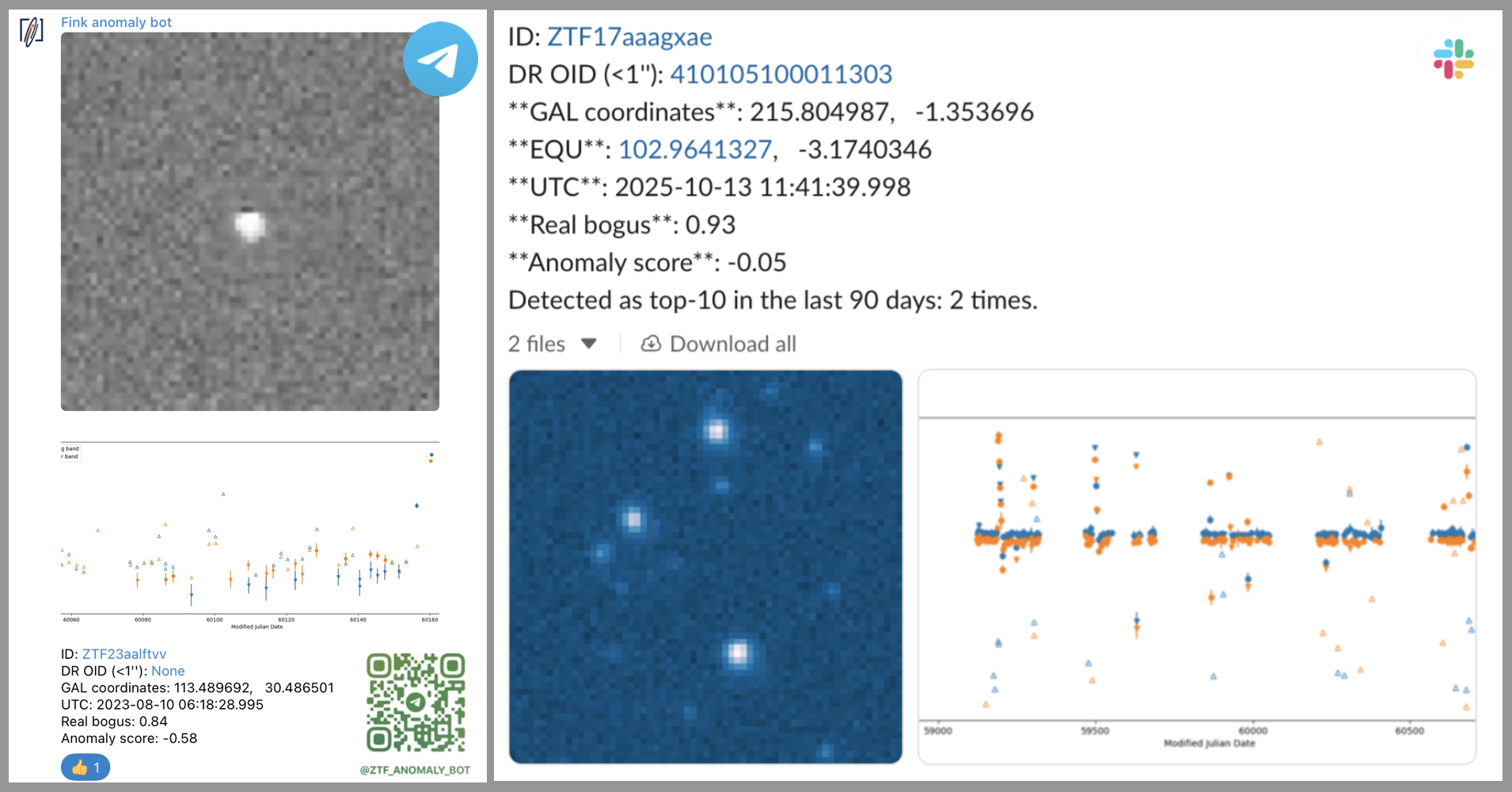}
\caption{Example of a Telegram (left) and Slack (right) notification showing one of the top-10 anomaly candidates with science image cutout, light curve in difference magnitudes, and alert parameters.}
\label{fig:notification}
\end{figure*}

Additionally, each alert includes a cutout of the science image and the Fink light curve in difference magnitudes. Examples of the notifications are presented in Fig.~\ref{fig:notification}.

\section{Results}
\label{Results}

The Fink \texttt{AD module} has been in operation since 25 January 2023. In this section, we present the results obtained during its first year of activity. We first discuss selected high-interest anomalies that triggered dedicated photometric and/or spectroscopic follow-up observations, and then summarize the main classes of objects and contaminants in the anomaly candidates.


\subsection{Selected anomalies}


\subsubsection{Fink J062452.88+020818.3 -- third WZ Sge-type object in AM CVn stars}
\label{AT2023awt}

\begin{figure*}
    \centering
    \includegraphics[width=0.8\linewidth]{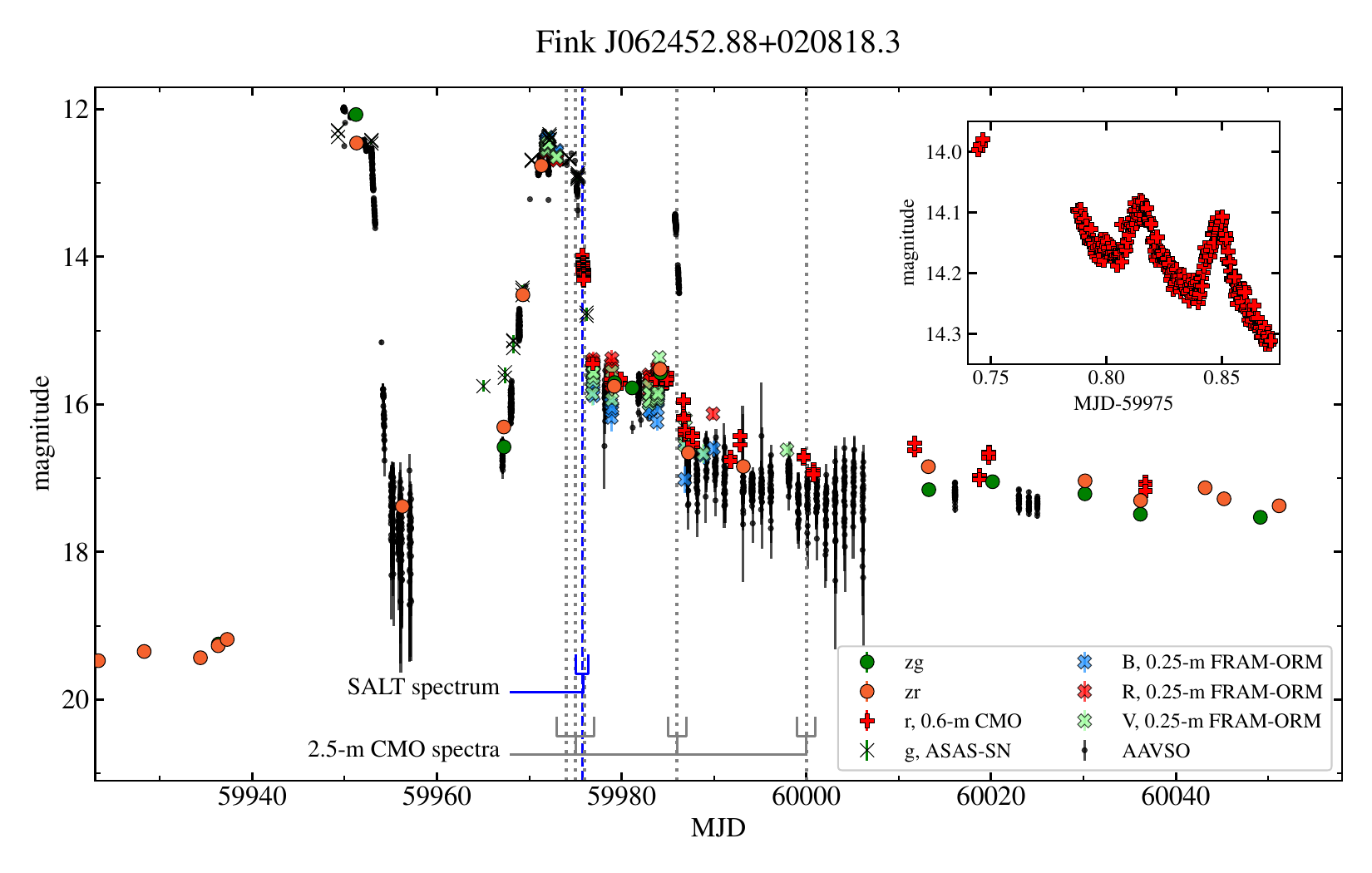}
    \caption{Composite light curve for Fink J062452.88+020818.3. The dashed lines denote moments when spectra were obtained. The inset plot shows the 31 January 2023 light curve segment with prominent superhumps, used to estimate the superhump period $P_{\mathrm{sh}} = 0.032(3)$~d.}
    \label{fig:AT2023awt_lc}
\end{figure*}

During its first night of operation (25 January 2023), the Fink \texttt{AD module} triggered the alert  corresponding to ZTF23aaaatwl. At the time of detection, the object was undergoing a superoutburst with an amplitude of 7.6$^m$ in $zr$-band. Archival data show that the activity began in early January 2023 and exhibited a complex double superoutburst structure,  as observed previously in several known WZ~Sge-type dwarf novae (Fig.~\ref{fig:AT2023awt_lc}).

\begin{figure*}
    \centering
    \includegraphics[width=0.8\linewidth]{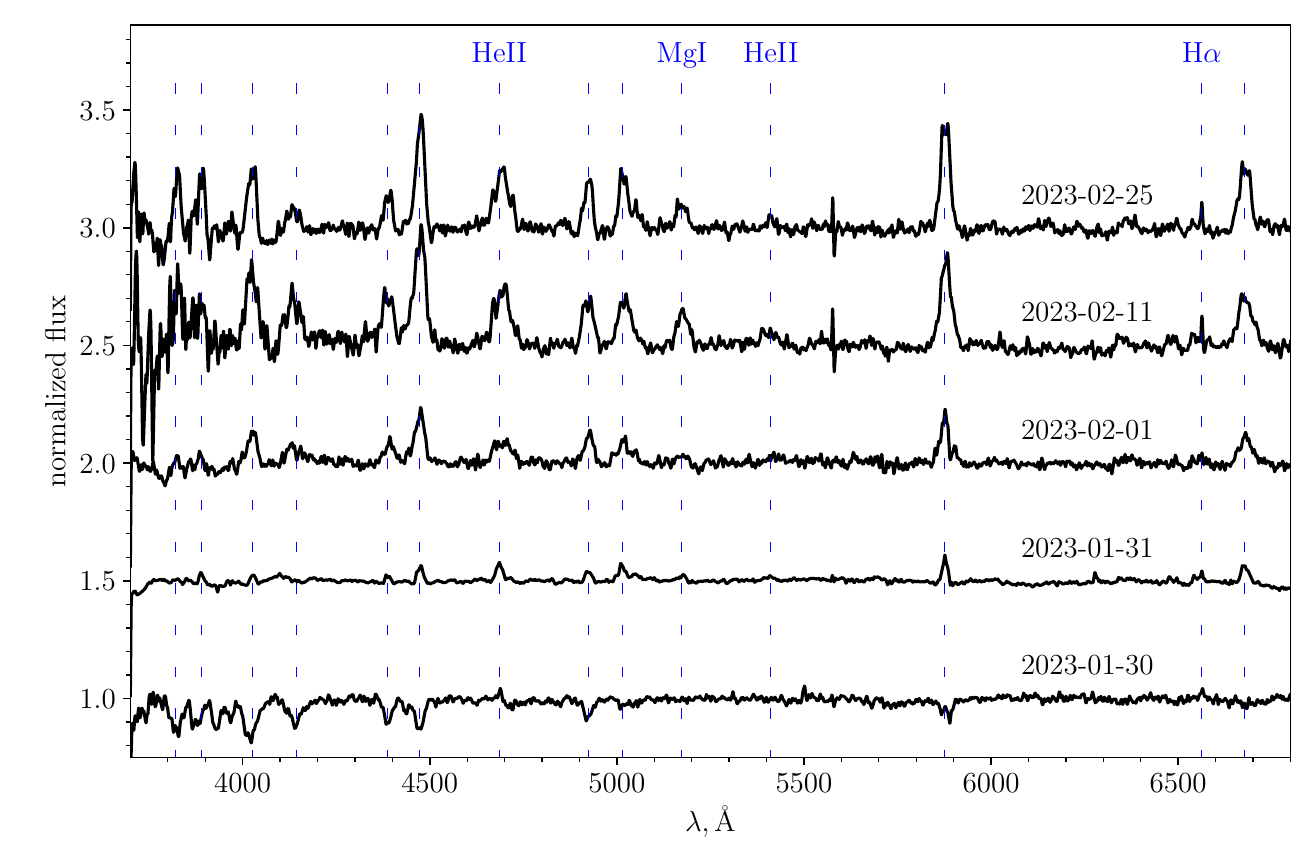}
    \caption{Evolution of nightly averaged spectra of Fink J062452.88+020818.3 from the 2.5-m CMO SAI MSU telescope. Most of the marked lines, except explicitly labeled, are neutral helium lines.}
    \label{fig:AT2023awt_spectra}
\end{figure*}

\begin{figure}
    \centering
    \includegraphics[width=0.7\linewidth]{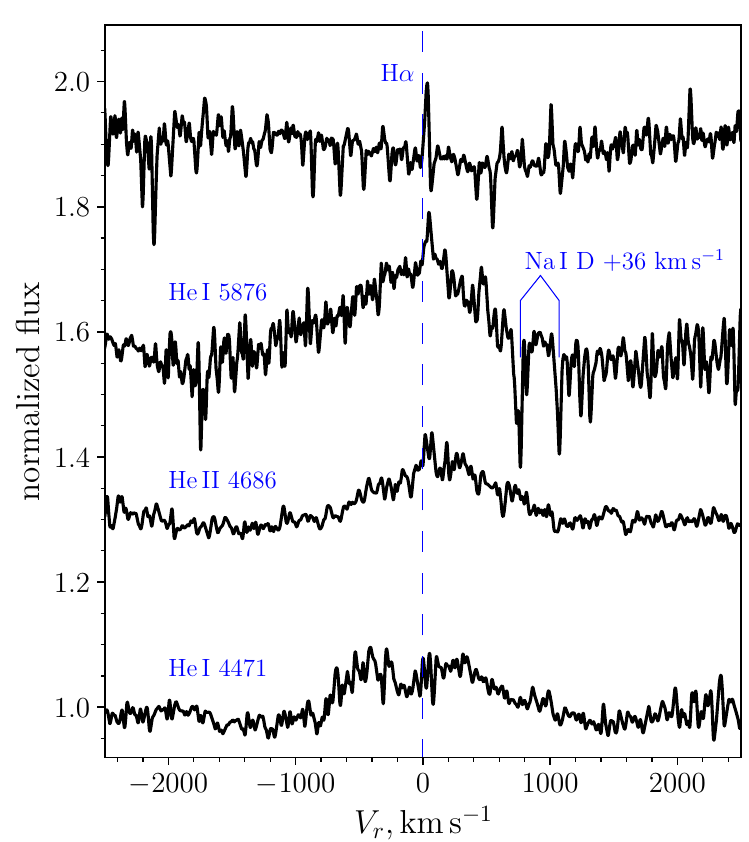}
    \caption{Profiles of some characteristic lines in the Fink J062452.88+020818.3 SALT spectrum acquired on 31 January 2023. The radial velocity is shown in the barycentric rest frame. The profiles are shifted vertically by 0.3 for convenience.}
    \label{fig:SALT}
\end{figure}

The first outburst was discovered on 5 January 2023 by Tomoo Kato\footnote{\url{http://cbat.eps.harvard.edu/unconf/followups/J06245297+0208207.html}, see also vsnet-alert 27282, 27287, 27296, 27303, 27342, 27351, 27353, 27354,  27368} with last non-detection on 4 January 2023 (limiting magnitude 14.0$^m$). The object was independently detected by the All-Sky Automated Survey for Supernovae (ASAS-SN;~\citealt{2014ApJ...788...48S,2017PASP..129j4502K}) and designated ASASSN-23ac/PNV~J06245297+0208207.

The Fink \texttt{AD module} subsequently reported a second outburst of this source. We submitted the event to the Transient Name Server\footnote{\url{https://www.wis-tns.org/}} (TNS)
 as AT~2023awt \citep{2023TNSTR.212....1B} and assigned it the internal Fink designation Fink~J062452.88+020818.3.

Initial spectroscopic observations, obtained on 6 January 2023, revealed peculiar properties of the system \citep{2023ATel15849....1M}. The spectrum has blue continuum and is dominated by numerous He\,I emission lines, indicating that the object is a helium dwarf nova (AM~CVn; ~\citealt{2005ASPC..330...27N}).

To reveal the physical origin of Fink~J062452.88+020818.3, we conducted its photometric monitoring starting from 31 January 2023 with the 0.6-m ASA RC600 telescope~\citep{2020ARep...64..310B} at the Caucasus Mountain Observatory of the Sternberg astronomical institute of Lomonosov Moscow State University (CMO SAI MSU; \citealt{2020gbar.conf..127S}). Observations were obtained in the Sloan $r$-band with exposures adjusted according to the brightness of the object. The frames were processed using the \texttt{VaST} software~\citep{2018A&C....22...28S}. Additional monitoring was performed with the Moravian Instruments G2-1600 CCD mounted on the 0.25-m FRAM-ORM telescope at the Roque de los Muchachos Observatory (La Palma). The data were acquired in the Johnson–Cousins $B$, $V$, and $R$ bands with 120-s exposures and automatically reduced using a dedicated Python pipeline based on the \texttt{STDPipe} package~\citep{stdpipe}. The reduction procedure included bias and dark current subtraction, flat-fielding, cosmic ray removal, astrometric calibration, aperture (with a 5-pixel radius) photometry, and photometric calibration using synthetic photometry based on Gaia DR3 low-resolution XP spectra~\citep{gaiadr3syn}. The combined light curve, incorporating our observations together with publicly available data from the American Association of Variable Star Observers (AAVSO), ASAS-SN, and ZTF, is presented in Fig.~\ref{fig:AT2023awt_lc}.

To investigate possible periodic variability, we analysed the high-cadence observations obtained with the 0.6-m ASA RC600 telescope. The search for periodicities was performed using the Deeming method~\citep{1975Ap&SS..36..137D}, as implemented in the \texttt{WINEFK}\footnote{\url{http://www.vgoranskij.net/software/}} package developed by Dr.\ V.\ P.\ Goranskij. Unfortunately, only one relatively long dataset exhibiting prominent superhumps, obtained on 31 January 2023, was suitable for estimating the photometric period. From this dataset, we derive an approximate superhump period of $P_{\rm sh}=0.032(3)$~d. By analogy with the double-superoutburst behavior observed in NSV~1440, where \citet{2019PASJ...71...48I} demonstrated that ordinary superhumps are present during the second superoutburst, we interpret the brightness modulations of Fink~J062452.88+020818.3 observed on 31 January 2023 (during the second superoutburst) as ordinary superhumps.

In addition to photometry, we obtained 46 spectra over five nights between 30 January and 25 February 2023 with the Transient Double-beam Spectrograph (TDS; \citealt{2020AstL...46..836P}) mounted on the 2.5-m telescope of CMO SAI MSU. The typical exposure time was 300~s, and a narrow 1\arcsec slit was used to improve radial-velocity accuracy. Due to the low signal-to-noise ratio ($\sim$10 in a faint state) of individual spectra, we present only the averaged evolution of the spectra in Fig.~\ref{fig:AT2023awt_spectra}, however, based on 20 spectra obtained on 25 February 2023, a periodic variability of the radial velocities of the helium emission lines can be suspected on timescales of approximately 40 minutes.

A high-resolution spectrum was also acquired with the 11-m South African Large Telescope (SALT; \citealt{2006SPIE.6267E..0ZB}) on 31 January 2023. This spectrum was used to examine the detailed structure of the line profiles (Fig.~\ref{fig:SALT}). The helium-dominated spectra with double-peaked emission-line profiles, together with their temporal evolution following the superoutburst, are characteristic features of AM~CVn systems. In addition to the typical AM CVn spectrum, weak narrow components are seen in the H$\alpha$ line and in some strong helium lines (see Fig.\,\ref{fig:SALT}), which probably originate in a circumstellar matter.

Overall, the combined spectroscopic and photometric analysis indicates that Fink~J062452.88+020818.3 is an AM~CVn star of the WZ~Sge type \citep{2026PASJ...78..199K}. Using the AAVSO International Variable Star Index (VSX; \citealt{Watson_2006}) type definitions\footnote{\url{https://vsx.aavso.org/index.php?view=about.vartypes}} (UGWZ/IBWD) for this group of objects, we claim that Fink~J062452.88+020818.3 is the third reliably classified object of this kind after NSV~1440 \citep{2019PASJ...71...48I} and V451~Boo\footnote{\url{http://ooruri.kusastro.kyoto-u.ac.jp/mailarchive/vsnet-alert/22181}},\footnote{\url{http://ooruri.kusastro.kyoto-u.ac.jp/mailarchive/vsnet-alert/22186}}.


\subsubsection{SN 2023mtp -- supernova with precursor}
\label{sn2023mtp}

On 10 August 2023 at 06:18:29 UTC, the \texttt{AD module} triggered a new transient event, subsequently designated as SN~2023mtp (ZTF23aalftvv).

\begin{figure}
    \centering
    \includegraphics[trim={0cm 0cm 0 0cm},clip,width=0.8\linewidth]{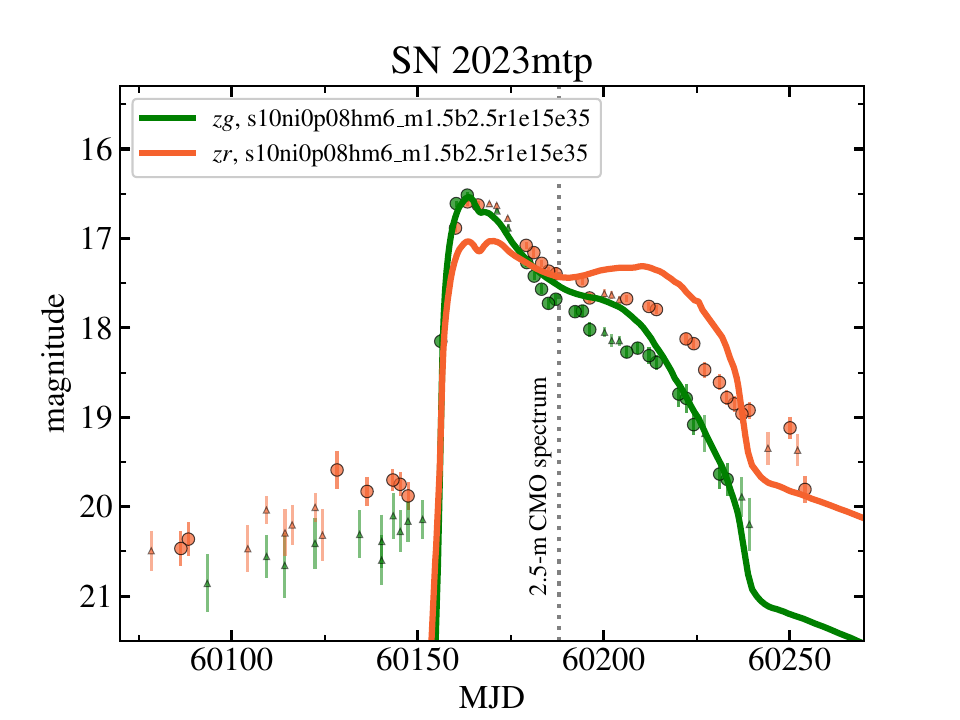}
    \caption{ZTF light curves of SN~2023mtp (circles denote good-quality measurements, triangles correspond to bad-quality measurements) together with the best-fit \texttt{STELLA} models from the grid of \citet{2023PASJ...75..634M}. The photometric data are corrected for the line-of-sight Galactic reddening \citep{2011ApJ...737..103S}. \textit{Note.} In the Fink broker, an alert is classified as \texttt{valid} if it passes the quality cuts, and as \texttt{badquality} otherwise. The quality cuts are defined as $(\texttt{rb} \geq 0.55)\ \&\ (\texttt{nbad} = 0)$, where \texttt{rb} is the real--bogus score and \texttt{nbad} is the number of bad pixels within a $5\times5$ pixel region around the transient.}
    \label{fig:2023mtp_lc}
\end{figure}

SN~2023mtp was first detected by ZTF on 14 May 2023 at 09:48:49 UTC, corresponding to a precursor activity preceding the main explosion~\citep{2023TNSTR1631....1F}. The main SN outburst occurred about 2.5~months later and is characterized by a rapid rise to maximum light (see Fig.~\ref{fig:2023mtp_lc}).

A possible host galaxy is a low-surface-brightness galaxy located at
$\mathrm{R.A.}=258.5400^\circ$, $\mathrm{Dec.}=81.0754^\circ$.
According to the the DESI Legacy Imaging Surveys\footnote{\url{https://www.legacysurvey.org/viewer/}} \citep{2019AJ....157..168D} DR10 catalogue, the galaxy has magnitudes $g=21.09^m$, $r=20.64^m$, and $z=20.45^m$, and a photometric redshift $z_{\rm ph}=0.065\pm0.034$ (DR9 Photo-z)~\citep{2021MNRAS.501.3309Z}.

The first publicly available spectrum of SN~2023mtp was obtained with the 2.5-m Isaac Newton Telescope on 17~August 2023 at 23:25:07~UTC, about 11~days after maximum light. As reported by \citet{2023TNSCR2014....1O}, the spectrum shows a hot, blue continuum with only tentative broad features around $6000~\AA$ and $6600~\AA$, and no clear emission or absorption lines. Approximately one week later, an additional spectrum was obtained with the LCO 2-m Faulkes Telescope North~\citep{2023TNSCR2213....1F}. This spectrum still exhibits a blue, nearly featureless continuum, but a broad H$\alpha$ feature becomes visible. 

\begin{figure*}
\centering
\includegraphics[width=\textwidth]{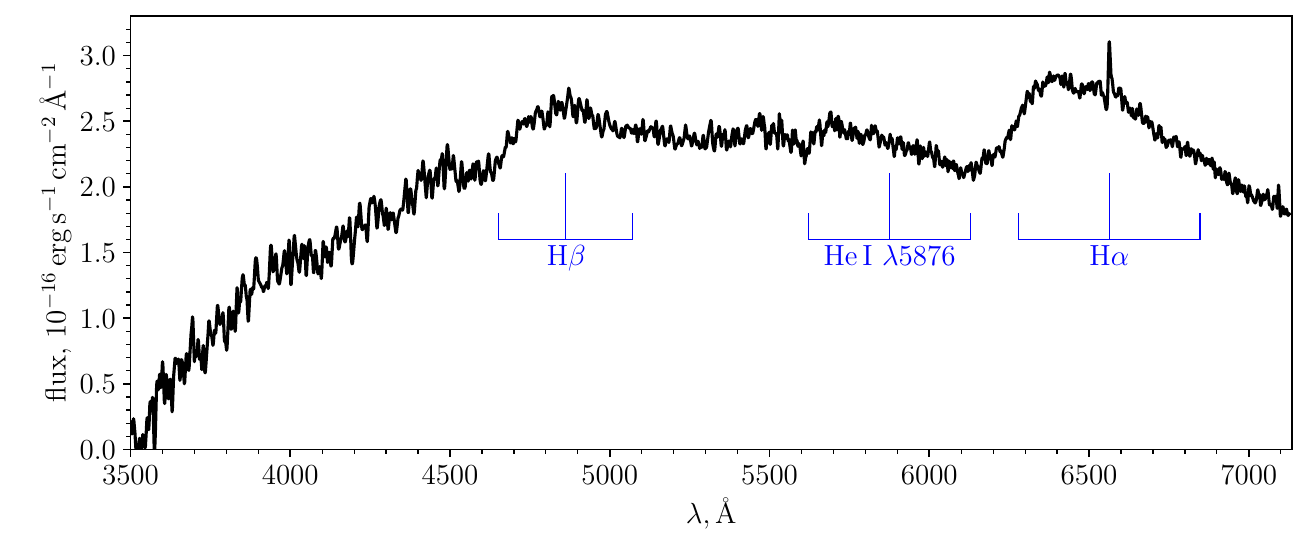}
\caption{Spectrum of SN~2023mtp obtained with the 2.5-m CMO telescope on 1 September~2023~\citep{2023TNSCR2494....1D}.}
\label{fig:ZTF23aalftvv}
\end{figure*}

To further clarify both the classification and the redshift, we obtained another spectrum on 1~September 2023 at 23:47:10~UTC with the 2.5-m CMO SAI MSU telescope (Fig.~\ref{fig:ZTF23aalftvv}; \citealt{2023TNSCR2494....1D}). The strong flux loss in the blue part of the spectrum is caused by atmospheric dispersion, as the slit was not aligned with the parallactic angle. The spectrum clearly reveals a set of Balmer lines and He features, allowing us to estimate a redshift of $z\simeq0.04$.

In addition, a fourth spectrum is available via the TNS \citep{2024TNSCR3189....1S}, obtained on 5~October~2023 at 05:23:49~UTC (approximately 59 days after maximum) with the Palomar 5.1-m telescope.

We performed spectral classification of SN~2023mtp based on four publicly available spectra using \texttt{SNID}~\citep{2007ApJ...666.1024B} with an extended template library~\citep{2025RNAAS...9...78M}. By specifying the age limits for each spectrum, the best-matching solutions correspond to Type~IIb, Type~IIP, and Type~IIn supernovae. However, no single template, nor any individual SN within a given subclass, is able to reproduce all four spectra simultaneously. The best-fitting templates for each epoch are presented in Fig.~\ref{fig:2023mtp_snid}.

Interestingly, for two of the spectra the formally best \texttt{SNID} matches correspond to TDE templates. Nevertheless, a TDE interpretation is disfavoured by the overall photometric evolution.

The presence of hydrogen lines in the spectra, together with the pronounced precursor activity, may indicate a core-collapse supernova of Type~IIn. A defining spectroscopic signature of SNe~IIn is the presence of narrow Balmer emission lines, often showing a narrow component (sometimes with a P~Cygni profile) superimposed on a broader base, tracing strong interaction between the supernova ejecta and a dense circumstellar medium. 
In the 2.5-m CMO spectrum, the H$\alpha$ profile shows a weak narrow component with a P Cygni-like structure, superimposed on a broader feature (also marginally visible in the P200 spectrum; Fig.~\ref{fig:2023mtp_snid}). However, other lines do not exhibit comparable narrow components, and the overall spectral appearance is not fully consistent with that of a classical SN~IIn.

In many known SNe~IIn, the progenitors exhibit strong photometric variability, including recurrent outbursts, during the years preceding the explosion. The peak absolute magnitudes of such pre-explosion outbursts typically range between $-11.5^m$ and $-15^m$, and their durations span from several weeks to a few years (e.g., \citealt{2024A&A...686A.231R}).

In contrast, the precursor detected for SN~2023mtp is significantly more luminous, with an absolute $zr$ magnitude of about $-16.3^m$, while its color during the precursor phase, $g-r\simeq0.5^m$, is comparable to those observed in known SN~IIn precursors.

We also attempted to reproduce the light curve using a grid of one-dimensional hydrodynamical core-collapse SN models computed with the \texttt{STELLA} code \citep{1998ApJ...496..454B,2006A&A...453..229B}. This grid comprises 228\,016 synthetic explosion models and spans a wide range of pre-supernova parameters, i.e. progenitor mass $M_{\rm ZAMS}\in[10,18]$~$M_\odot$, explosion energy $E_{\rm exp}\in[0.5,5.0]$~Bethe, $^{56}$Ni mass $M_{\rm ^{56}Ni}\in [0.001,0.3]$~$M_\odot$, mass-loss rate $\dot{M}\in[10^{-5.0},10^{-1.0}]$~$M_\odot/yr$ with a wind velocity of 10 km~s$^{-1}$, circumstellar medium (CSM) density structure $\beta\in[0.5,5.0]$, and CSM radius $R_{\rm CSM}\in[10^{14},10^{15}]$~cm~\citep{2023PASJ...75..634M}. However, none of the models provide a satisfactory fit to the observed light curve of SN~2023mtp. The best-fitting model is shown in Fig.~\ref{fig:2023mtp_lc}, the parameters of the model are $M_{\rm ZAMS}=10~M_\odot$, $M_{\rm ^{56}Ni}=0.08~M_\odot$, $\dot{M}=10^{-1.5}$~$M_\odot/yr$, $\beta=2.5$, $R_{\rm CSM}=10^{15}$~cm,  $E_{\rm exp}=3.5$~Bethe. 

Overall, both the photometric and spectroscopic behavior of SN~2023mtp appear unusual and cannot be readily explained within a single standard supernova scenario.


\subsubsection{Fink J222324.32+744222.0 -- UX Ori-type star}

\begin{figure*}
    \centering
    \includegraphics[width=1\linewidth]{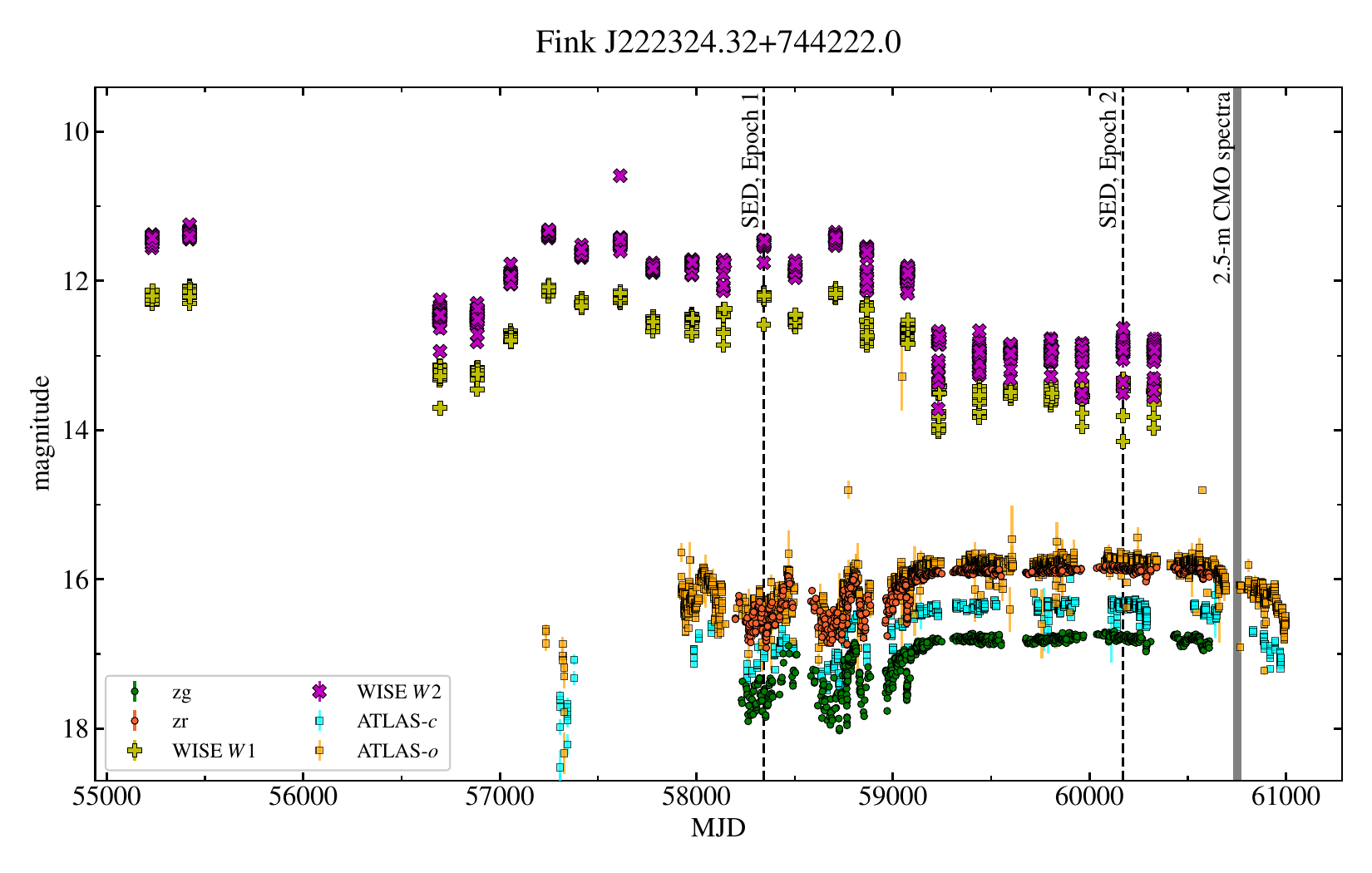}
    \caption{Multicolor light curve of Fink~J222324.32+744222.0 based on ATLAS (ATLAS-$c$, ATLAS-$o$), WISE ($W1$, $W2$), and ZTF DR23 ($zg$, $zr$) data. Black vertical dashed lines mark the epochs used for SED construction: Epoch 1 (MJD~$\simeq$~58344) and Epoch 2 (MJD~$\simeq$~60170). The grey shaded band indicates the time interval during which the spectra were obtained with the 2.5-m CMO telescope.}
    \label{fig:ZTF18abgpgxt-lc}
\end{figure*}

ZTF18abgpgxt was detected by the \texttt{AD module} on 19~October 2023 at 07:48:17 UTC. Its light curve exhibits a sharp increase in brightness by approximately 1$^m$ near MJD~$\simeq59070$ (from $\sim$17.6$^m$ to $\sim$16.8$^m$ in the $zg$-band and from $\sim$16.6$^m$ to $\sim$15.9$^m$ in the $zr$-band), followed by a prolonged plateau phase that lasted for almost 4 years (see Fig.\ref{fig:ZTF18abgpgxt-lc}). Based on these photometric characteristics, we initially classified it as an FU Ori-type variable and added to the AAVSO VSX database\footnote{\url{https://www.aavso.org/vsx/index.php?view=detail.top&oid=2388343}}. Since it was not present in any variable star catalogues, it was assigned a Fink name: Fink~J222324.32+744222.0.

\begin{figure}
    \centering
    \includegraphics[width=0.6\linewidth]{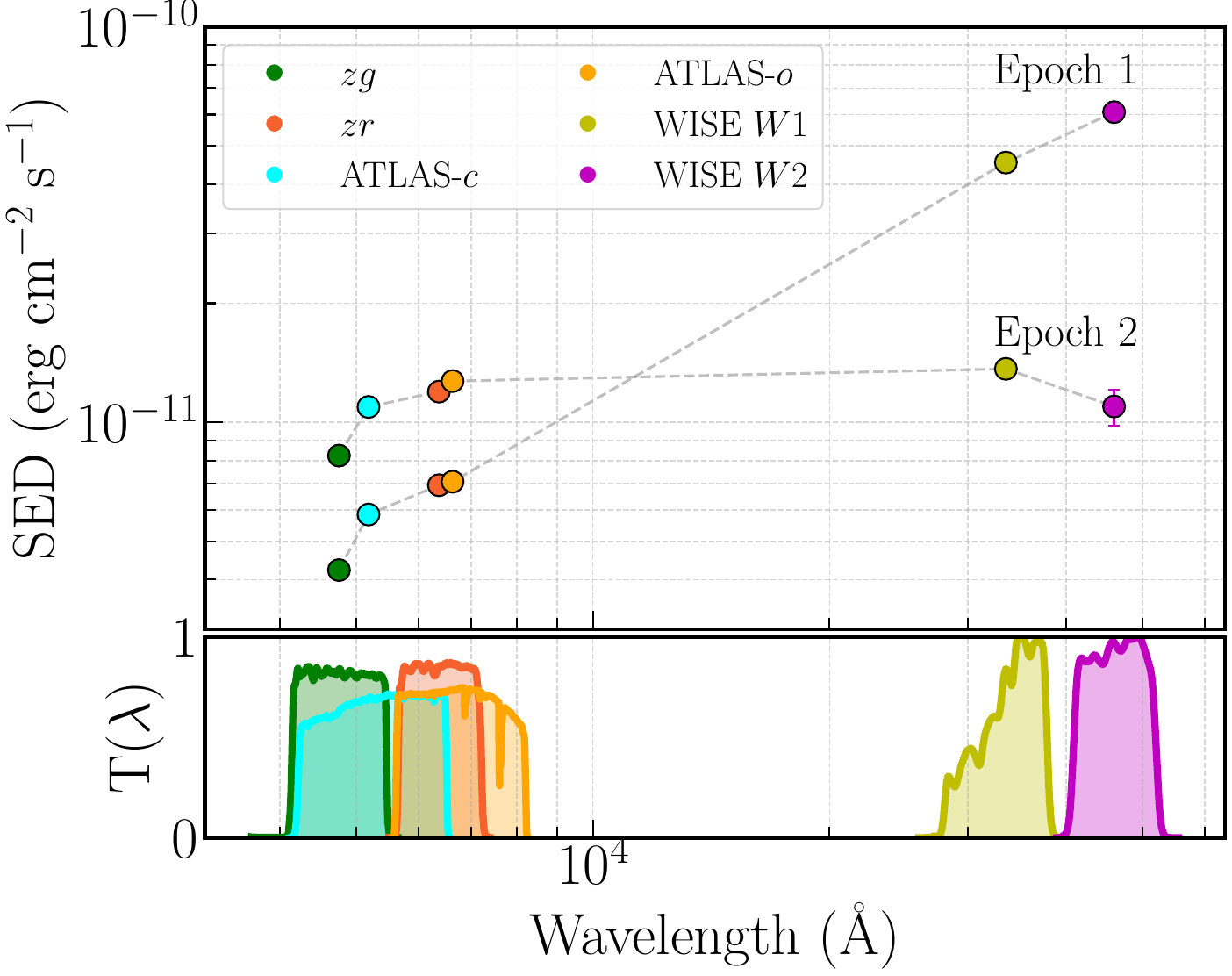}
    \caption{Top: spectral energy distributions for Fink~J222324.32+744222.0 plotted for two epochs, Epoch 1 (MJD~$\simeq$~58344) and Epoch 2 (MJD~$\simeq$~60170). Bottom: filter transmissions.}
    \label{fig:ZTF18abgpgxt-sed}
\end{figure}

Additionally, we constructed a multicolor light curve combining optical photometry from ZTF DR and ATLAS~\citep{2018PASP..130f4505T}, as well as infrared photometry from WISE~\citep{2010AJ....140.1868W}. The full light curve is shown in Fig.~\ref{fig:ZTF18abgpgxt-lc}. A noticeable anticorrelation is observed between the optical and infrared passbands: episodes of optical fading coincide with brightenings in the infrared, and vice versa. To investigate this further, we selected two epochs,  MJD$\simeq$58344 (Epoch~1) and MJD$\simeq$60170 (Epoch~2), and constructed spectral energy distributions (SEDs) at these times. As shown in Fig.~\ref{fig:ZTF18abgpgxt-sed}, the comparison clearly demonstrates a redistribution of energy between the optical and infrared diapason for two moments. This behavior is indicative of changes in the circumstellar environment, likely due to variations in dust geometry or composition. Such variability is commonly observed in UX~Ori-type stars and is often attributed to clumps in a circumstellar disk passing through the line of sight~\citep{1994A&A...292..165G, 1994AJ....108.1906H, 1997ApJ...491..885N}.

\begin{figure}
    \centering
    \includegraphics[width=0.6\linewidth]{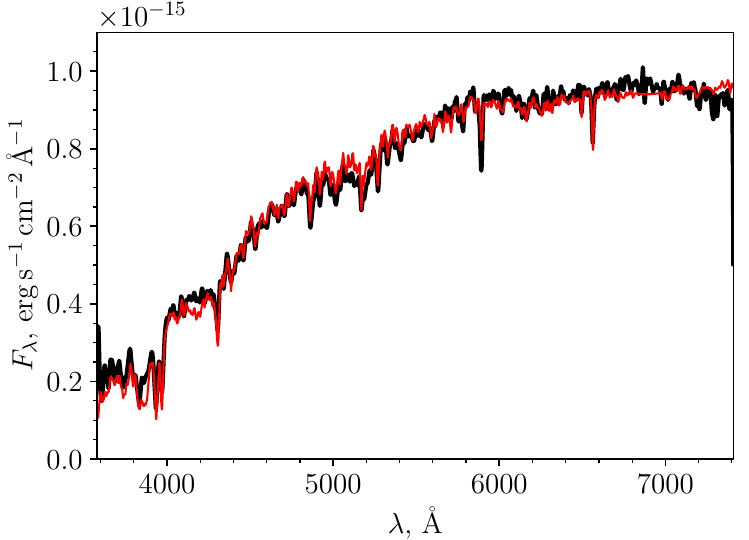}
    \caption{Spectrophotometric observation of Fink J222324.32+744222.0 (black curve) obtained on 2.5-m CMO telescope with a 10\arcsec slit width. The red curve shows a G2IV standard star spectrum from the \citet{1998PASP..110..863P} atlas, reddened by $A_V = 1.67^m$ ($R_V = 3.1$) using the standard extinction law of~\citet{1989ApJ...345..245C}.}
    \label{figure:g2}
\end{figure}

To further investigate the nature of Fink~J222324.32+744222.0 we performed follow-up spectroscopic observations  using the 2.5-m CMO SAI MSU telescope (see logs of observations in Table~\ref{SpecObs}). Spectral fitting with a stellar spectral flux library \citep{1998PASP..110..863P} suggests a spectral type of G2 with an extinction value of $A_V = 1.67^m$ (Fig.~\ref{figure:g2}) assuming $R_V = 3.1$ with the standard extinction law~\citep{1989ApJ...345..245C}.

\begin{table}
   \caption[]{Logs of spectroscopic observations of Fink~J222324.32+744222.0 with 2.5-m CMO SAI MSU telescope.}
   \label{SpecObs}
   $$ 
   \begin{array}{lccc}
      \hline
      \noalign{\smallskip}
      \text{Date} & \text{HMJD} & \text{Exposure [s]} & \text{Slit [\arcsec]} \\
      \noalign{\smallskip}
      \hline
      \noalign{\smallskip}
      \text{03-03-2025} & 60738.0173 & 4800 & 1  \\
      \text{24-03-2025} & 60759.0401 & 2400 & 1  \\
      \text{24-03-2025} & 60759.0593 & 600  & 10 \\
      \text{25-03-2025} & 60759.8494 & 4800 & 1  \\
      \text{29-03-2025} & 60763.9511 & 3600 & 1  \\
      \noalign{\smallskip}
      \hline
   \end{array}
   $$ 
\end{table}

\begin{figure*}
    \centering
    \includegraphics[width=0.8\linewidth]{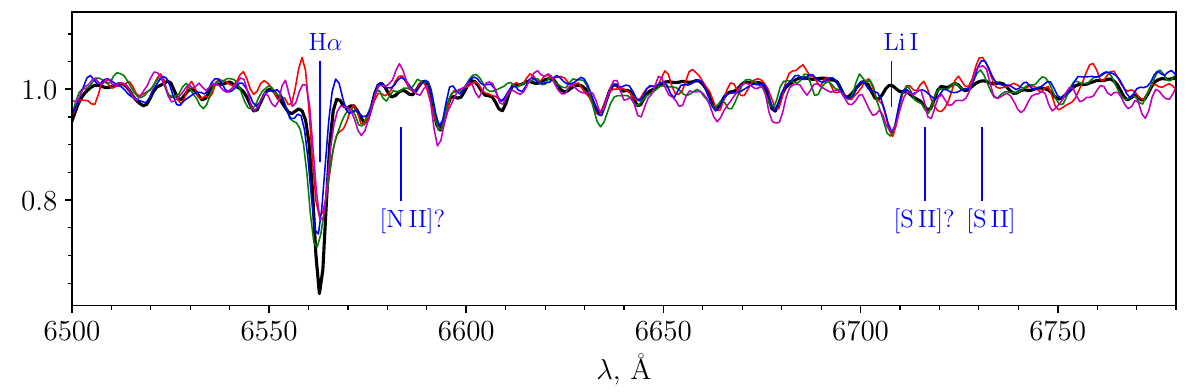}
    \caption{Spectra of Fink~J222324.32+744222.0 obtained with the 2.5-m CMO telescope. Colored lines represent different epochs: red -- HMJD~$\simeq$~60738.02, green -- HMJD~$\simeq$~60759.04, blue -- HMJD~$\simeq$~60759.85, and magenta -- HMJD~$\simeq$~60763.95. The black line shows the spectrum of the comparison star HD~120787.}
    \label{figure:ZTF18abgpgxt_sp}
\end{figure*}

We obtained 4 spectra with the narrow 1\arcsec slit to achieve better resolving power $R\sim2000$. First of all, these spectra show clear variability in H$\alpha$ line, see Fig~\ref{figure:ZTF18abgpgxt_sp}. To highlight peculiarities in the spectra of the object, we also obtained spectra of a few bright stars with spectral types G0 -- G5 and luminosity classes of IV -- V. The best match was found for HD~120787 (G3V), spectrum of which is plotted in Fig.~\ref{figure:ZTF18abgpgxt_sp} with the black line. Now, it becomes clear that the object has a weak variable emission component in H$\alpha$ line, the Li\,I 6708{\AA} line is greatly enhanced in comparison with HD 120787, and forbidden line [S\,II] 6731{\AA} is present in all spectra nearly at the stellar velocity ($\sim-20~$km\,s$^{-1}$). The second line [S\,II] 6716{\AA} is weaker, if present at all, therefore it indicates a relatively dense gas with $n_{\rm e}>10^3$ cm$^{-3}$. Emission in [N\,II] 6583.45{\AA} line can also be suspected, the lines [O\,I6300/6362\AA] have not been detected. Observed spectroscopic properties allow us to classify Fink~J222324.32+744222.0 as weak-line T Tauri star with UX Ori-type photometric activity.

Given this irregular variability, an alternative explanation is that Fink~J222324.32+744222.0 could be a long-period variable (LPV). The significant photometric changes and slow evolution over time resemble semi-regular or Mira-like variables. Another possibility is that the observed variations are caused by a dust-related event, such as an eclipse by circumstellar material. It is also possible that the star is an evolved system with residual circumstellar dust, potentially from past mass-loss episodes or an undetected binary companion undergoing interaction. The high total extinction observed toward the object further supports the presence of circumstellar material.  

To sum up, the spectral features and observed color excess suggest that the detected anomaly corresponds to a G2-type main-sequence star exhibiting irregular variability in optical and infrared passbands.


\subsubsection{Fink J042203.10+362318.7 -- flaring M-dwarf}

Among the objects identified as anomalous by the Fink pipeline, we discovered a flaring M-dwarf Fink~J042203.10+362318.7 (ZTF20aahjjjm). 

Flares on M dwarfs provide key insight into stellar magnetism and high-energy processes in fully convective stars. Although most systematic flare studies rely on space-based missions, recent work shows that ground-based wide-field surveys, such as ZTF, can also provide valuable constraints on stellar activity, flare energetics, and their distribution across the Galaxy \citep{2024MNRAS.533.4309V,2026MNRAS.tmp..153L}.

\begin{figure}
    \centering
    \includegraphics[trim={0cm 0cm 0 0cm},clip,width=0.8\linewidth]{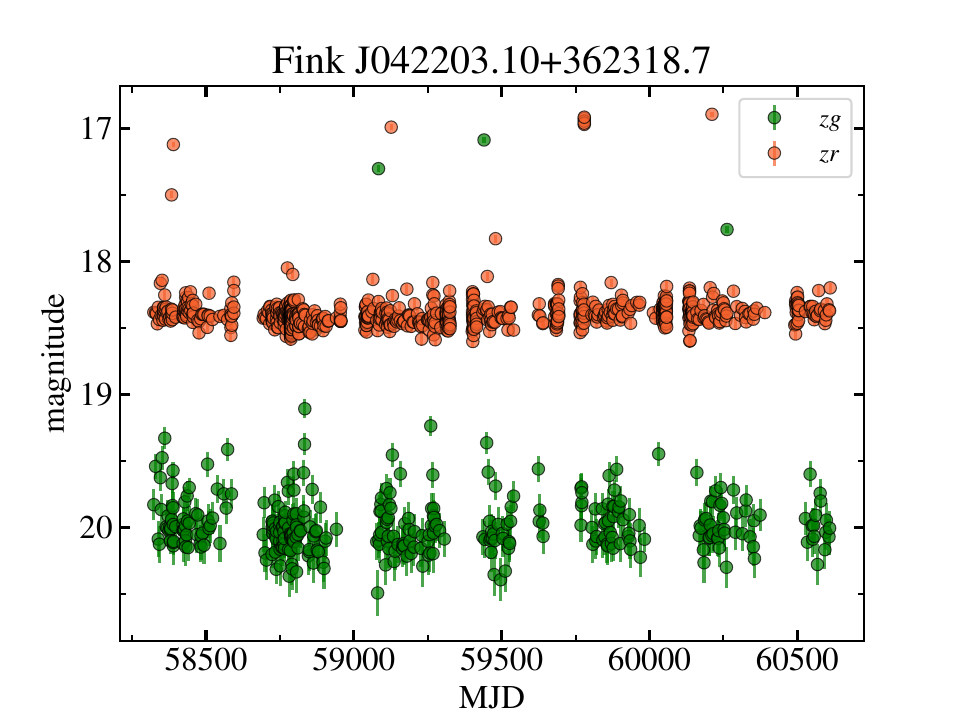}
    \caption{ZTF DR23 light curves of flaring M-dwarf Fink~J042203.10+362318.7. The photometric data are taken from the SNAD Viewer~\citep{2023PASP..135b4503M} and not corrected for the line-of-sight Galactic reddening.}
    \label{fig:m_dwarf}
\end{figure}

\begin{figure}
    \centering
    \includegraphics[trim={0.8cm 0cm 0 0cm},clip,width=0.8\linewidth]{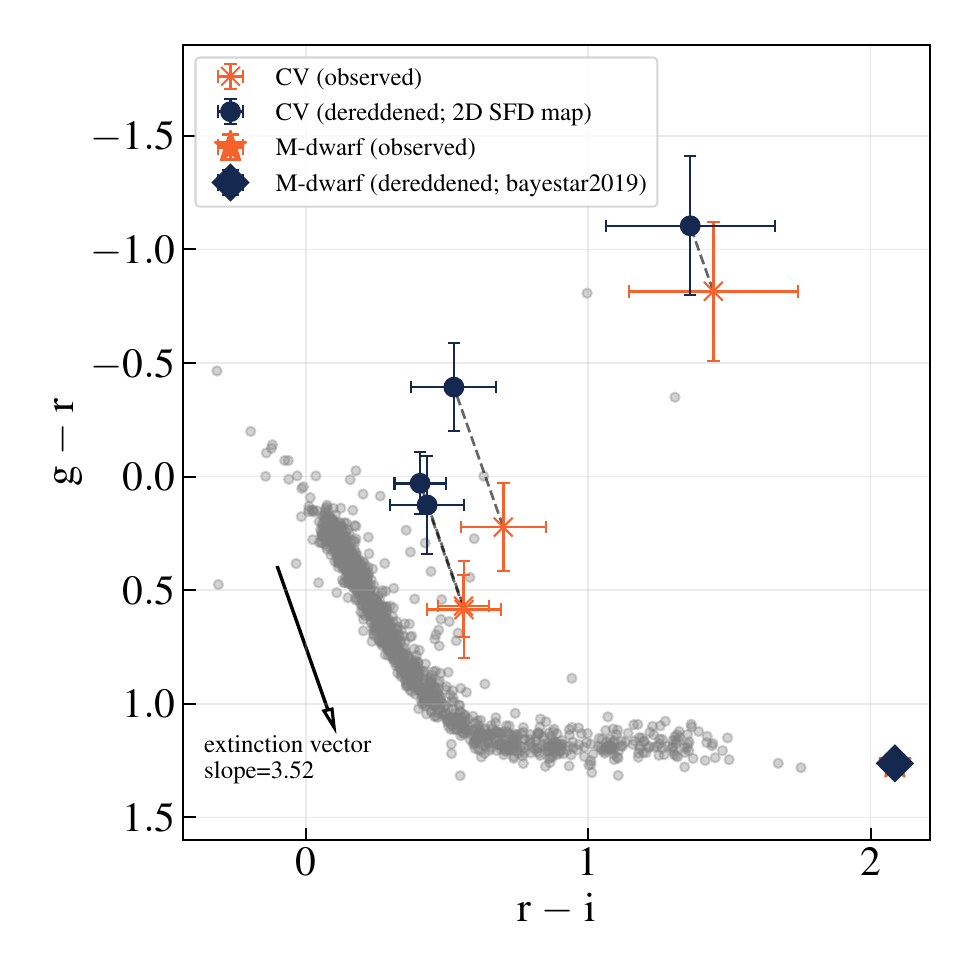}
    \caption{Color–color diagram showing dwarf novae Fink~J061603.51+080222.8, Fink~J015434.19+555025.6, Fink~J073518.75-092114.3, Fink~J040855.54+174554.2 and flaring M-dwarf Fink~J042203.10+362318.7 as observed (orange crosses) and after dereddening (blue circles). The stellar locus is constructed using stars in the vicinity of WD~1657$+$343. Photometry is taken from Pan-STARRS DR2.}
    \label{fig:cc_cv}
\end{figure}

The ZTF light curve of Fink~J042203.10+362318.7 exhibits recurrent short-duration flares typical of magnetically active M dwarfs (Fig.~\ref{fig:m_dwarf}). The source has a counterpart in Gaia with a photogeometric distance of $70.464^{+0.647}_{-0.652}$~\citep{2021AJ....161..147B}; according to three-dimensional dust maps~\citep{2019ApJ...887...93G}, the line-of-sight reddening is negligible, with $E(B-V)\simeq0$. Its position on the $(g-r)$-$(r-i)$ color-color diagram is fully consistent with the locus of late-type stars (Fig.~\ref{fig:cc_cv}).

\subsection{Dwarf novae}
\label{DNe}

Among the discoveries, we identified nine new cataclysmic variables. All of them were submitted to the AAVSO VSX and assigned internal Fink identifiers. We consider all these objects to be dwarf novae, i.e. U Geminorum type stars (UG), for two of them the classification is uncertain (see Table~\ref{tab:UG}). Multicolor ZTF light curves of the candidates are shown in~Fig.~\ref{fig:UG}.

\begin{table*}[h]
\centering
 \footnotesize
\begin{tabular}{|l|r|r|l|l|l|l|l|l|l|}
\hline
\textbf{Fink name} & \textbf{R.A.} & \textbf{Dec.} & \textbf{Type} & \textbf{$m^{\rm PS}_g$} & \textbf{$m^{\rm PS}_r$} & \textbf{$m^{\rm PS}_i$} & \textbf{Sep$''$}  & \textbf{ZTF ID} \\
\hline
Fink J061603.51+080222.8 & 94.01462 & $+$8.03967& UG & 22.690$\pm$0.116 & 22.120$\pm$0.072 & 21.562$\pm$0.055 & 0.105 & ZTF17aaagtdb \\ 
Fink J194741.57+223702.3 & 296.92321 & $+$22.61731& UG & 20.432$\pm$0.015 & 18.762$\pm$0.004 & 17.963$\pm$0.004 & 0.952 & ZTF18aaudzmj \\ 
Fink J203440.21+504800.8 & 308.66754 & $+$50.80022& UG: & --  & -- & --  & -- & ZTF18aayoyvo \\ 
Fink J015434.19+555025.6 & 28.64246 & $+$55.84047& UG & 22.846$\pm$0.180 & 22.261$\pm$0.116 & 21.701$\pm$0.059 & 0.108 & ZTF18abjcuhv \\ 
Fink J073518.75-092114.3 & 113.82812 & $-$9.35397& UG & 22.805$\pm$0.107 & 23.621$\pm$0.286 & 22.178$\pm$0.090 & 0.091 & ZTF18acrvccs \\ 
Fink J052649.27+603520.4 & 81.70529 & $+$60.58900& UG & 20.519$\pm$0.024 & 19.616$\pm$0.010 & 19.003$\pm$0.008 & 0.018 & ZTF18acsobip \\ 
Fink J032911.36+284009.7 & 52.29733 & $+$28.66936& UG: & --  & -- & -- & -- & ZTF18adnjmbk \\ 
Fink J052132.94+540444.6 & 80.38725 & $+$54.07906& UG & 20.637$\pm$0.022 & 20.297$\pm$0.018 & 20.510$\pm$0.018 & 0.066 & ZTF19aagndmr \\ 
Fink J040855.54+174554.2 & 62.23142 & $+$17.76506& UG & 23.169$\pm$0.146 & 22.947$\pm$0.127 & 22.248$\pm$0.081 & 0.111 & ZTF19aagxcga \\ 
\hline
\end{tabular}
\caption{List of Fink CVs. For each object, the nearest Pan-STARRS source is identified, and its magnitudes from the PS DR2 stacked PSF photometry is given, along with the angular separation.}
\label{tab:UG}
\end{table*}

To address whether these systems are observed in a quiescent state, we examined Pan-STARRS1 (PS1;~\citealt{2016arXiv161205560C}) data, which provide deeper photometry than ZTF. Four objects (Fink~J061603.51+080222.8, Fink~J015434.19+555025.6, Fink~J073518.75-092114.3, Fink~J040855.54+174554.2) are detected in quiescence in PS1 and show dwarf-nova-like light curves with outburst amplitudes $\gtrsim4$~mag.

Two sources (Fink~J203440.21+504800.8 and Fink~J032911.36+284009.7) appear to be below the Pan-STARRS detection limits. For Fink~J052132.94+540444.6, PS1 photometry is likely contaminated by an outburst, and the quiescent state cannot be reliably identified.

Two other candidates (Fink~J194741.57+223702.3 and Fink~J052649.27+603520.4) represent a different and particularly interesting case. A PS1 object is present within $<1''$ of the ZTF position; however, if this object were identified as the quiescent counterpart, the resulting outburst amplitudes would be smaller than 2$^m$ in $zr$-band, which is atypically low for a CV. We therefore interpret these PS1 sources as unrelated background stars. 

\begin{figure*}
    \centering
    \includegraphics[width=0.48\linewidth]{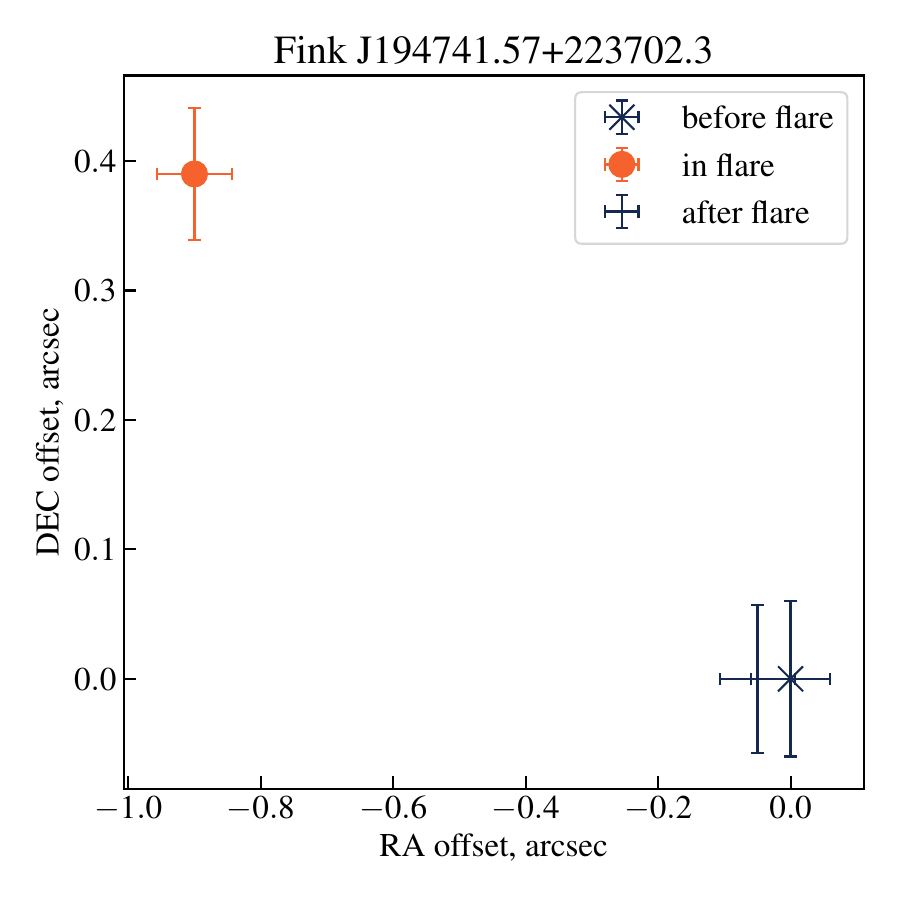}\hfill
    \includegraphics[width=0.48\linewidth]{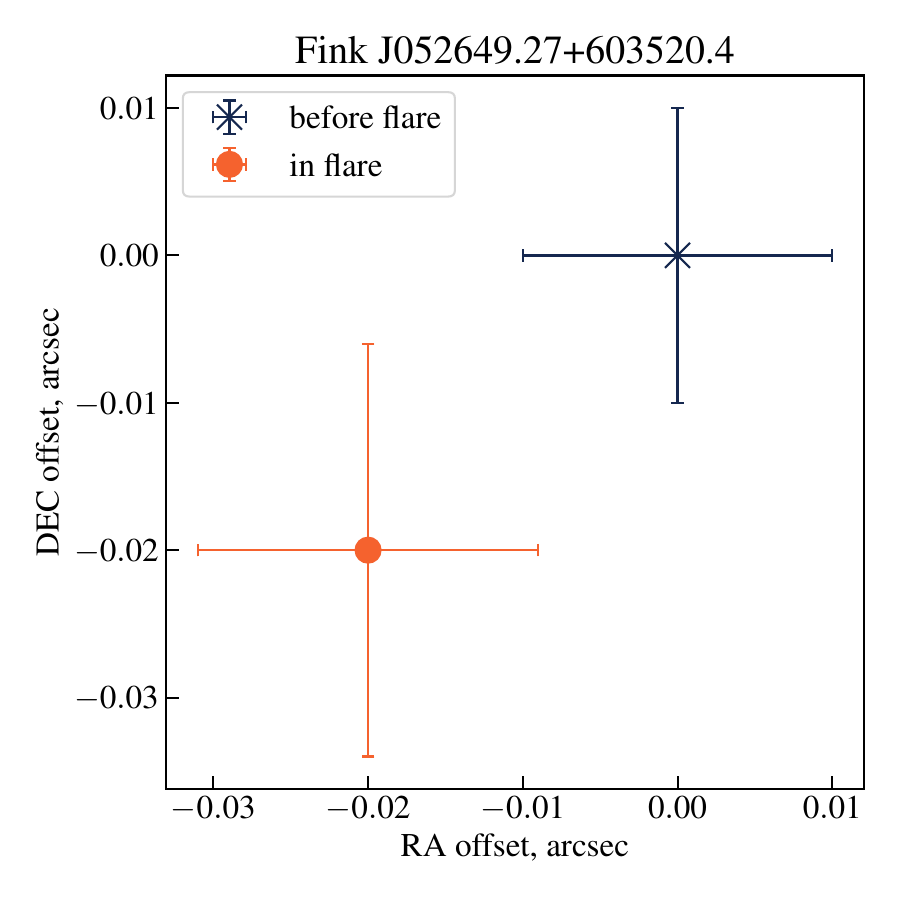}
    \caption{Positional offsets between the outburst and the quiescent source positions. The positions were measured from stacked ZTF $zr$-band images using two-dimensional Gaussian fitting.}
    \label{fig:sep}
\end{figure*}

To verify this interpretation, we measured the positional offsets between the outburst and the quiescent source positions (see Fig.~\ref{fig:sep}). The positions were derived from stacked images, and a two-dimensional Gaussian model was used to determine the centroids. The measured offsets exceed the $16\sigma$ positional uncertainty for Fink~J194741.57+223702.3 and are at the level of $1\sigma$ for Fink~J052649.27+603520.4.

\begin{figure*}
    \centering
    \includegraphics[width=\linewidth]{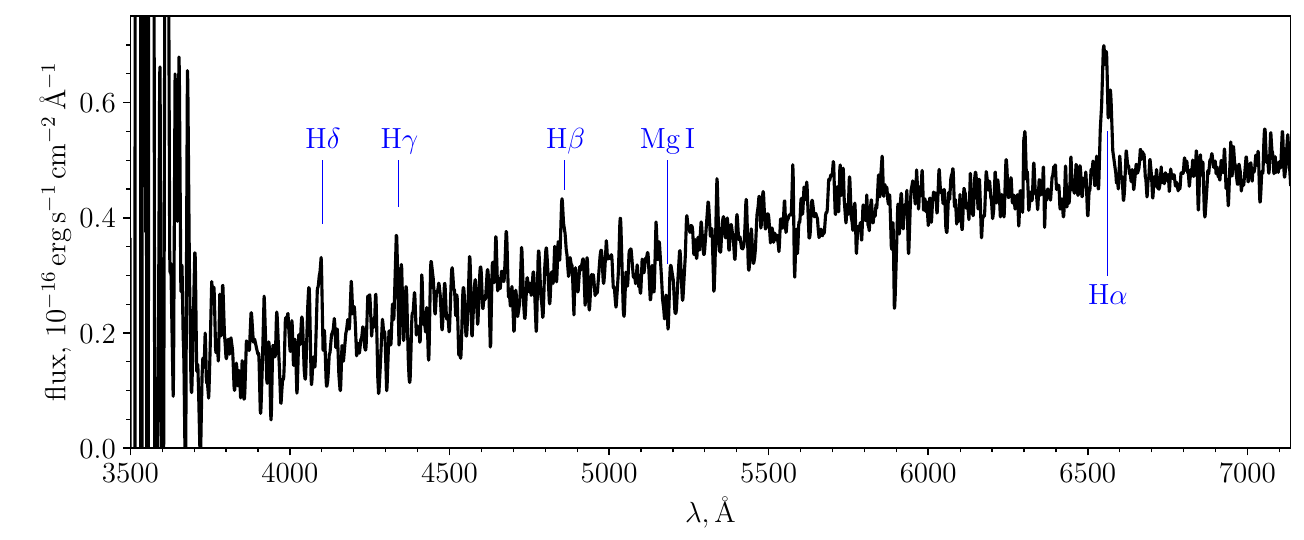}
    \caption{Spectrum of Fink~J052649.27+603520.4 obtained with the 2.5-m CMO telescope. It suggests a superposition of two sources: a cataclysmic variable and a background star. Emission lines correspond to a radial velocity of approximately $-550~\mathrm{km\,s^{-1}}$, while absorption features around the Mg\,I yield a velocity of about $+50~\mathrm{km\,s^{-1}}$.}
    \label{fig:ZTF18acsobip}
\end{figure*}

Since the angular separation does not allow a definitive conclusion about the presence of two distinct objects in the case of Fink~J052649.27+603520.4, we obtained optical spectroscopy with the 2.5-m CMO telescope on 7~January 2026. The spectrum reveals two distinct sets of spectral features, supporting the interpretation of a superposition of sources: the emission lines are strongly blueshifted, corresponding to a radial velocity of approximately $-550~\mathrm{km\,s^{-1}}$, whereas the absorption features indicate a much smaller redshift of about $+50~\mathrm{km\,s^{-1}}$ (see Fig.~\ref{fig:ZTF18acsobip}).

We show the positions of the four dwarf novae (Fink~J061603.51+080222.8, Fink~J015434.19+555025.6, Fink~J073518.75-092114.3, Fink~J040855.54+174554.2) detected in quiescent state on the 
($g$-$r$)-($r$-$i$) color-color diagram (Fig.~\ref{fig:cc_cv}). Their location is broadly consistent with the region occupied by WZ Sge-type dwarf novae in SDSS color-color diagrams \citep{2012PASJ...64...63K}. We used Pan-STARRS DR2 colors, and to construct the stellar locus we selected stars in the vicinity of WD~1657$+$343, a field characterized by the lowest Galactic extinction among the standard fields \citep{2012ApJ...750...99T}. The observed colors (orange crosses) are compared with extinction-corrected colors (blue circles). The correction was applied along the extinction vector using the total line-of-sight reddening from two-dimensional dust maps~\citep{1998ApJ...500..525S} and therefore represents an upper limit on the possible dereddening for Galactic sources.  

The positions of the objects cannot be explained solely by interstellar 
reddening and indicate intrinsically non-stellar spectral energy distributions. 
In particular, Fink~J073518.75$-$092114.3 is located far from the stellar locus. 
Its spectral energy distribution, constructed from Pan-STARRS DR2 and 2MASS~\citep{2006AJ....131.1163S} 
photometry, shows a bimodal shape with a local minimum in the $zr$-band and 
a maximum in the 2MASS $J$-band, possibly indicating non-thermal cyclotron 
emission, producing humps in the $zg$-band and the near-infrared.


\subsection{Supernovae}
\label{SN}

Among the detected anomalies, we identified 33 supernova candidates, 30 of which had not been previously reported and were thus discovered for the first time by the \texttt{AD module}. All new candidates were reported to TNS. Table~\ref{tab:sn_list} summarizes the main properties of the SN candidates. Column~1 gives the ZTF ID, and columns~2 and~3 the equatorial coordinates (right ascension and declination, respectively). Supernova type is given in column~4. Column~5 lists the Pan-STARRS $r$-band magnitude of the closest object within $5\arcsec$, while column~6 provides the angular separation to this Pan-STARRS source. Column~7 contains the corresponding TNS identifier. Additional remarks concerning the host-galaxy identification and supernova type are given in the last column.

\begin{table*}[h]
\centering
\begin{tabular}{|l|r|r|l|l|l|l|l|}
\hline
\textbf{TNS name} & \textbf{R.A.} & \textbf{Dec.} & \textbf{Type$^*$} & \textbf{$m^{\rm PS, <5''}_r$} & \textbf{Sep$''$} & \textbf{ZTF ID} & \textbf{Remarks} \\
\hline
\multicolumn{8}{|c|}{DR photometry}\\
\hline
AT 2018mne & 241.421986 & $+$54.606085  & Ia & -- & -- & ZTF18aaivvoq & hostless, SLSN candidate \\
AT 2018mln & 341.251031 & $+$24.542285 & Ia & 22.325$\pm$0.076 & 3.585 & ZTF18abccpml &  \\
AT 2018mng & 348.887450 & $+$27.895208 & Ia & 22.598$\pm$0.120 & 3.516 & ZTF18abfedgq &  \\
AT 2018mlo & 1.256827 & 33.670149 & IIn & 24.657$\pm$0.509 & 3.209 & ZTF18abgtpsq & SLSN candidate \\
AT 2018mfk & 335.041400 & $+$11.040446 & IIP & 21.234$\pm$0.033 & 2.043 & ZTF18abjgyiu &  \\
\multirow{2}{*}{AT 2018mku} & \multirow{2}{*}{18.315273} & \multirow{2}{*}{17.176923} & \multirow{2}{*}{IIP} & \multirow{2}{*}{--} & \multirow{2}{*}{--} & \multirow{2}{*}{ZTF18abncrhy} & possible host in $\sim15''$ \\
 &  &  &  &  &  &  & 2MASS J01131613+1710507 \\
AT 2018mmg & 32.652406 & 20.464304 & Ia & 18.883$\pm$0.006 & 4.838 & ZTF18abotarq &  \\
AT 2018mkw & 28.711464 & 15.442162 & Ia & 21.752$\pm$0.061 & 2.919 & ZTF18abtmazz &  \\
AT 2018mmw & 39.848330 & $-$10.681329 & IIn & -- & -- & ZTF18abtmtpy & SLSN candidate \\
AT 2018mlm & 20.183191 & 12.729882 & Ia & 21.014$\pm$0.028 & 3.188 & ZTF18abtrrke & \\
AT 2018mmz & 43.627334 & -2.932137 & Ia & 21.641$\pm$0.041 & 0.310 & ZTF18abtslnm & \\
AT 2018mkp & 307.848279 & $-$15.955488 & Ia & 23.229$\pm$0.307 & 0.177 & ZTF18abtvicq & \\
AT 2023tup & 65.307016 & $-$4.959305 & Ia & 23.856$\pm$0.916 & 3.562 & ZTF18abuipuh & SLSN candidate \\
AT 2018mll & 33.965192 & $-$1.436017 & Ia & 21.612$\pm$0.050 & 0.758 & ZTF18abuiwyz & \\
AT 2018mkt & 30.255298 & $+$18.210990 & Ia & 22.680$\pm$0.137 & 0.291 & ZTF18abvkvsk & \\
AT 2018mmy & 75.433699 & $-$4.172180 & IIP & 22.663$\pm$0.141 & 0.175 & ZTF18abyjrdx & \\
\multirow{2}{*}{AT 2018mmx} & \multirow{2}{*}{57.504595} & \multirow{2}{*}{$-$15.963332} & \multirow{2}{*}{Ia} & \multirow{2}{*}{--} & \multirow{2}{*}{--} & \multirow{2}{*}{ZTF18acckshw} & possible host in $\sim8''$ \\
 &  &  &  &  &  &  & WISEA J035000.72-155754.6 \\
AT 2018mkq & 13.902170 & $-$10.703639 & IIP & 21.308$\pm$0.030 & 1.368 & ZTF18acaeopi & \\
AT 2018mmk & 50.534618 & $+$14.965834 & IIn & -- & -- & ZTF18acmzvif & hostless, SLSN candidate \\
\multirow{4}{*}{AT 2018mme} & \multirow{4}{*}{67.476578} & \multirow{4}{*}{$+$3.451243} & \multirow{4}{*}{Ia} & \multirow{4}{*}{22.927$\pm$0.156} & \multirow{4}{*}{3.022} & \multirow{4}{*}{ZTF18acskrxt} & possible host in $\sim26''$ \\
 &  &  &  &  &  &  & LEDA 3680353 \\
 &  &  &  &  &  &  & ($z=0.04482\pm0.00011$, E), \\
 &  &  &  &  &  &  & ``pure'' SN Ia \\
AT 2018mly & 38.299830 & $+$17.950827 & Ia & -- & -- & ZTF18acsymtk & \\
AT 2018mmv & 150.066874 & $+$67.000437 & IIP & 22.299$\pm$0.123 & 3.294 & ZTF18acvvxge & \\
\multirow{3}{*}{AT 2018mkv} & \multirow{3}{*}{44.195781} & \multirow{3}{*}{$+$10.043904} & \multirow{3}{*}{Ia} & \multirow{3}{*}{--} & \multirow{3}{*}{--} & \multirow{3}{*}{ZTF18adogfjl} & possible host in $\sim18''$ \\
 &  &  &  &  &  &  & LEDA 1373761 \\
  &  &  &  &  &  &  & ($z=0.06671\pm0.00013$) \\
AT 2018mma & 52.813138 & $+$13.399680 & IIL & -- & -- & ZTF19aacpbew & hostless \\
AT 2018mmj & 51.475445 & $+$15.637542 & Ia & 24.968$\pm$1.139 & 3.537 & ZTF19aainpwu & hostless \\
\multirow{4}{*}{AT 2018mlz} & \multirow{4}{*}{27.469425} & \multirow{4}{*}{$+$48.658704} & \multirow{4}{*}{IIL} & \multirow{4}{*}{--} & \multirow{4}{*}{--} & \multirow{4}{*}{ZTF19abkdlht} & possible host in $\sim26''$ \\
 &  &  &  &  &  &  & LEDA 2322429 \\
  &  &  &  &  &  &  & ($z=0.02096\pm0.00008$, S0), \\
  &  &  &  &  &  &  & ``pure'' SN Ia  \\
AT 2019aauc & 348.077725 & $+$82.484227 & IIP & 23.330$\pm$0.342 & 2.154 & ZTF19abuaxkc & hostless, SLSN candidate \\
AT 2019aauv & 119.119806 & $+$82.797799 & IIn & 23.817$\pm$0.835 & 4.021 & ZTF19acfzwva & SLSN candidate \\
AT 2018mmf & 30.435008 & $+$48.233737 & Ia & 21.933$\pm$0.078 & 2.650 & ZTF20abwfnmq & \\
\hline
\multicolumn{8}{|c|}{Difference photometry}\\
\hline
\multirow{3}{*}{SN 2023lid} & \multirow{3}{*}{248.450117} & \multirow{3}{*}{$+$48.418628} & \multirow{3}{*}{Ia} & \multirow{3}{*}{23.198$\pm$0.215} & \multirow{3}{*}{4.948} & \multirow{3}{*}{ZTF23aaoxngr} & possible host in $\sim9''$ \\
 &  &  &  &  &  &  & LEDA 2317068 \\
  &  &  &  &  &  &  & ($z=0.08436\pm0.00001$, S0-a) \\
AT 2023iro & 209.664807 & $+$44.183844 & Ia  & -- & -- & ZTF23aajaimx & hostless \\
SN 2023mtp & 258.535673 & $+$81.074837 & IIL & 22.078$\pm$0.096 & 1.673 & ZTF23aalftvv & SN with precursor \\
AT 2023qxw & 30.783982 & $-$8.722530  & Ia & 19.398$\pm$0.007 & 0.127 & ZTF23aazsjal & \\
\hline
\end{tabular}
\caption{List of supernova candidates discovered by the Fink \texttt{AD module}. $^*$Column Type lists the \texttt{sncosmo} photometric type.}
\label{tab:sn_list}
\end{table*}

The majority ($N$=29) of these events exhibit nearly ``flat'' light curves in difference magnitude, making them appear anomalous compared to typical variable and transient sources. This behavior is explained by the fact that the ZTF template image was taken during a supernova event itself. As a result, the object manifests as a negative-flux source in subsequent difference images, giving the impression of a disappearing transient. Such cases are expected in the early stages of a survey, as the sky reference inevitably includes some on-going transients. However, this effect is not limited to the early phase of the ZTF survey in 2018; similar cases were observed later -- for example, AT~2019aauv exploded in September 2019. Since these transients were detected after the event had faded, spectroscopic confirmation was not possible. Nevertheless, the module also detected four on-going supernovae, enabling a more detailed study (see Section~\ref{sn2023mtp}).

To provide preliminary classification, particularly for the post-transient discoveries, we performed multi-band light curve fitting using the Python library \texttt{sncosmo}\footnote{\url{https://sncosmo.readthedocs.io/en/stable/}}. The light curves were fitted with Peter Nugent's supernova models\footnote{\url{https://c3.lbl.gov/nugent/nugent_templates.html}}, which cover the main SN types (Ia, Ib/c, IIP, IIL, IIn) and SALT2 model of SNe~Ia~\citep{2014A&A...568A..22B}. Each model is characterized by a set of parameters.

Nugent's templates are simple spectral time series which can be scaled up and down. The free parameters of the fit were the redshift $z$, the observer-frame time corresponding to the source's zero phase, $t_0$, and the amplitude. The zero phase is defined relative to the explosion moment and the observed time $t$ is related to phase via $t = t_0 + {\rm phase} \times (1 + z)$.

The SALT2 model is more sophisticated and contains the parameters that affect the shape of the spectrum at each phase. In addition to the redshift, $t_0$, and amplitude, the light curves are also characterized by $x_1$ (stretch) and $c$ (color) parameters. The $x_1$ parameter describes the time-stretching of the light curve. The $c$ parameter corresponds to the color offset with respect to the average at the date of maximum luminosity in $B$-band, i.e. $c = (B - V)_{\rm max} - \langle B - V \rangle$. In SALT2 model, the zero phase is defined relative to the maximum in $B$-band.

To perform the fits, we used difference photometry for normally subtracted supernovae and data-release  photometry for the ``flat'' cases.
For difference photometry we used all available passbands in the Fink broker ($zg$ and $zr$), while for DR photometry we used all available passbands ($zg$, $zr$, $zi$) from ZTF DR23. 
By construction, difference photometry does not include the background (host galaxy), so we used these fluxes as they are. In contrast, DR photometry contains the host contribution. In this case, we accounted for the host galaxy contamination by subtracting the reference magnitude from the ZTF DR light curves. The reference magnitudes were retrieved from the ZTF archival data\footnote{\url{https://irsa.ipac.caltech.edu/Missions/ztf.html}}. We also corrected all light curves for the line-of-sight reddening in the Milky Way using the estimates from \citet{2011ApJ...737..103S}.

We adopted $[-15^m; -22^m]$ as an acceptable range for the supernova absolute magnitude \citep{2014AJ....147..118R} and, using the maximum apparent magnitude, roughly transformed it into the corresponding redshift range. In some cases, spectroscopic redshifts of the probable host galaxies are available; however, we kept the redshift as a free parameter since the host association was not obvious. For example, for AT~2018mkv the center of the possible host LEDA~1373761 is located at a distance of $\sim 20''$ from the transient, while there is also a closer faint galaxy in the same region. Since the association is ambiguous, we allowed the redshift to vary: the redshift obtained from the SN light-curve fit is $0.059 \pm 0.013$, while LEDA~1373761 has $z_{\rm sp} = 0.06671 \pm 0.00013$~\citep{2004ApJ...607..202M}. The small difference in $z$ estimates supports LEDA~1373761 as the most likely host of AT~2018mkv. The only exceptions are two spectroscopically confirmed supernovae, SN~2023lid and SN~2023mtp, for which we fixed the redshift to the TNS value, and AT~2018mmx, for which we adopted the photometric redshift of the host $z_{\rm ph} = 0.165\pm0.015$ from the Legacy Surveys~\citep{2021MNRAS.501.3309Z}.

We applied a $\chi^2$ criterion to choose the best-fit model for each object. Results of the light curve fit are given in Appendix~\ref{ap:sn_fit}, the best-fit model for each  transient is listed in column 5 of Table~\ref{tab:sn_list}, yielding 20 SNe~Ia and 13 core-collapse SNe.

Among the candidates, we identify seven objects (see remarks in Table~\ref{tab:sn_list}) with peak absolute magnitudes brighter than $-21^m$, the commonly adopted threshold for superluminous supernovae \citep{2012Sci...337..927G}. For three of these objects (AT~2018mne, AT~2018mmk, and AT 2019aauc), the \texttt{sncosmo} fits are unreliable because the rising part of the light curve is absent. For the remaining four candidates (AT~2018mlo, AT~2018mmw, AT~2023tup, and AT~2019aauv), their high luminosities are further supported by photometric redshifts from the Legacy Surveys \citep{2021MNRAS.501.3309Z} for sources located at the transient positions and considered as their likely host galaxies.

In addition, we identify two promising candidates to so-called ``pure'' SNe~Ia~\citep{2011AstL...37..663P}. ``Pure'' supernovae are transients exploding in elliptical galaxies or in the halos of spiral galaxies. Such environments are characterized by low dust extinction, low metallicity, and old stellar populations. Moreover, these events are commonly associated with the double-degenerate progenitor channel~\citep{2011NewA...16..250L}. These properties make ``pure'' SNe a more homogeneous sub-class and therefore particularly suitable for cosmological applications.

The first ``pure'' SN  candidate, AT~2018mme, is well fitted by the \texttt{SALT2} SN~Ia model. Its possible host galaxy is LEDA~3680353 --- elliptical galaxy located at an angular separation of $\sim26''$, with a spectroscopic redshift of $z=0.04482\pm0.00011$ \citep{2012ApJS..199...26H}, corresponding to a projected physical separation of approximately 23~kpc. Using the host redshift, we obtain a peak absolute magnitude of $M_{zr,~\rm max}=-18.64^m\pm0.04$ for AT~2018mme.

The second candidate, AT~2018mlz, is formally classified as a SN~IIL based on the \texttt{sncosmo} fit; however, only the declining part of the light curve is covered by the available data. Therefore, a Type~Ia classification cannot be excluded. In particular, the event is not associated with star-formation regions of spiral galaxies, where core-collapse supernovae typically occur. A possible host galaxy, LEDA~2322429, is located at an angular separation of $\sim26''$, with a spectroscopic redshift of $z=0.02096\pm0.00008$, corresponding to a projected physical separation of approximately 11~kpc. The host is classified as an S0 galaxy \citep{2012ApJS..199...26H}.

This photometric classification should be treated with caution, as our goal is not precise typing but rather to demonstrate that the candidate light curves can be reasonably described by supernova models. Some ``flat'' candidates, with missed actual explosion, nonetheless have well-sampled early-time light curves; these are valuable for studying the explosion physics and pre-supernova parameters~\citep{1998ApJ...496..454B,2006A&A...453..229B}. Identifying such cases is also important for improving future survey strategies and reducing the number of similarly missed events.


\subsection{Hostless transients}

A significant fraction of SN candidates detected by the Fink \texttt{AD module} does not show an obvious host galaxy in the ZTF images. However, for some of them a faint host becomes visible in deeper imaging data from the Pan-STARRS1 or the Legacy Surveys. Only six transients remain hostless (see remarks in Table~\ref{tab:sn_list}). For four of them, no Legacy Surveys coverage is currently available. For the remaining two objects, AT~2018mne and AT~2023iro, Legacy Surveys images are available, but no source is detected at or near the transient position.

The apparent absence of a host galaxy can be caused not only by observational limitations of the survey, but also by more complex astrophysical scenarios. In particular, a host galaxy may remain undetected because of its intrinsically low surface brightness (e.g.,~\citealt{2019MNRAS.485..796M}).  More exotic possibilities are also possible, such as a progenitor star being ejected from its host galaxy and exploding as a supernova far from its birthplace. Such run-away stars can be produced either by a supernova explosion in a binary system \citep{1961BAN....15..265B} or through dynamical interaction in proto-stellar clusters~\citep{1967BOTT....4...86P}. The fastest among these objects, hypervelocity stars, are thought to originate from the disruption of stellar binaries by the supermassive black hole in the center of galaxies \citep{1988Natur.331..687H}, although other scenarios have also been proposed \citep{2003ApJ...599.1129Y}. In addition, star formation may occur in tidal debris left after galaxy interactions, leading to supernova explosions in intergalactic medium.

In all these cases, the absence of a visible host galaxy provides a valuable opportunity to investigate unusual progenitor environments and to constrain the astrophysical scenarios that give rise to such hostless transients \citep{pessi2024}.


\section{Discussion}
\label{sec:discussion}


\subsection{Misclassification among candidates in anomalies}

During the visual inspection of candidates in anomalies, we identified and corrected several misclassifications that had propagated into public databases.
As the volume of time-domain data continues to grow, the demand for rapid and fully automated alert classification increases accordingly. While such approaches are essential for handling large data streams, they can also lead to systematic contamination of databases when classifications are accepted without human verification, ultimately reducing their scientific reliability.

A representative example is AT~2018eup/ZTF18abyxzgl, which was automatically classified as a cataclysmic variable and added to the AAVSO VSX with this classification\footnote{\url{https://vsx.aavso.org/index.php?view=detail.top&oid=2225346}. The classification has been updated later by us.}. However, the shape of its light curve is fully consistent with that of a Type~Ia supernova, making the CV classification incorrect.

Conversely, AT~2023lwt/ZTF23aapkxeb was automatically reported to the TNS as a possible supernova\footnote{\url{https://www.wis-tns.org/object/2023lwt}}, whereas its light curve clearly indicates a cataclysmic variable. These examples illustrate that automated classification pipelines, while powerful, still benefit significantly from targeted expert validation, particularly for objects flagged as anomalous.


\subsection{Contaminants}
\label{HPM}

It is important to note that anomalous photometric behavior does not necessarily imply an astrophysically peculiar origin. A ``weird'' light curve can be a result of instrumental issues such as CCD defects, optical system aberrations, poor weather conditions, or pipeline processing issues \citep{2019MNRAS.489.3582D,2026NewA..12202466S}.

It is worth stressing that at early stages of the anomaly detection module, a significant fraction of the detected anomalies were associated with moving objects, primarily known asteroids. However, this is no longer the case in the current version of the module, as we now require a minimum of four photometric points per filter for an alert to be processed. Most asteroids typically produce only one detection due to their apparent motion.

\begin{figure*}
    \centering
    \includegraphics[trim={0 5cm 0 5cm},clip, width=0.8\linewidth]{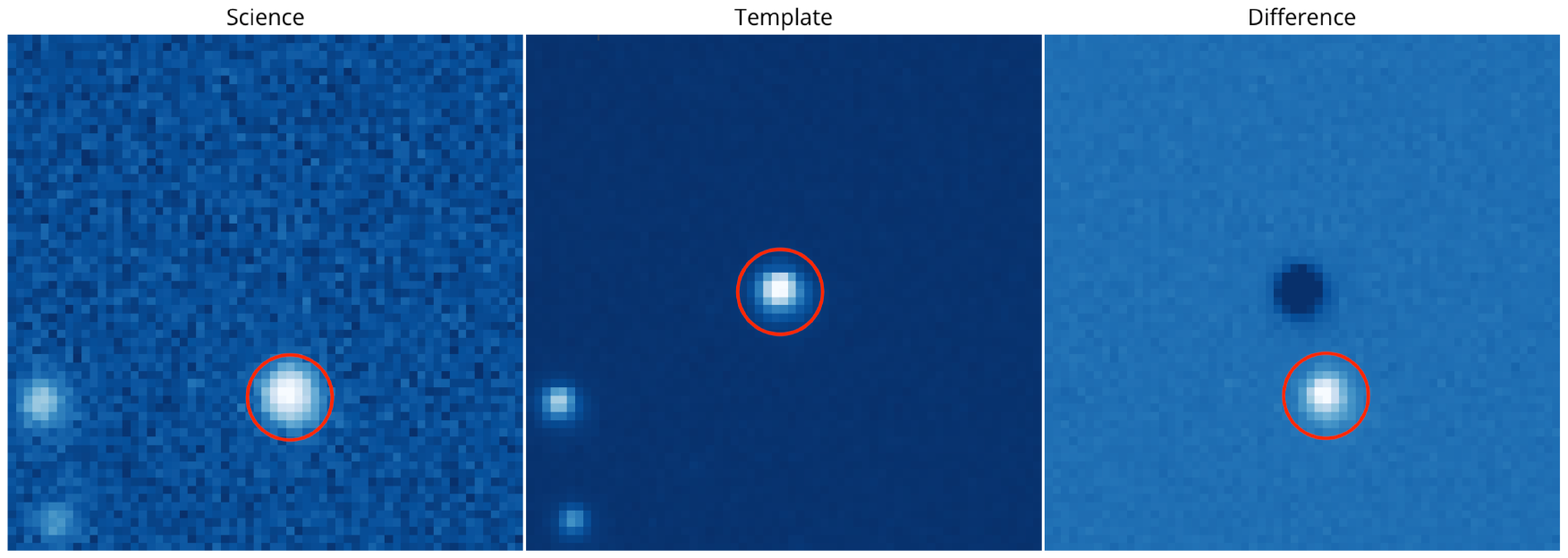}
    \caption{Three image cutouts for ZTF18abkalak: science image obtained on 14 January 2026 at 12:17:12 UTC, the reference template, and the difference between both. The red circle indicates the position of HPM star G~260-1.}
    \label{fig:hpm}
\end{figure*}

Another substantial fraction of such non-astrophysical contaminants includes ``flat'' supernova (see Section~\ref{SN}) and high proper motion (HPM) stars. In the case of HPM stars, the object is present on the reference image but moves significantly over time, triggering alerts each time the telescope revisits the same position (see Fig.~\ref{fig:hpm} for an example). A non-exhaustive list of anomalous alerts associated with such HPM objects is presented in Table~\ref{tab:high_pm}.

\begin{table}
    \centering
    \caption{ZTF objects ranked among the top-10 anomalies by the pipeline and associated with high proper motion stars.}
    \begin{tabular}{ll}
        \hline
        \textbf{ZTF ID} & \textbf{HPM star} \\
        \hline
        ZTF18abkalak & G 260-1 \\
        ZTF18adgsbul & G 135-67 \\
        ZTF19abhyflg & LP 624-55 \\
        ZTF23aacafew & G 260-1 \\
        ZTF23aajfetj & Wolf 611 \\
        \hline
    \end{tabular}
    \label{tab:high_pm}
\end{table}

While many classes of contaminants can be mitigated through carefully designed selection criteria and automated filters, others remain challenging due to their diversity and context-dependent interpretation. In particular, the notion of what constitutes an ``interesting'' or ``anomalous'' object is inherently subjective and strongly depends on the scientific goals of the researcher (see, e.g., \citealt{pruzhinskaya2025}). This motivates the inclusion of expert knowledge directly into the anomaly detection process (see Section~\ref{CSM}).


\subsection{Citizen science model}
\label{CSM}

The incorporation of knowledge to the AD process can be done using the so-called active anomaly detection approach. Unlike passive methods, active AD incorporates feedback from domain experts to iteratively refine the model. 

As a first step in this direction, we developed a citizen science model. The core idea of this approach is to collect expert assessments of candidate in anomalies and use these evaluations to retrain the model. As an algorithm we implemented the Active Anomaly Discovery (AAD; \citealt{Das2017}). AAD extends the standard IF framework by incorporating expert feedback, allowing the model to adapt its decision boundaries iteratively. A detailed description of the algorithm applied to astronomical transient data is provided in \citealt{2021A&A...650A.195I}.

\begin{figure*}
    \centering
    \includegraphics[width=\linewidth]{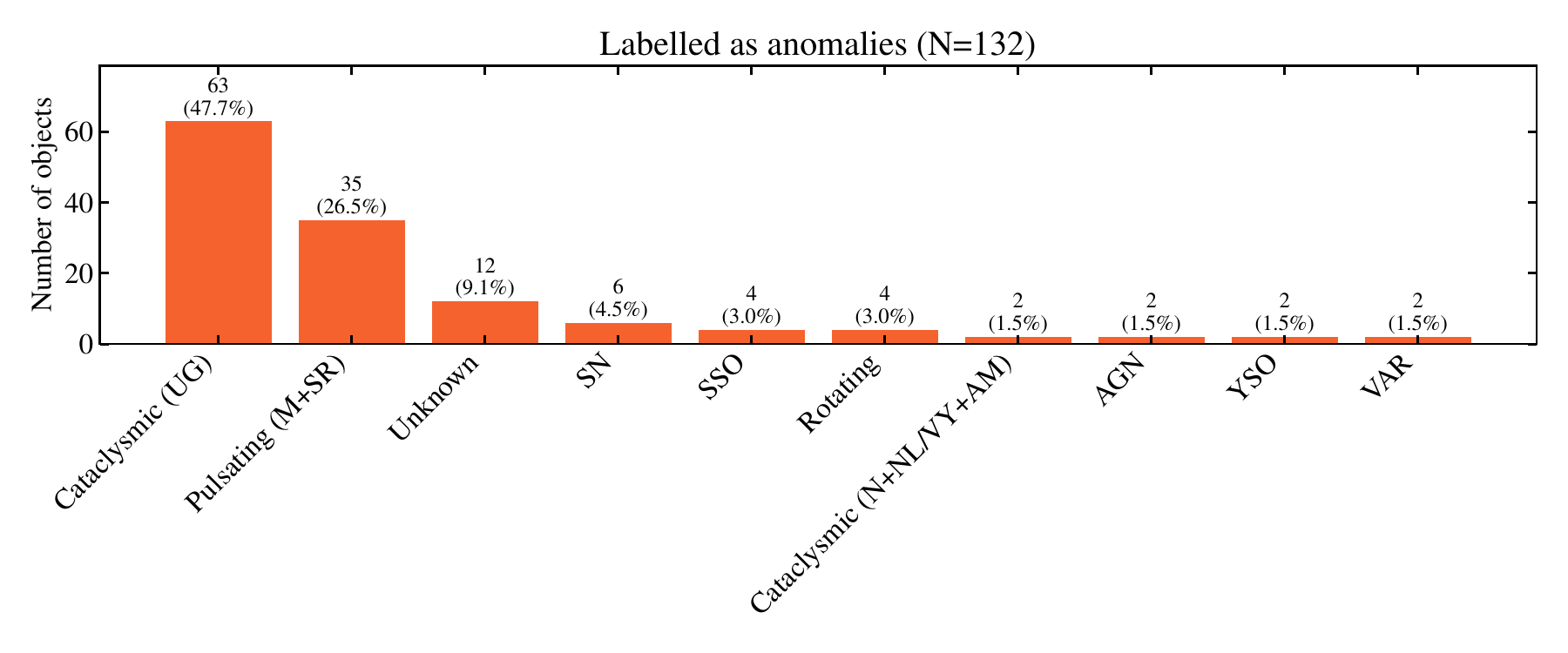}\hfill
    \includegraphics[width=\linewidth]{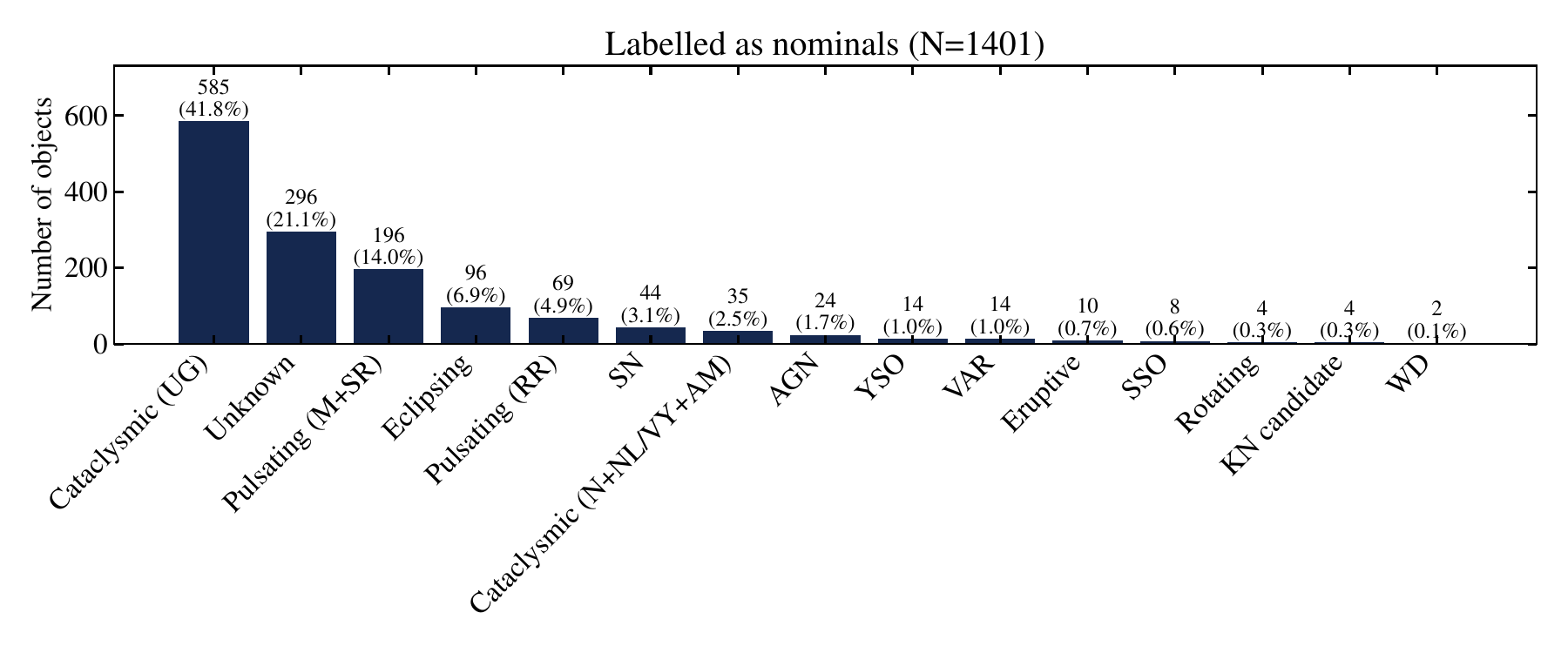}
    \caption{Distribution of alerts with expert feedback in the citizen science model by astronomical object type.}
    \label{fig:csm}
\end{figure*}

In the citizen science model, expert feedback is collected via a dedicated public Telegram bot\footnote{\url{https://t.me/ZTF_anomaly_bot}} (Fig.~\ref{fig:notification}). A ``thumbs up'' reaction indicates that an object was considered anomalous, while a ``thumbs down'' reaction classifies it as nominal. The model is periodically retrained using the accumulated expert feedback. As of 27 January 2026, the citizen science model has collected 789 expert reactions for the $zr$-band alerts and 744 reactions for the $zg$-band alerts from 31 volunteers. The distribution of anomalous and nominal objects by types is shown in Fig.~\ref{fig:csm}.
 
\begin{figure*}
    \centering
    \includegraphics[width=0.48\linewidth]{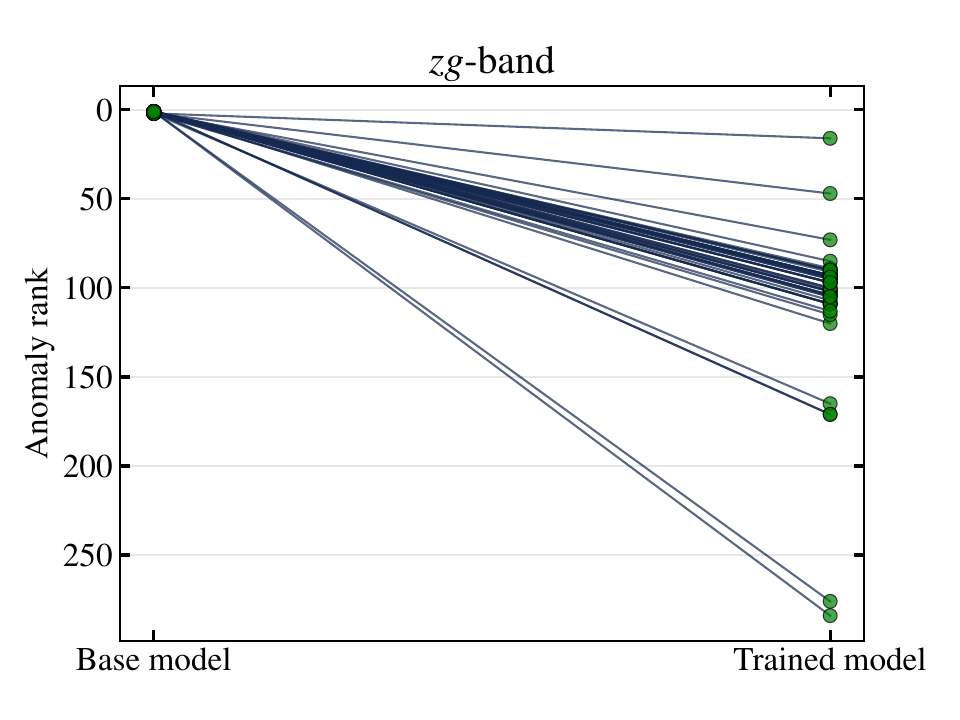}\hfill
    \includegraphics[width=0.48\linewidth]{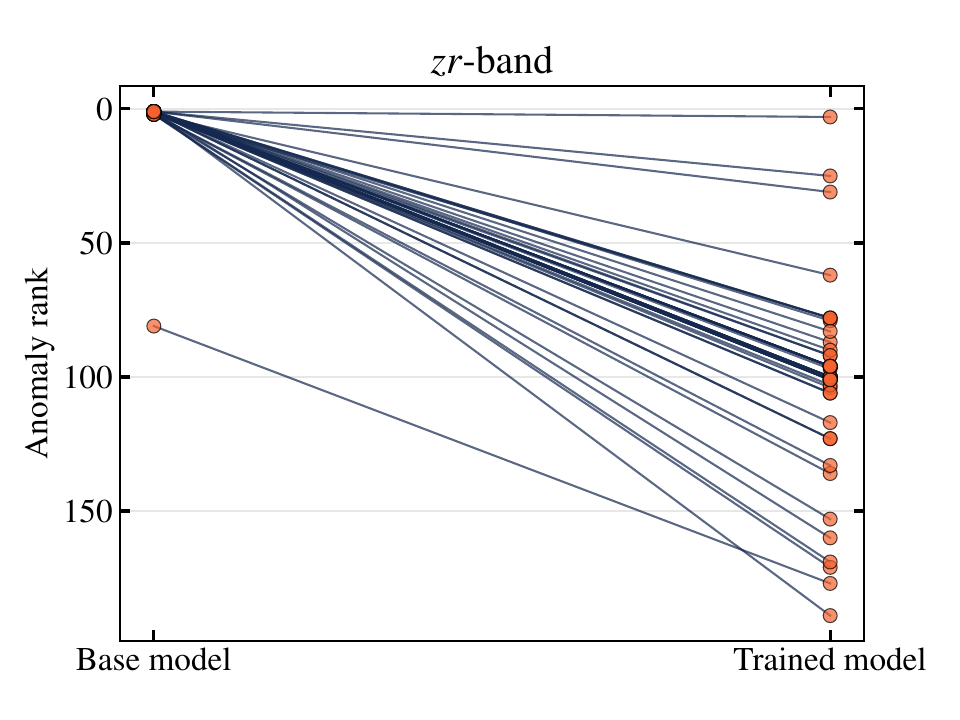}
    \caption{Anomaly rank for all alerts -- eclipsing binaries -- marked by the experts as nominals in the initial Isolation Forest (\texttt{Base model}) and in the AAD-trained forest (\texttt{Trained model}). }
    \label{fig:EB_rank}
\end{figure*}

To illustrate the impact of expert feedback on anomaly ranking, we selected all alerts classified as eclipsing binaries -- the only class with sufficient number of reactions and consistently labelled as nominal by all the experts.
For these objects, we computed anomaly rank using the \texttt{Base model} (Section~\ref{sec:anomaly_detection}) and compared them with the scores obtained after processing the same objects through the AAD-trained forest -- \texttt{Trained model}. We find that, in the \texttt{Trained model}, eclipsing binaries are systematically shifted toward lower anomaly ranks and the majority of them drop out of the top-ranked anomalous candidates, in agreement with expert expectations (see Fig.~\ref{fig:EB_rank}).

A caveat of this approach is the lack of a universally agreed definition of an anomaly among experts. As illustrated in Fig.~\ref{fig:csm}, the majority of classes receive mixed classifications. For example, dwarf novae constitute the largest class among anomalous objects, but they are also the most frequent class among nominal ones. This behavior is expected, as the definition of an anomaly is inherently expert-dependent~\citep{pruzhinskaya2025}. To address this limitation, the next step in the development of the module involves the creation of personalized anomaly detection models tailored to the scientific interests of individual experts (Pshenichniy et al., in prep.).


\section{Conclusion}
\label{sec:conclusions}

We presented the first-year results of the anomaly detection pipeline operating within the Fink broker since 25 January 2023. The pipeline combines feature-based Isolation Forest ranking the most anomalous alerts to experts via Slack/Telegram by the end of each observational night, enabling fast contextual assessment and rapid response.

During the first year, the AD stream proved capable of recovering both unique astrophysical systems and new representatives of rare classes. We highlighted several individual high-interest discoveries and performed dedicated photometric and spectroscopic follow-up campaigns. 

The most notable individual anomalies include:

\begin{itemize}

\item The rare AM~CVn system Fink~J062452.88+020818.3, whose photometric properties indicate a WZ~Sge-type object. This system represents one of the very few known members of this rare subclass.

\item The unusual transient SN~2023mtp, characterized by a luminous precursor and spectra that cannot be consistently matched by templates of a single supernova type across all epochs.

\item The UX~Ori-type star Fink~J222324.32+744222.0, whose optical and infrared variability indicates changes in the circumstellar dust environment.

\item The flaring M-dwarf Fink~J042203.10+362318.7, exhibiting recurrent short-duration flares typical of magnetically active late-type stars.

\end{itemize}

In addition, the AD module enabled the discovery, photometric classification, and reporting to TNS of 30 previously unreported supernovae. Among them are candidates for SLSNe, ``pure'' SNe~Ia, and hostless transients. Furthermore, nine new cataclysmic variables were identified and reported to the AAVSO VSX database as dwarf novae.

These case studies illustrate an important practical point for time-domain anomaly detection: building an AD pipeline is not sufficient on its own. To convert ``anomaly scores'' into astrophysical insight, the detected candidates must be actively analysed, contextualised, and in many cases followed up with additional observations. We emphasise that follow-up is not only initiated internally: several targets attracted additional community attention after our reports to TNS.

We also demonstrated the value of incorporating expert feedback into the AD loop. Using a public Telegram bot, we collected reactions to alerts and retrained the model using the AAD algorithm, which  shifts consistently nominal classes (e.g. eclipsing binaries) towards lower anomaly ranks. At the same time, the feedback statistics highlight a fundamental limitation: the notion of an ``anomaly'' is expert-dependent. This motivates the next development step toward personalised AD models tuned to the scientific interests of individual users.

Finally, our results show that the Fink AD pipeline provides an effective bridge between large-scale survey streams and domain experts. By coupling machine-learning–based ranking with expert assessment and observational follow-up, anomaly detection evolves from a purely statistical exercise into a discovery-driven component of time-domain astronomy.

\begin{acknowledgements}
We thank Dr. Sergei Antipin (SAI MSU) and Dr. Oleg Malkov (INASAN) for valuable discussions on Fink J062452.88+020818.3 and Dr. T. Yu. Magakian (Byurakan Observatory NAS Armenia) for help with the interpretation of Fink J222324.32+744222.0. We thank the subscribers of the Fink anomaly detection citizen science channel for their contributions and feedback used to iteratively improve the model.

This work was developed within the Fink community and made use of the Fink
resources. Fink is supported by LSST-France and CNRS/IN2P3.

The work was carried out using equipment developed with the support of the M.V. Lomonosov Moscow State University Program of Development.

The computational resources were provided by Yandex Cloud\footnote{\url{https://yandex.cloud/ru/research-education-program}} as part of a grant for scientific research.

This work was co-funded by the European Union and supported by the Czech Ministry of Education, Youth and Sports (MEYS) (Project No. CZ.02.01.01/00/22\_008/0004632 -- FORTE). This research is partially based on the data acquired by the robotic telescope FRAM-ORM, which is supported by the grant of the MEYS LM2018102.

 A.\,K. acknowledges support from the National Research Foundation (NRF) of South Africa.

Anais Möller is supported by the ARC Discovery Early Research Award (DE230100055).

The Photometric Redshifts for the Legacy Surveys (PRLS) catalog used in this paper was produced thanks to funding from the U.S. Department of Energy Office of Science, Office of High Energy Physics via grant DE-SC0007914.

We gratefully acknowledge the contributions of the AAVSO observers community, whose photometric data and metadata resources were used in this study and made available through the AAVSO’s scientific
archives\footnote{\url{https://aavso.org/}}.

\end{acknowledgements}

\clearpage

\twocolumn
%
  \bibliographystyle{aa} 
  \bibliography{biblio} 
%

\appendix
\section{Features}
\label{sec:features}

In this appendix, we describe the photometric features computed from the light curves.  
For each feature, we provide its mathematical definition, the minimum number of data points required, and failure conditions.
\subsection{Mean}
The arithmetic mean of the magnitude values:
\begin{equation}
    \bar{m} = \frac{1}{N} \sum_{i=1}^{N} m_i
\end{equation}
Minimum points: $N \ge 1$.\\
Failure conditions: fails for $N = 0$.

\subsection{Weighted Mean}
The weighted mean of the magnitudes, where $w_i = 1/\sigma_i^2$:
\begin{equation}
    \bar{m}_w = \frac{\sum_{i=1}^{N} w_i m_i}{\sum_{i=1}^{N} w_i}
\end{equation}
Minimum points: $N \ge 1$.\\
Failure conditions: fails for $N = 0$ or if all $\sigma_i$ are zero or infinite.

\subsection{Standard Deviation}
The sample standard deviation of magnitudes:
\begin{equation}
    \sigma_m = \sqrt{\frac{1}{N-1} \sum_{i=1}^{N} (m_i - \bar{m})^2}
\end{equation}
Minimum points: $N \ge 2$.\\
Failure conditions: fails for $N < 2$.

\subsection{Median}
The median of the magnitudes:
\begin{equation}
    \tilde{m} = \mathrm{median}\{m_i\}
\end{equation}
Minimum points: $N \ge 1$.\\
Failure conditions: fails for $N = 0$.

\subsection{Amplitude}
Half amplitude of the magnitudes:
\begin{equation}
    A = \frac{1}{2} \left( \max\{m_i\} - \min\{m_i\} \right)
\end{equation}
Minimum points: $N \ge 1$.\\
Failure conditions: fails for $N = 0$.

\subsection{BeyondNStd}
The fraction of points beyond $k=1$ standard deviation from the mean:
\begin{equation}
    \frac{1}{N} \left| \{i : |m_i - \bar{m}| > k \cdot \sigma_m \} \right|
\end{equation}
Minimum points: $N \ge 2$.\\
Failure conditions: fails for $N < 2$. If $\sigma_m = 0$, the result is 0.

\subsection{Cusum}
The range of magnitude cumulative sums:
\begin{equation}\label{eq:cusum}
    \max_j(S_j) - \min_j(S_j),
\end{equation}
where
\begin{equation}
    S_j \equiv \frac1{N\,\sigma_m} \sum_{i=0}^j{\left(m_i - \bar{m}\right)},  ~~j \in \{1.. N - 1\} 
\end{equation}

Minimum points: $N \ge 2$.\\
Failure conditions: fails for $N < 2$ or if $\sigma_m = 0$.

\subsection{InterPercentile Range}
The inter-percentile range for $\mathrm{quantile}=0.1$ of magnitude sample:
\begin{equation}
    Q_{0.9} - Q_{0.1}
\end{equation}
Minimum points: $N \ge 1$.\\
Failure conditions: fails for $N = 0$.

\subsection{Kurtosis}
The sample excess kurtosis (Fisher's definition), corrected for bias:
\begin{equation}
    g_2 = \frac{N(N+1)}{(N-1)(N-2)(N-3)} \sum_{i=1}^{N} \left( \frac{m_i - \bar{m}}{\sigma_m} \right)^4 - \frac{3(N-1)^2}{(N-2)(N-3)}
\end{equation}
Minimum points: $N \ge 4$.\\
Failure conditions: fails for $N < 4$ or if $\sigma_m = 0$.

\subsection{Linear Trend}
Unweighted linear regression $m(t) = at + b$:
\begin{enumerate}
    \item Slope: $\hat{a}$.
    \item Error of slope: $\hat{\sigma}_{\hat{a}} = \sqrt{\hat{\sigma}_{\epsilon}^2 / S_{xx}}$, where $S_{xx} = \sum (t_i - \bar{t})^2$.
    \item Standard deviation of residuals: $\hat{\sigma}_{\epsilon} = \sqrt{\frac{1}{N-2} \sum (m_i - \hat{m}_i)^2}$.
\end{enumerate}
Minimum points: $N \ge 3$.\\
Failure conditions: fails for $N < 3$ or if all $t_i$ are identical.

\subsection{Linear Fit}
Weighted linear regression ($w_i = 1/\sigma_i^2$):
\begin{enumerate}
    \item Weighted slope: $\hat{a}_w$.
    \item Error of slope: $\hat{\sigma}_{\hat{a}_w} = 1/\sqrt{S_{xx,w}}$, where $S_{xx,w} = \sum w_i(t_i - \bar{t}_w)^2$.
    \item Reduced chi-squared: $\chi^2_{\mathrm{red}} = \frac{1}{N-2} \sum w_i (m_i - \hat{m}_{i,w})^2$.
\end{enumerate}
Minimum points: $N \ge 3$.\\
Failure conditions: fails for $N < 3$, if all $t_i$ are identical, or if any $\sigma_i$ is zero.

\subsection{Magnitude Percentage Ratio (1)}
Ratio of inter-percentile ranges:
\begin{equation}
    \frac{Q_{0.60} - Q_{0.40}}{Q_{0.95} - Q_{0.05}}
\end{equation}
Minimum points: $N \ge 1$.\\
Failure conditions: if denominator is zero.

\subsection{Magnitude Percentage Ratio (2)}
Ratio of inter-percentile ranges:
\begin{equation}
    \frac{Q_{0.8} - Q_{0.2}}{Q_{0.9} - Q_{0.1}}
\end{equation}
Minimum points: $N \ge 1$.\\
Failure conditions: if denominator is zero.

\subsection{Maximum Slope}
Maximum absolute slope between two sub-sequential observations:
\begin{equation}
    \max_{i=0..N-2} \left| \frac{m_{i+1} - m_i}{t_{i+1} - t_i} \right|
\end{equation}
Minimum points: $N \ge 2$.\\
Failure conditions: fails for $N < 2$.

\subsection{Median Absolute Deviation}
The median absolute deviation (MAD):
\begin{equation}
    \mathrm{MAD} = \mathrm{median}\{|m_i - \tilde{m}|\}
\end{equation}
Minimum points: $N \ge 1$.\\
Failure conditions: fails for $N = 0$.

\subsection{Median Buffer Range Percentage}
Fraction of points within a buffer zone around the median (buffer size = $0.1 \times$ Amplitude):
\begin{equation}
    \frac{1}{N} \left| \{ i : |m_i - \tilde{m}| < 0.1 \, A \} \right|
\end{equation}
Minimum points: $N \ge 1$.\\
Failure conditions: fails for $N = 0$.

\subsection{Percent Amplitude}
Maximum deviation from the median:
\begin{equation}
    \max(\max\{m_i\} - \tilde{m}, \tilde{m} - \min\{m_i\})
\end{equation}
Minimum points: $N \ge 1$.\\
Failure conditions: fails for $N = 0$.

\subsection{Mean Variance}
The coefficient of variation:
\begin{equation}
    CV = \frac{\sigma_m}{\bar{m}}
\end{equation}
Minimum points: $N \ge 2$.\\
Failure conditions: fails for $N < 2$ or if $\bar{m} = 0$.

\subsection{Anderson-Darling Normal}
The Anderson-Darling test statistic for normality:
\begin{equation}
    A^{*2} = A^2 \left(1 + \frac{4}{N} - \frac{25}{N^2}\right)
\end{equation}
Minimum points: $N \ge 1$.\\
Failure conditions: fails for $N = 0$.

\subsection{Reduced Chi-squared}
Reduced chi-squared with respect to the weighted mean:
\begin{equation}
    \chi^2_{\mathrm{red}} = \frac{1}{N-1} \sum_{i=1}^{N} \frac{(m_i - \bar{m}_w)^2}{\sigma_i^2}
\end{equation}
Minimum points: $N \ge 2$.\\
Failure conditions: Fails for $N < 2$ or if any $\sigma_i = 0$.

\subsection{Skew}
Sample skewness, corrected for bias:
\begin{equation}
    G_1 = \frac{N}{(N-1)(N-2)} \sum_{i=1}^{N} \left( \frac{m_i - \bar{m}}{\sigma_m} \right)^3
\end{equation}
Minimum points: $N \ge 3$.\\
Failure conditions: Fails for $N < 3$ or if $\sigma_m = 0$.

\subsection{Stetson K}
Stetson's K variability index:
\begin{equation}
    K = \frac{\sum_{i=1}^{N} |\delta_i|}{\sqrt{N \cdot \sum_{i=1}^{N} \delta_i^2}}, \quad \delta_i = \frac{m_i - \bar{m}_w}{\sigma_i}
\end{equation}
Minimum points: $N \ge 1$.\\
Failure conditions: Fails for $N = 0$, if all $\sigma_i = 0$, or if all $\delta_i = 0$.

\section{Spectral fits of SN~2023mtp}
\label{ap:sn2023mtp}

In this section, we present the best-fitting \texttt{SNID}~\citep{2007ApJ...666.1024B} templates for the four publicly available spectra of SN~2023mtp.

\begin{figure*}
    \centering
    \includegraphics[width=0.65\linewidth]{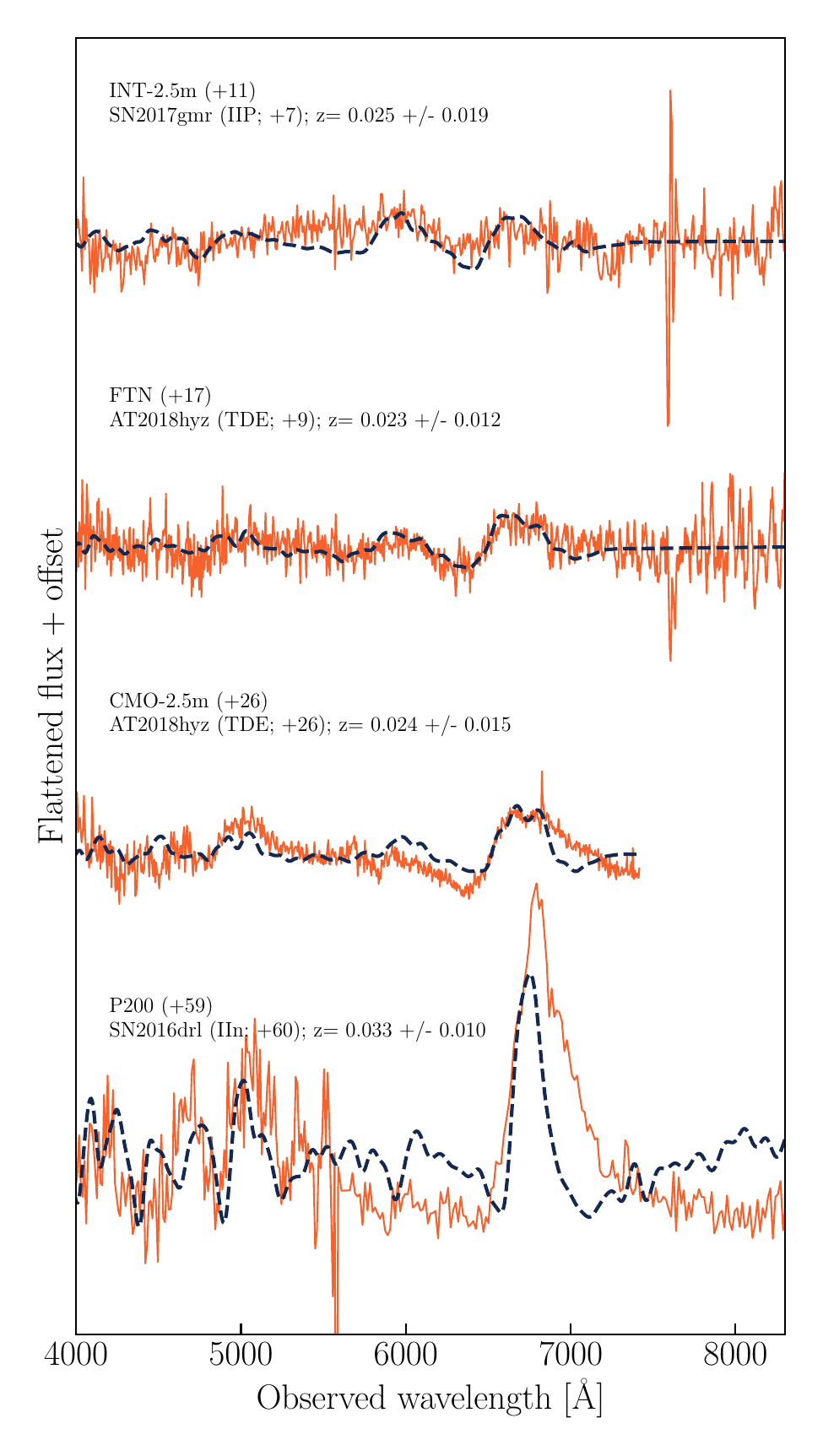}
    \caption{Publicly available spectra (solid orange lines) of SN~2023mtp obtained with the following instruments: the Intermediate Dispersion Spectrograph on the 2.5-m Isaac Newton Telescope (INT)~\citep{2023TNSCR2014....1O}, the FLOYDS spectrograph on the 2-m Faulkes Telescope North (FTN)~\citep{2023TNSCR2213....1F}, the Transient Double-beam Spectrograph on the 2.5-m CMO telescope~\citep{2023TNSCR2494....1D}, and the Double Spectrograph on the Palomar 5.1-m (P200) telescope \citep{2024TNSCR3189....1S}. Dashed dark-blue lines show the \texttt{SNID}~\citep{2007ApJ...666.1024B} best-fit template for each SN~2023mtp spectrum; the corresponding supernova name together with its type, phase relative to maximum and redshift are listed on the plot.}
    \label{fig:2023mtp_snid}
\end{figure*}

\section{Supernova light-curve fits}
\label{ap:sn_fit}

In this section, we present multicolor light-curve fits for supernova candidates using the \texttt{sncosmo}\footnote{\url{https://sncosmo.readthedocs.io/en/stable/}} library.

\begin{figure*}[t]
\centering
\includegraphics[width=0.48\linewidth]{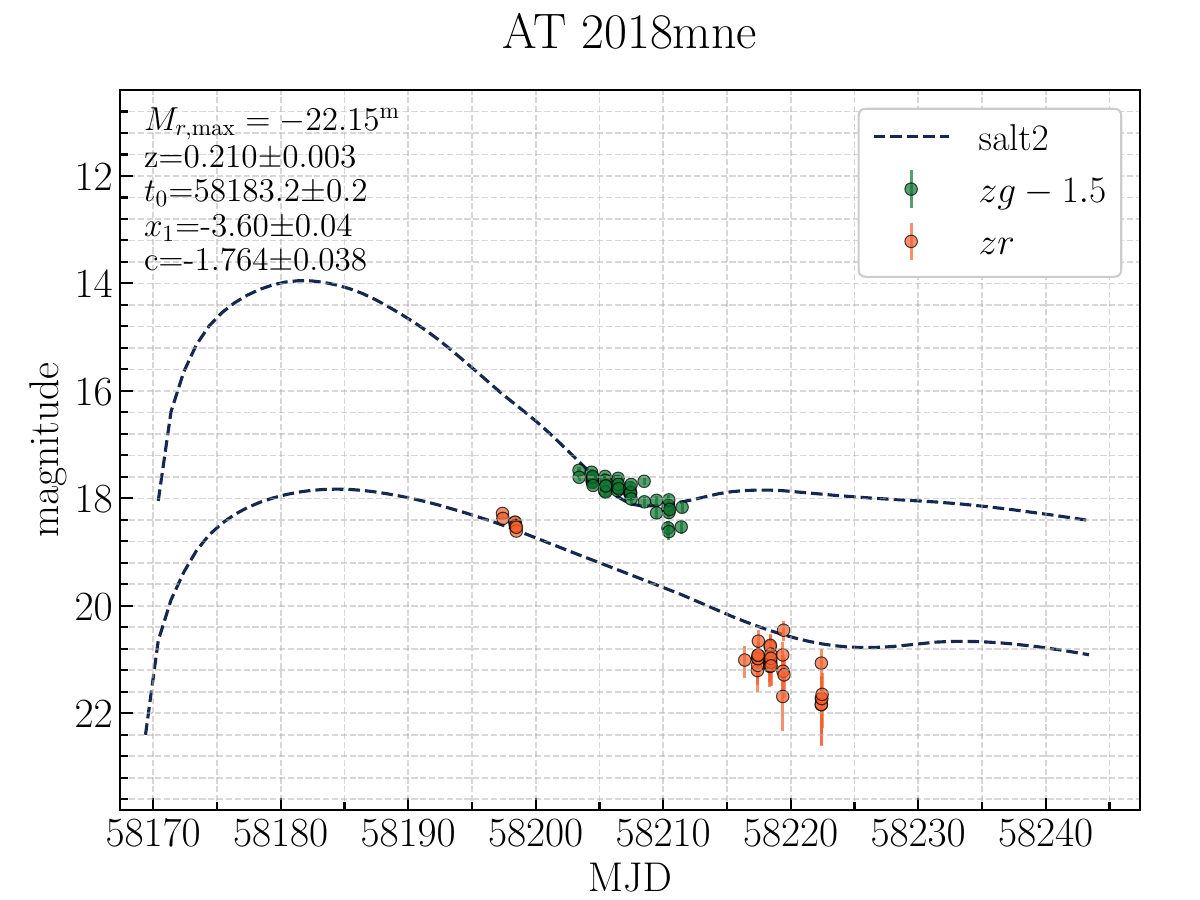}\hfill
\includegraphics[width=0.48\linewidth]{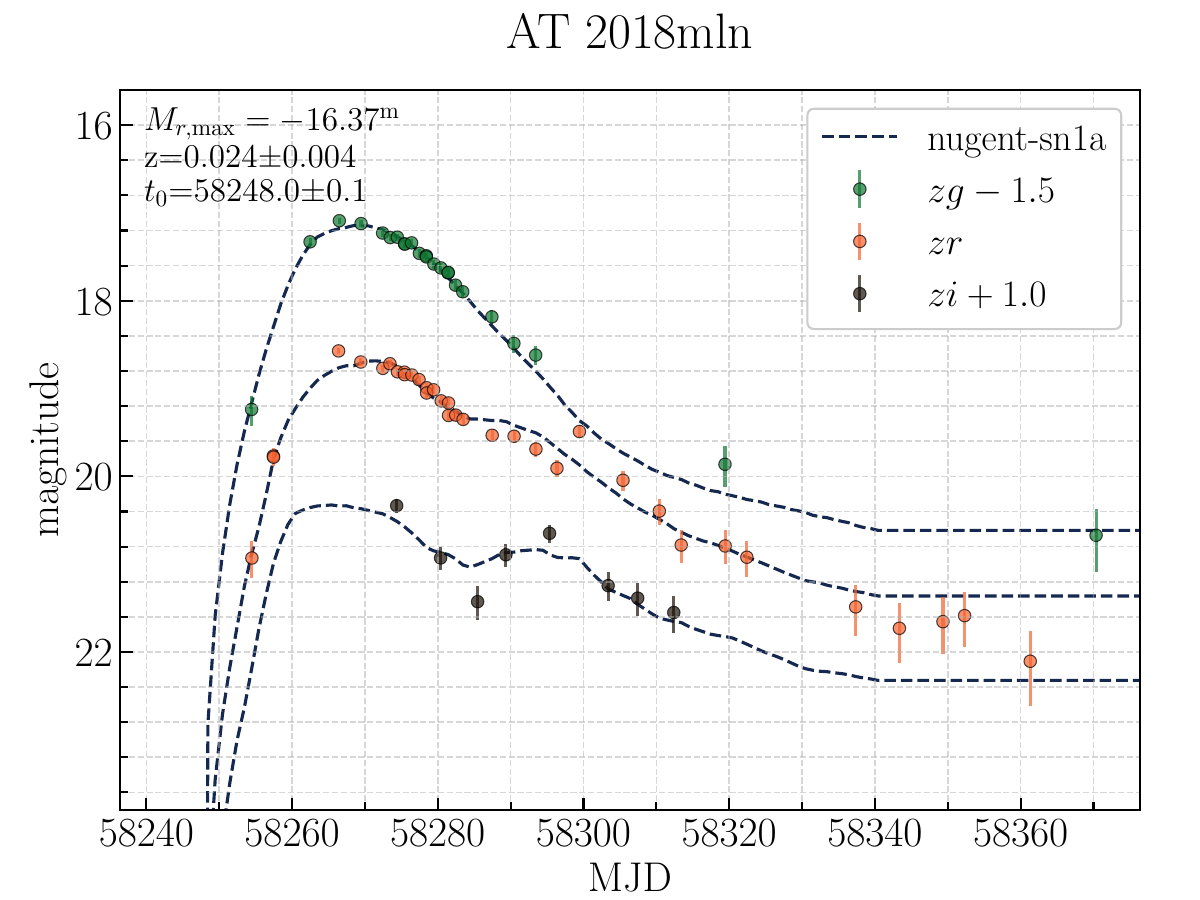}

\includegraphics[width=0.48\linewidth]{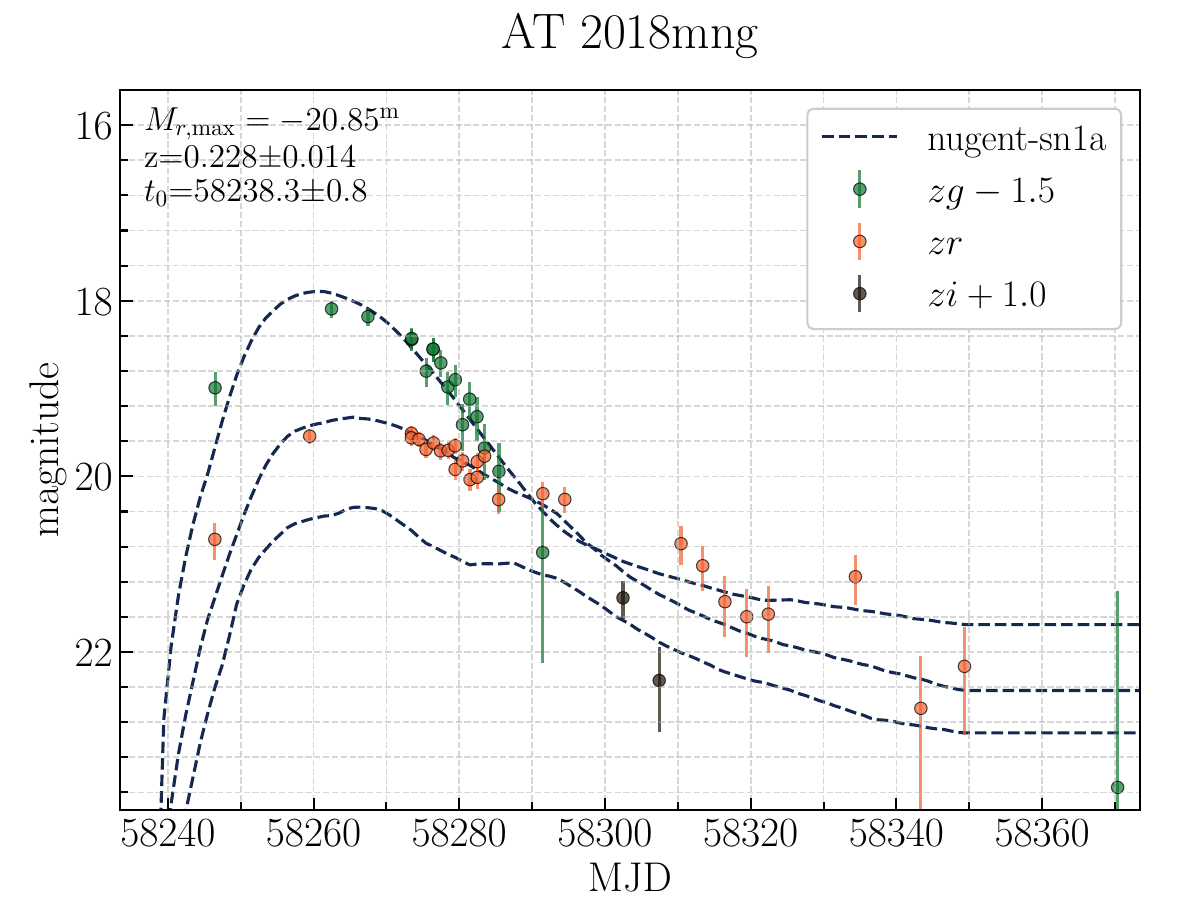}\hfill
\includegraphics[width=0.48\linewidth]{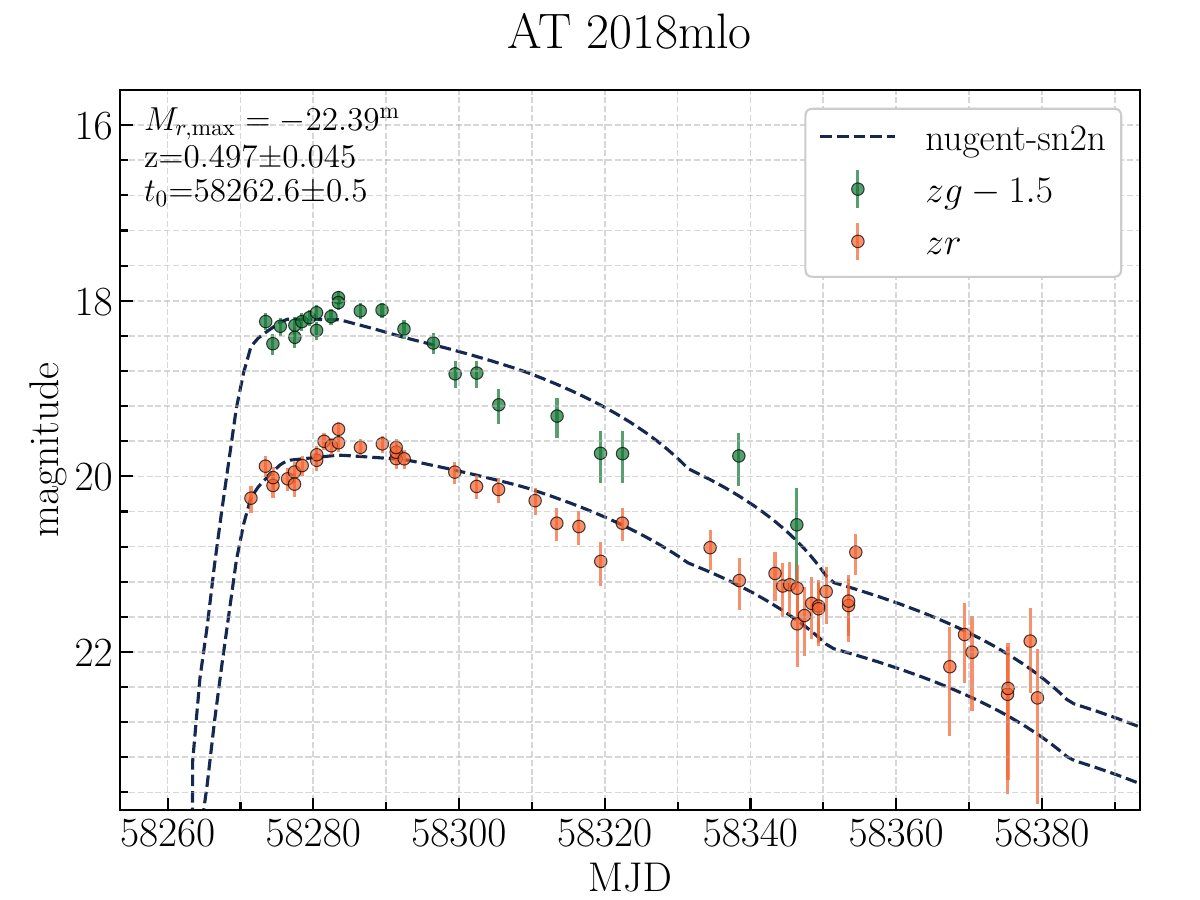}

\includegraphics[width=0.48\linewidth]{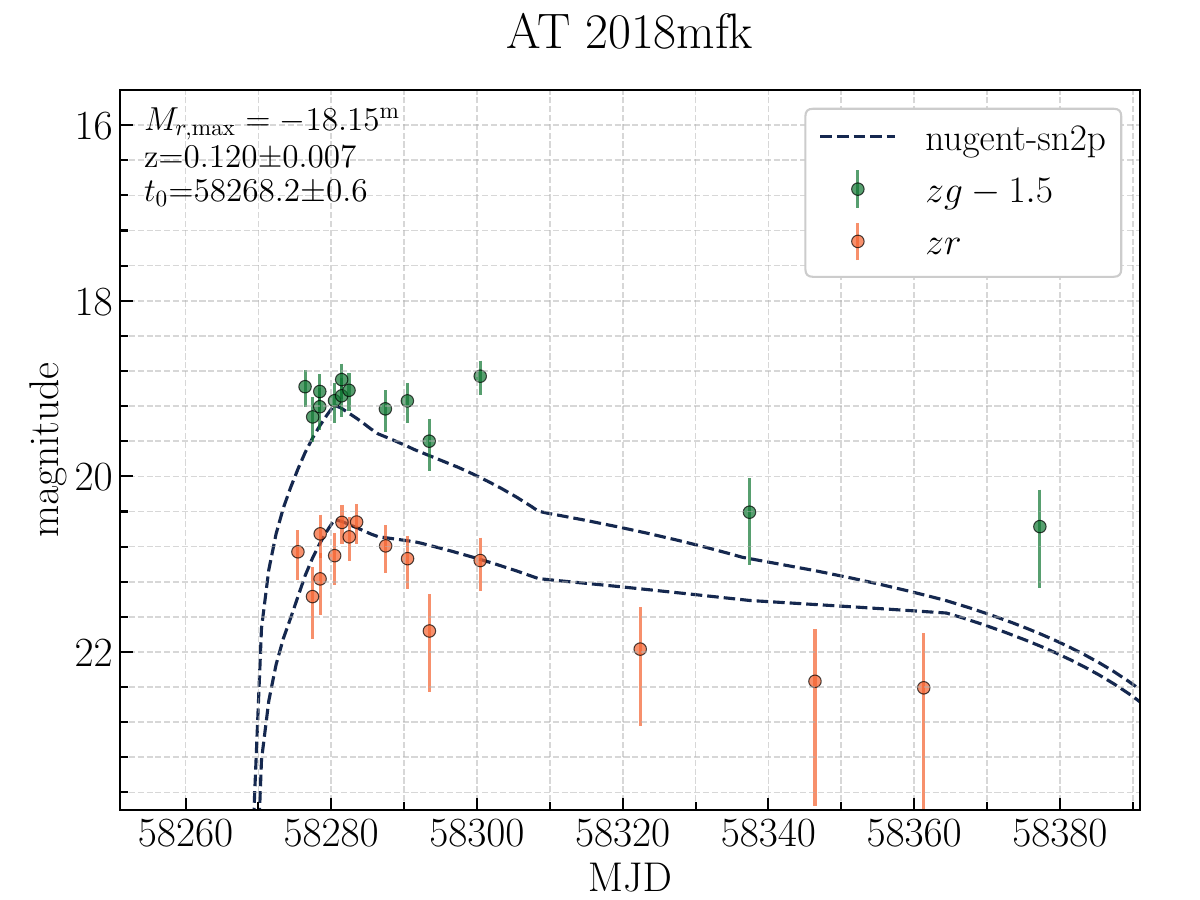}\hfill
\includegraphics[width=0.48\linewidth]{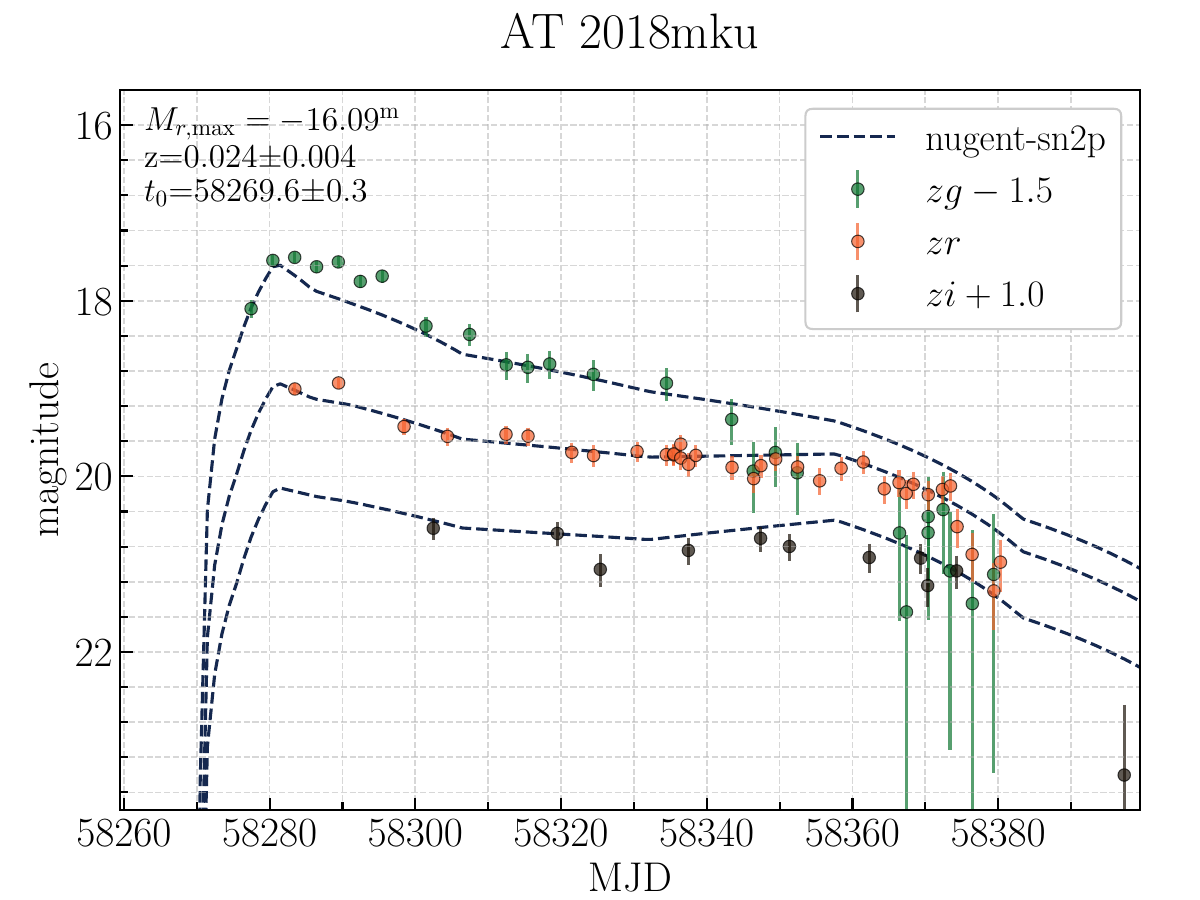}

\caption{ZTF light curves of supernova candidates found by the Fink anomaly detection pipeline, with best-model fits and fitted model parameters.}
\label{fig:ztf_lc_grid}
\end{figure*}

\begin{figure*}[t]\ContinuedFloat
\centering
\includegraphics[width=0.48\linewidth]{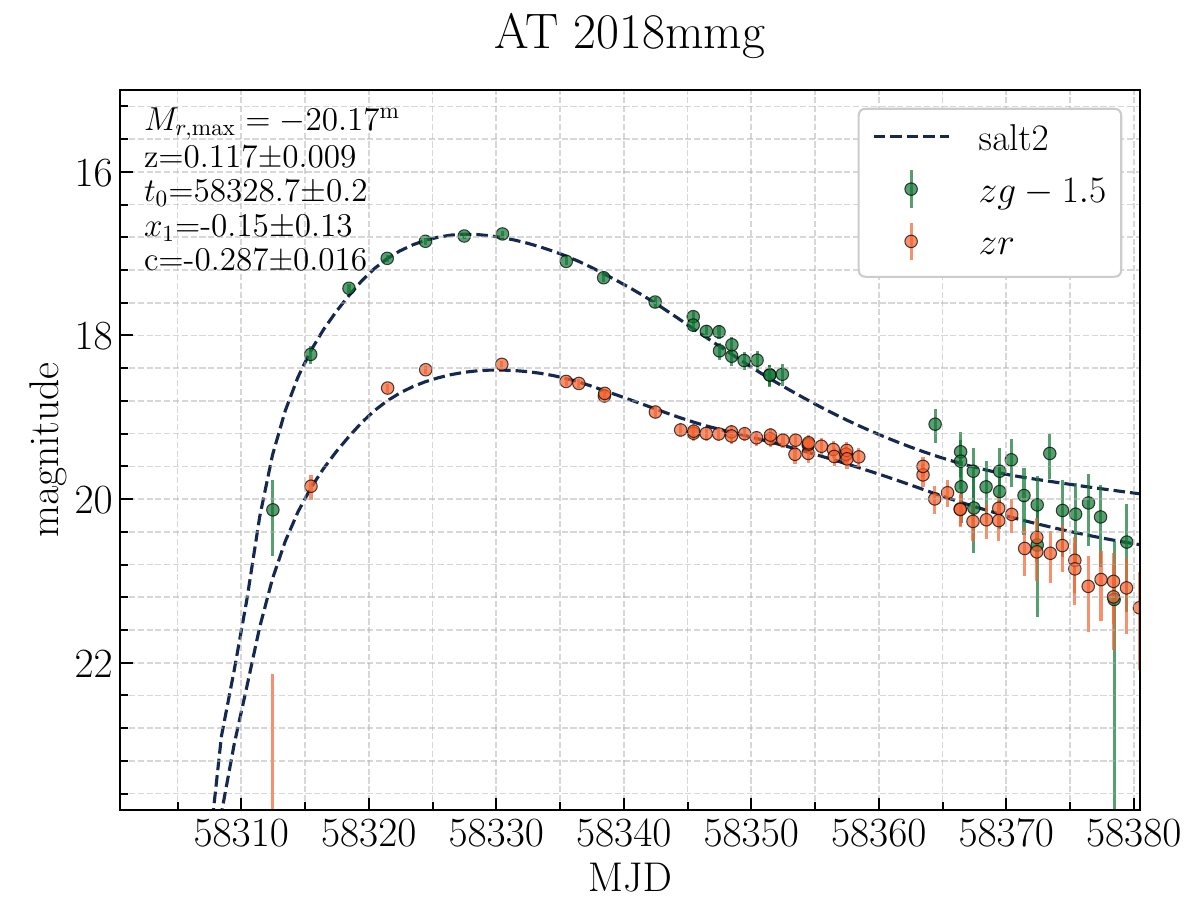}\hfill
\includegraphics[width=0.48\linewidth]{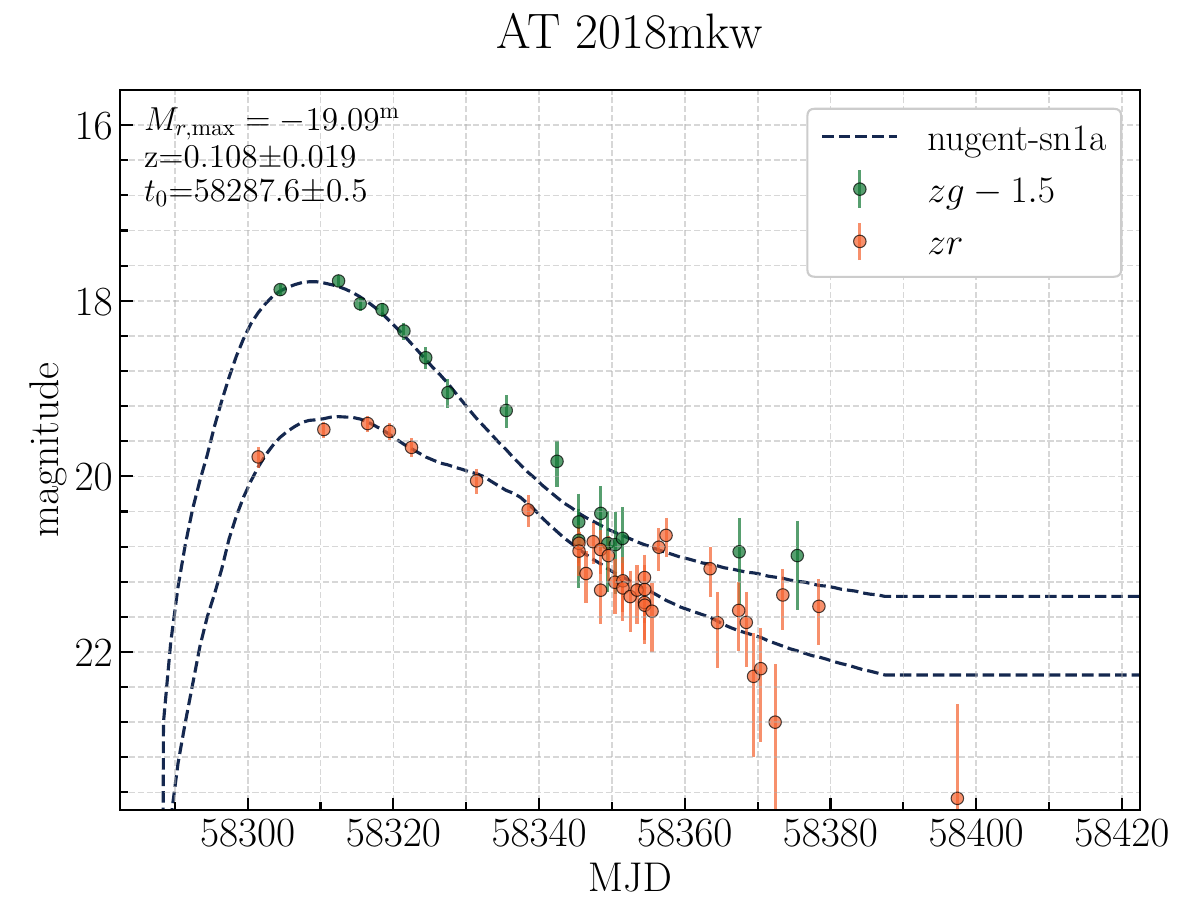}

\includegraphics[width=0.48\linewidth]{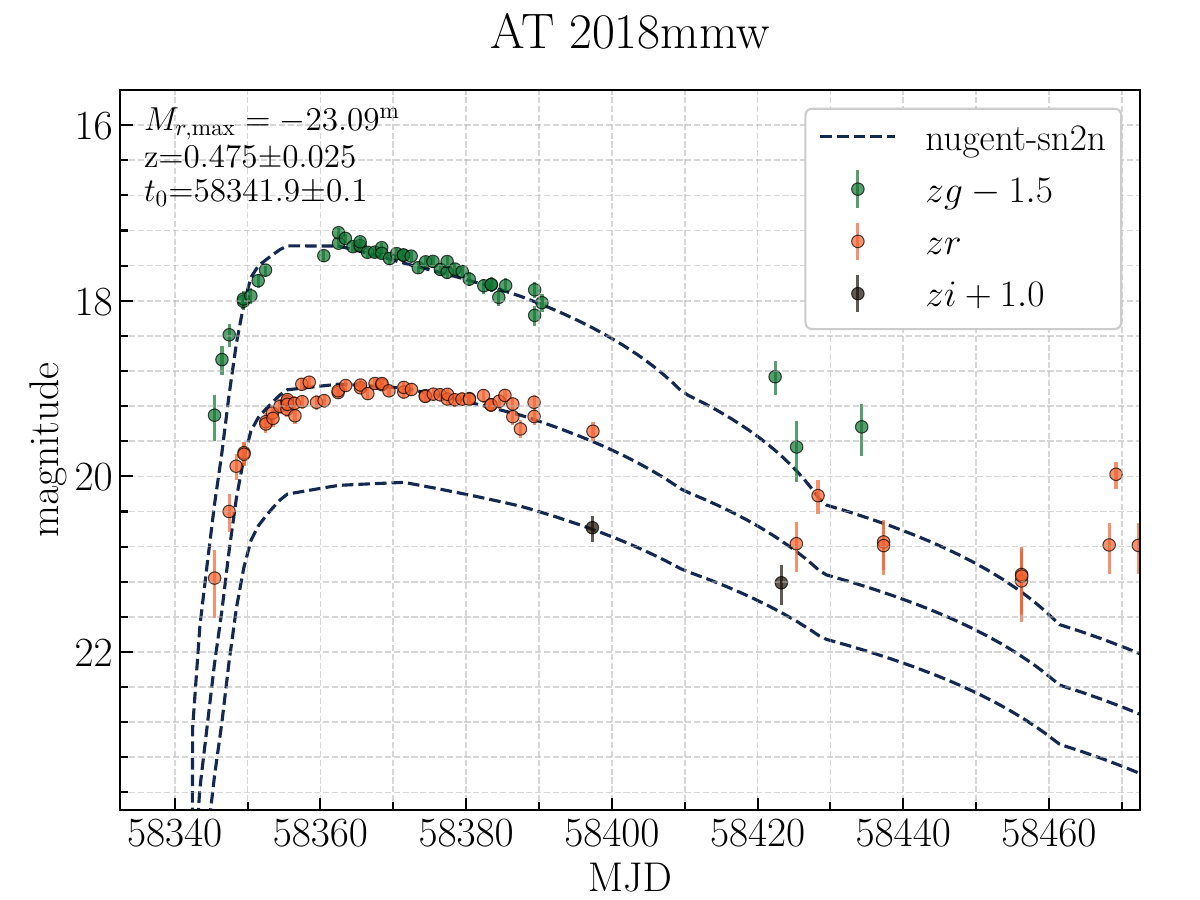}\hfill
\includegraphics[width=0.48\linewidth]{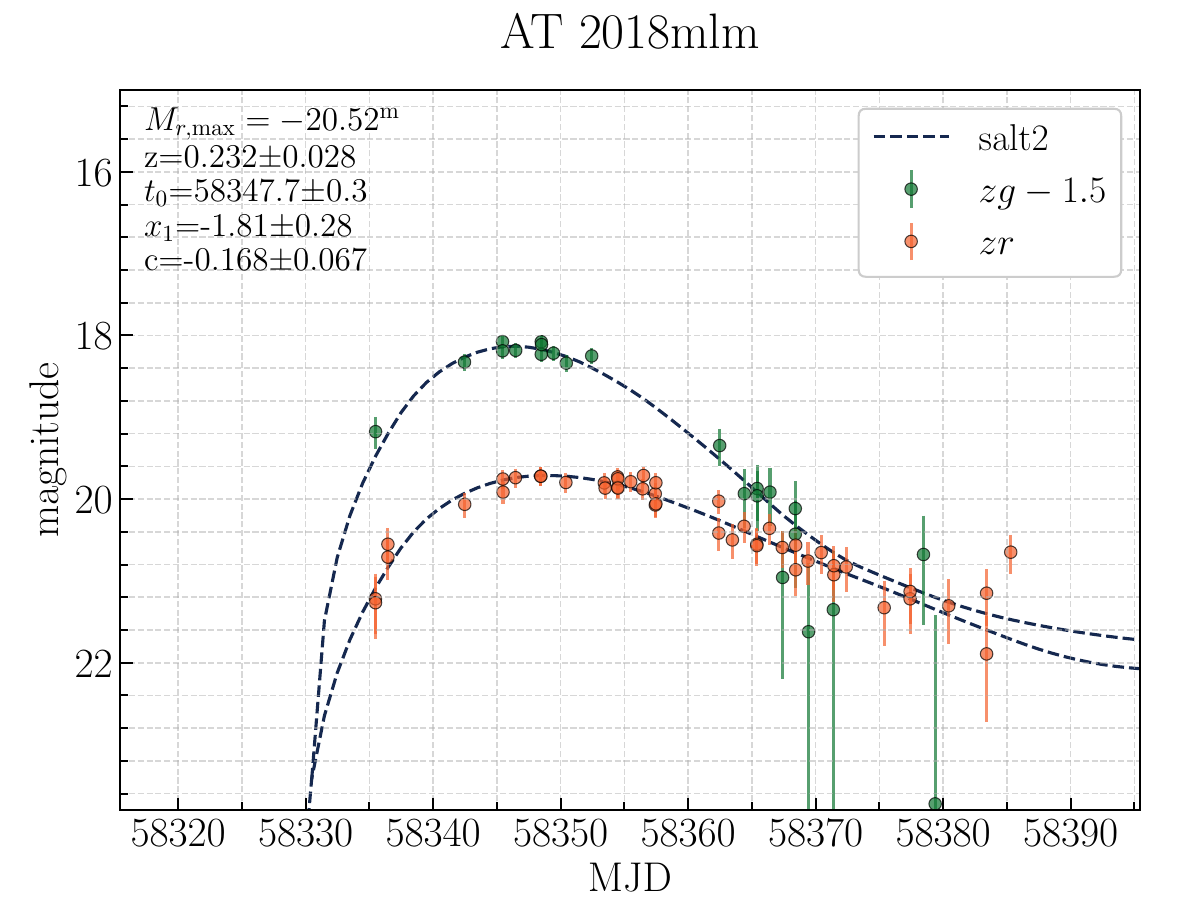}

\includegraphics[width=0.48\linewidth]{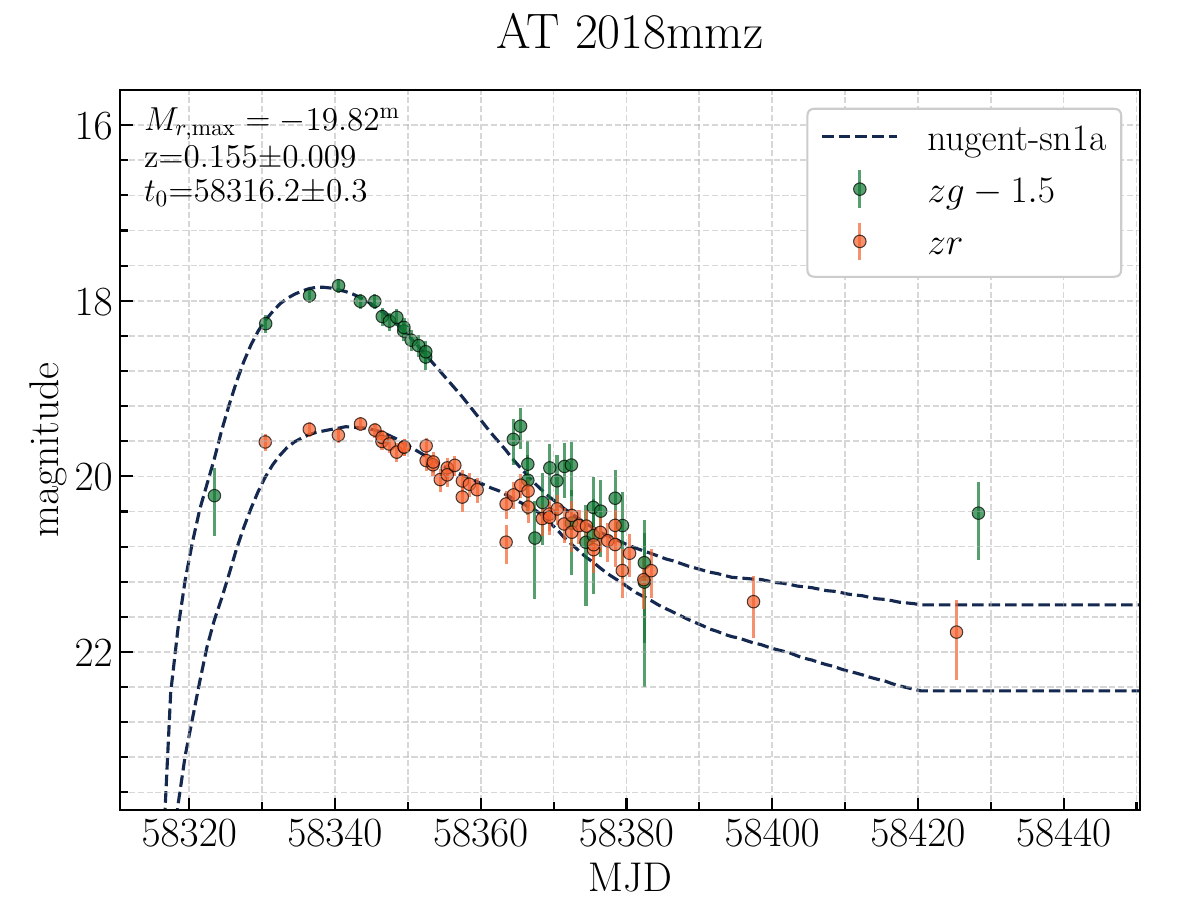}\hfill
\includegraphics[width=0.48\linewidth]{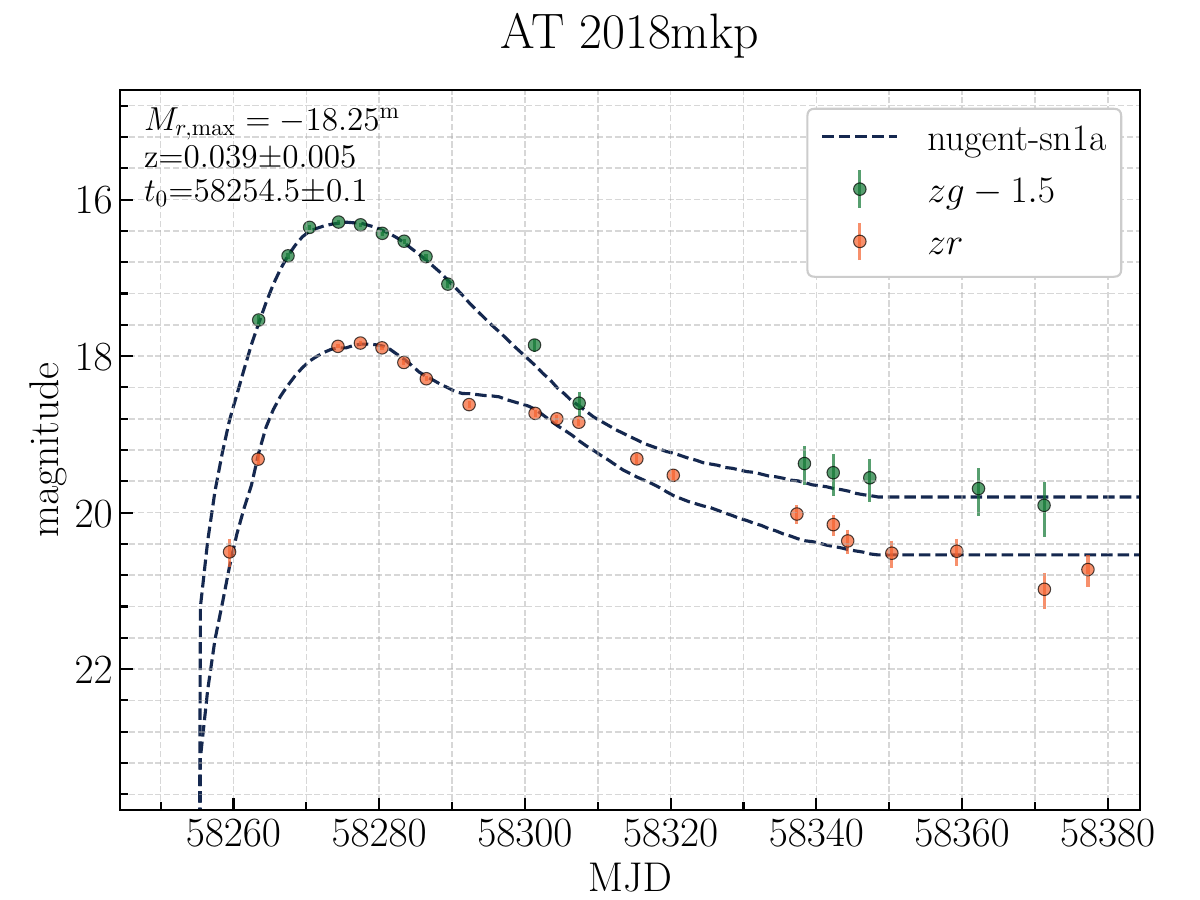}

\caption{ZTF light curves of supernova candidates found by the Fink anomaly detection pipeline, with best-model fits and fitted model parameters (continued).}
\end{figure*}

\begin{figure*}[t]\ContinuedFloat
\centering
\includegraphics[width=0.48\linewidth]{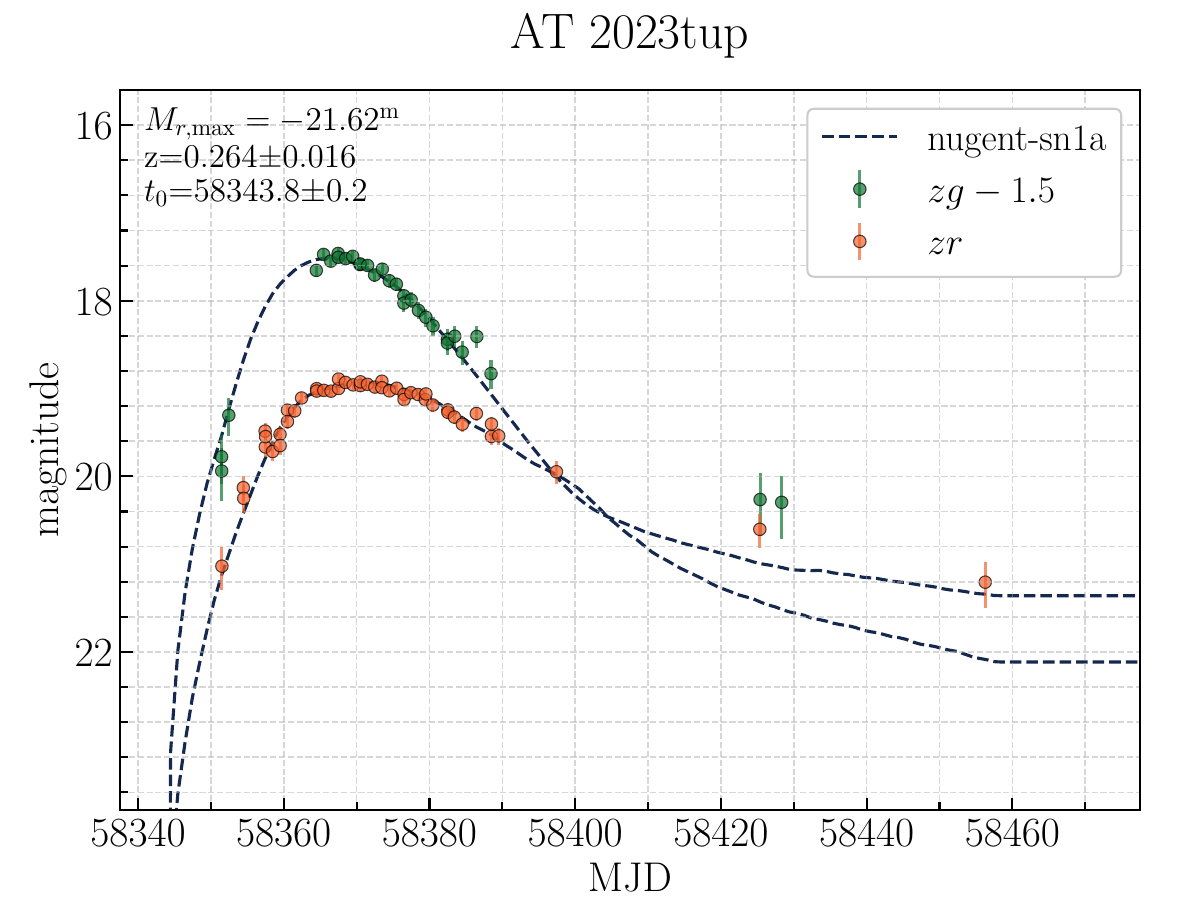}\hfill
\includegraphics[width=0.48\linewidth]{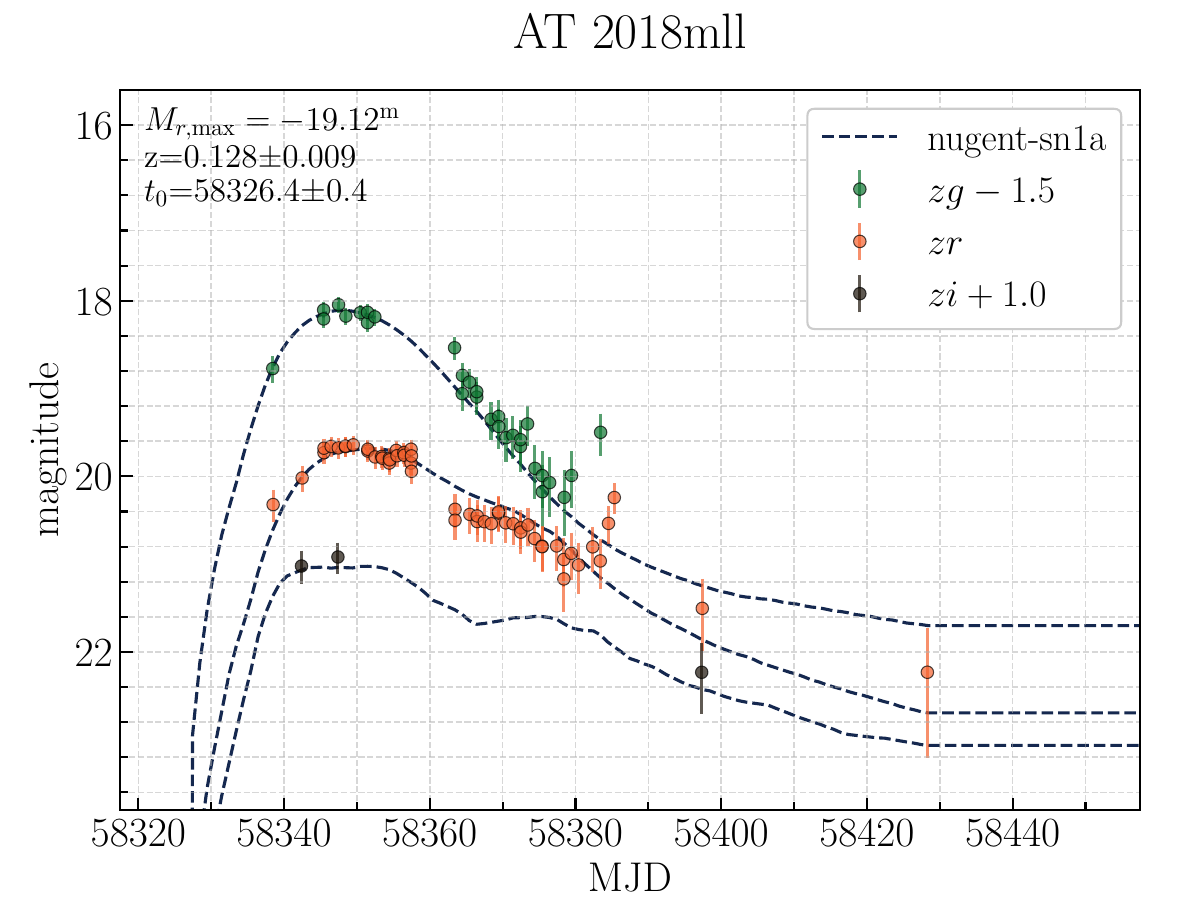}

\includegraphics[width=0.48\linewidth]{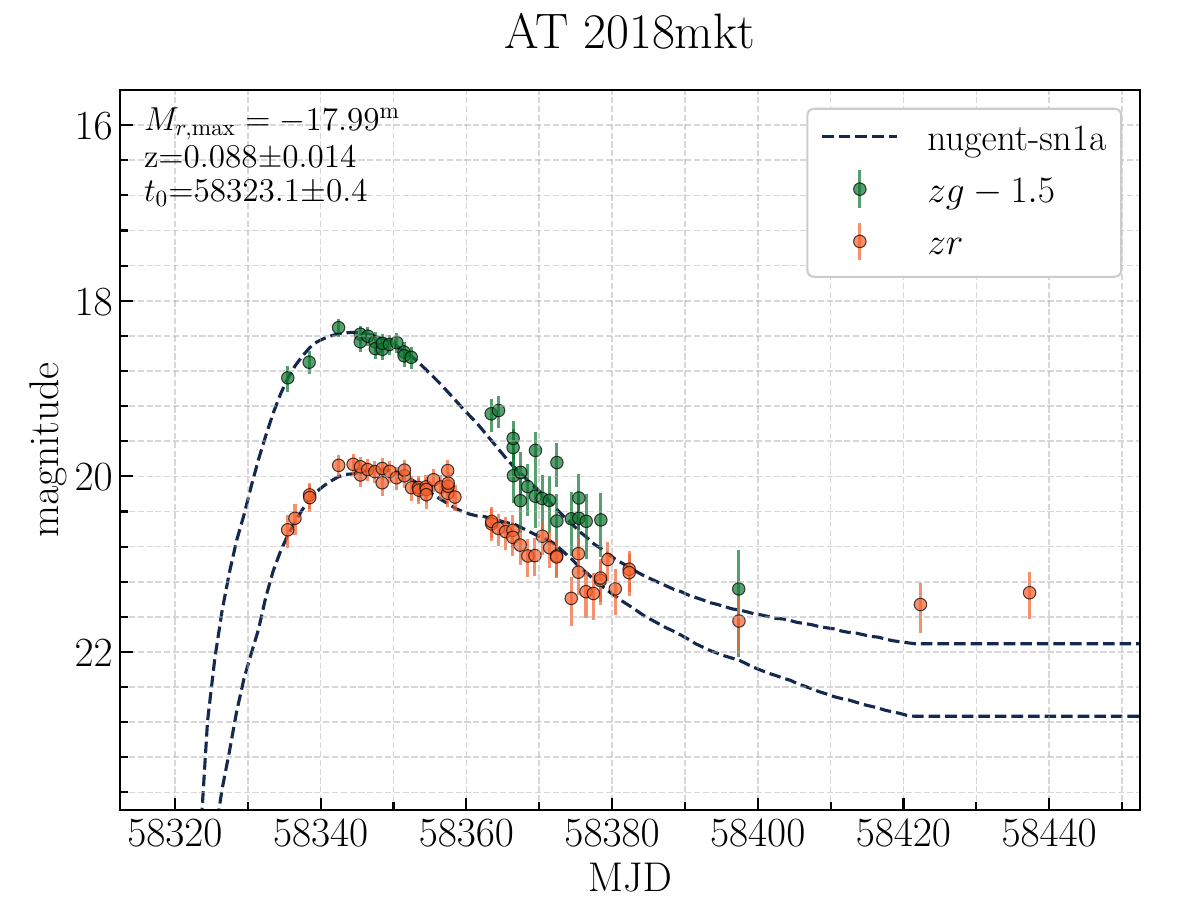}\hfill
\includegraphics[width=0.48\linewidth]{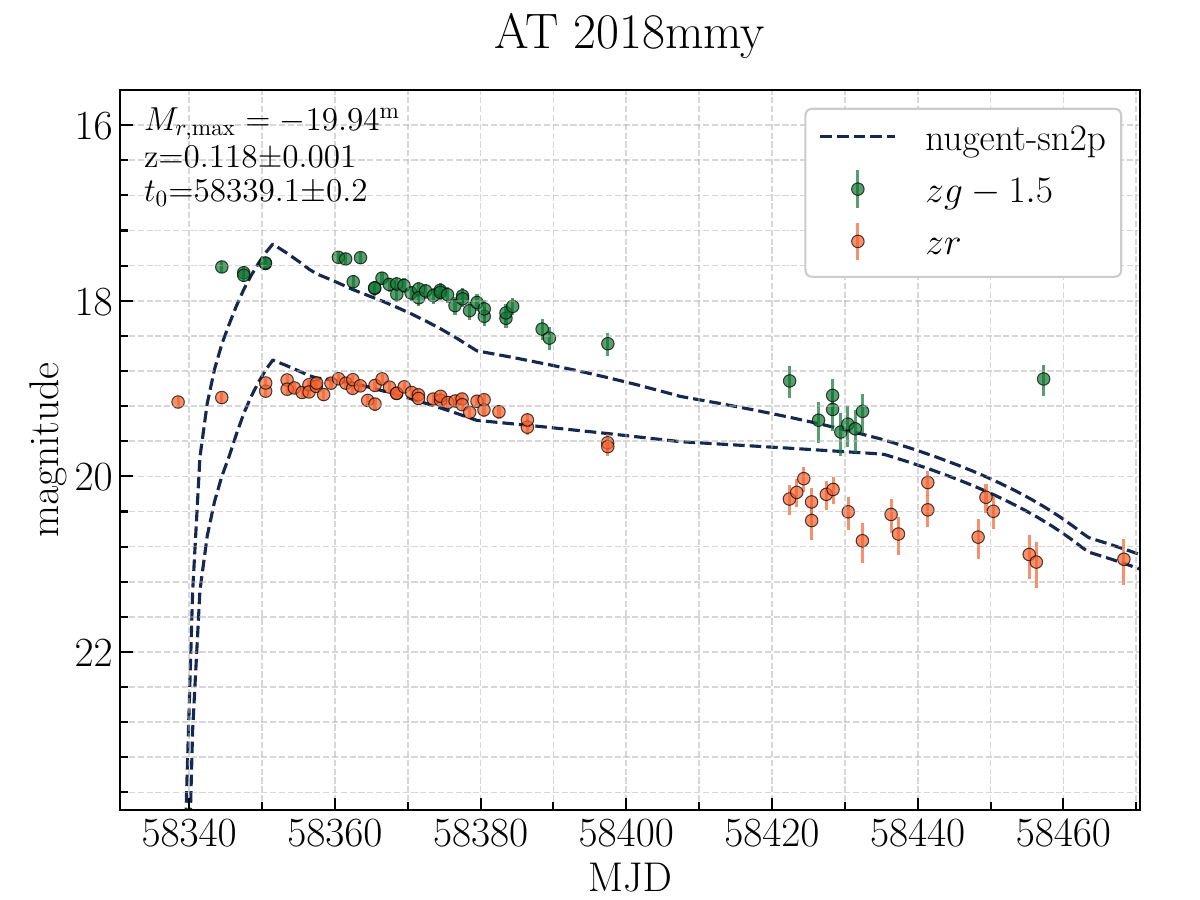}

\includegraphics[width=0.48\linewidth]{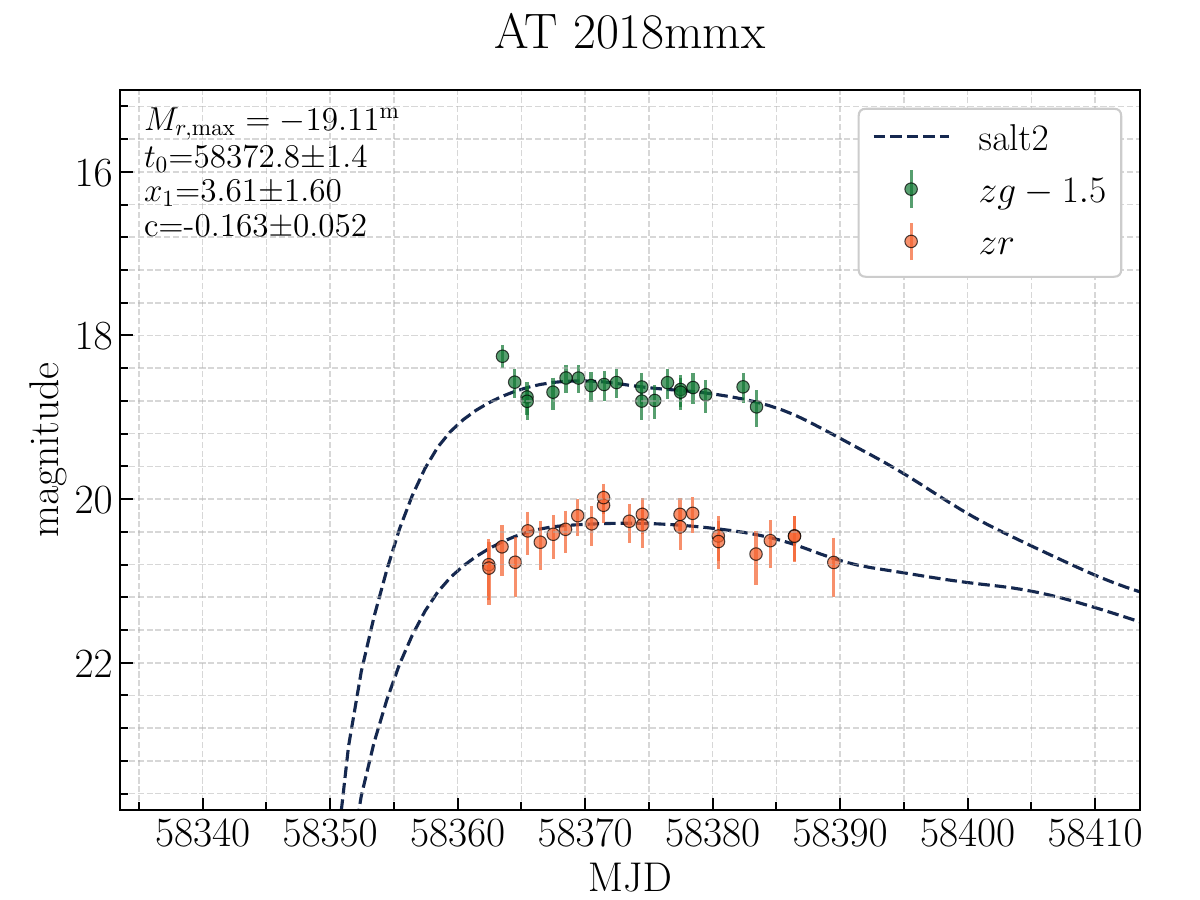}\hfill
\includegraphics[width=0.48\linewidth]{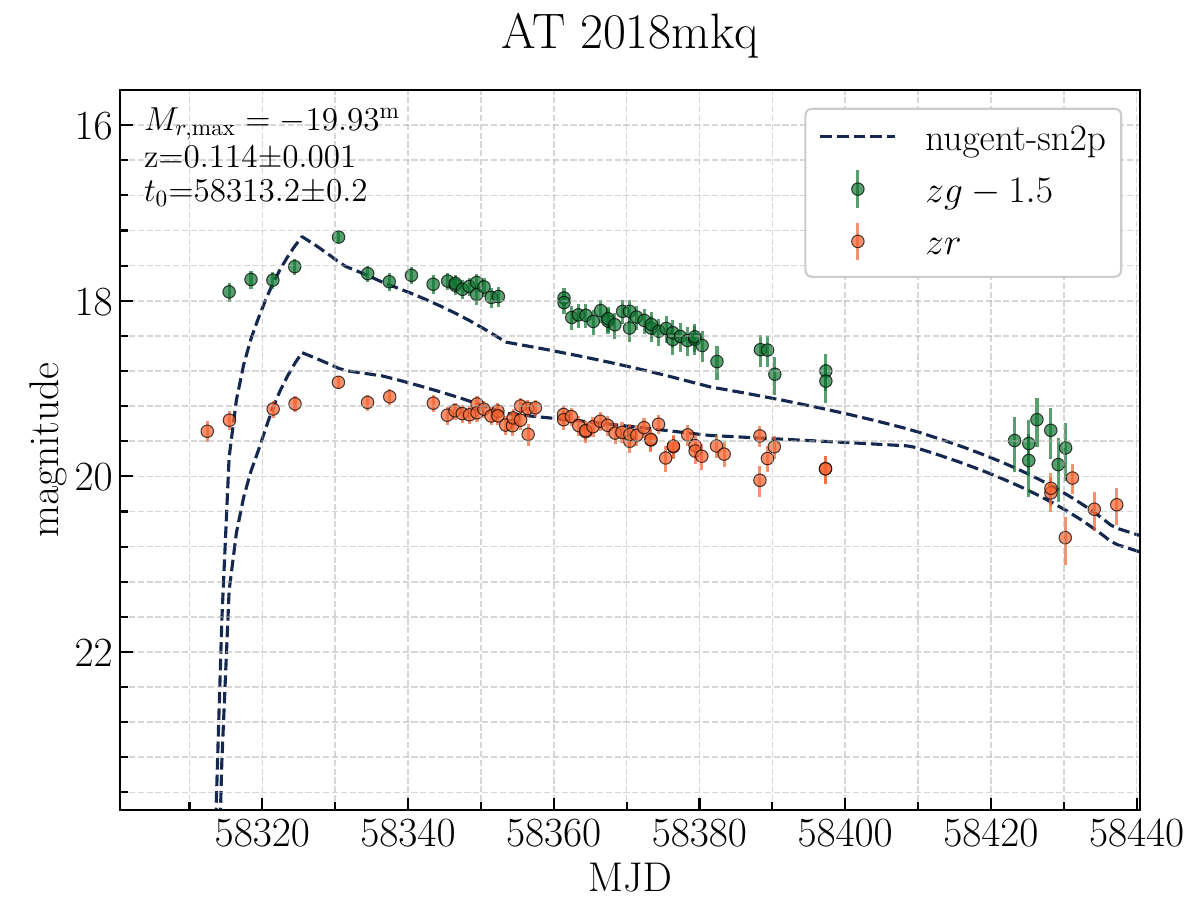}

\caption{ZTF light curves of supernova candidates found by the Fink anomaly detection pipeline, with best-model fits and fitted model parameters (continued).}
\end{figure*}

\begin{figure*}[t]\ContinuedFloat
\centering
\includegraphics[width=0.48\linewidth]{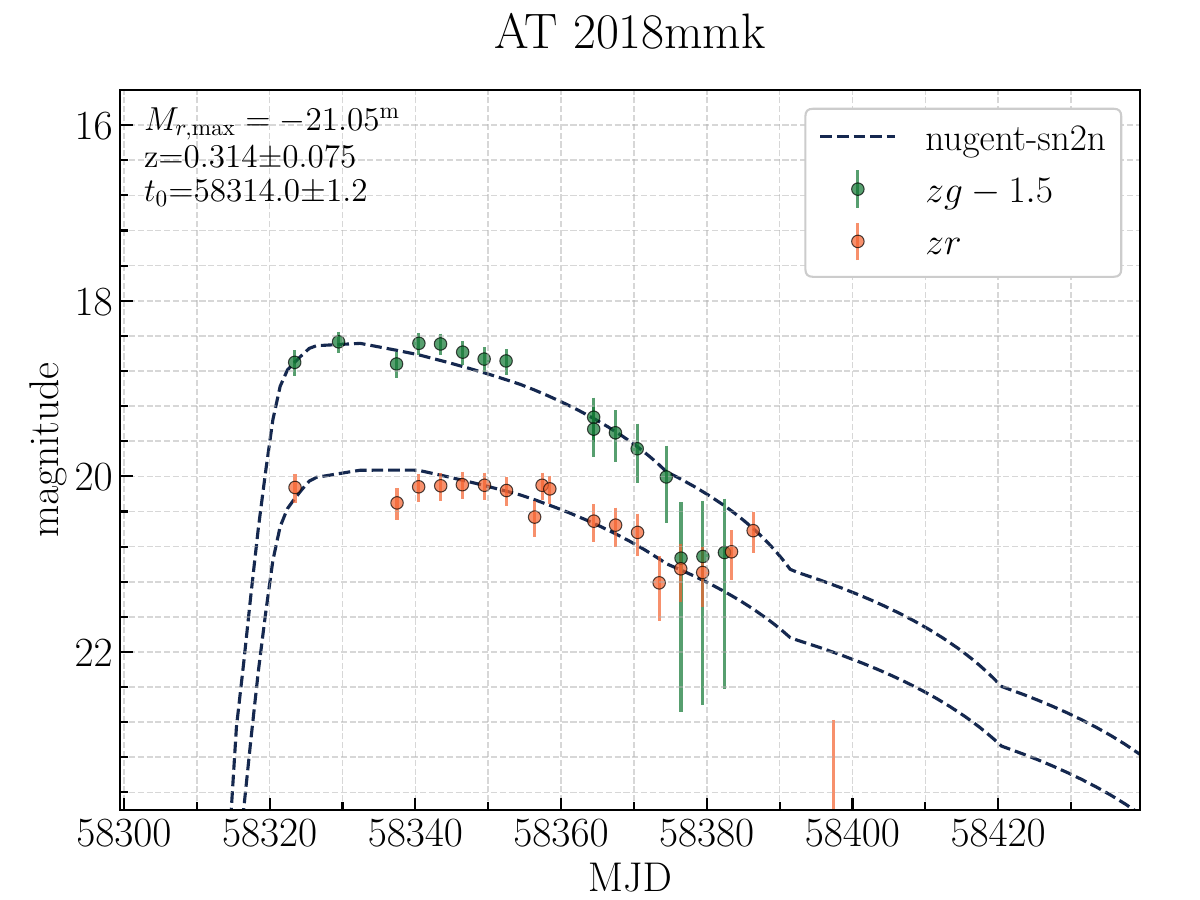}\hfill
\includegraphics[width=0.48\linewidth]{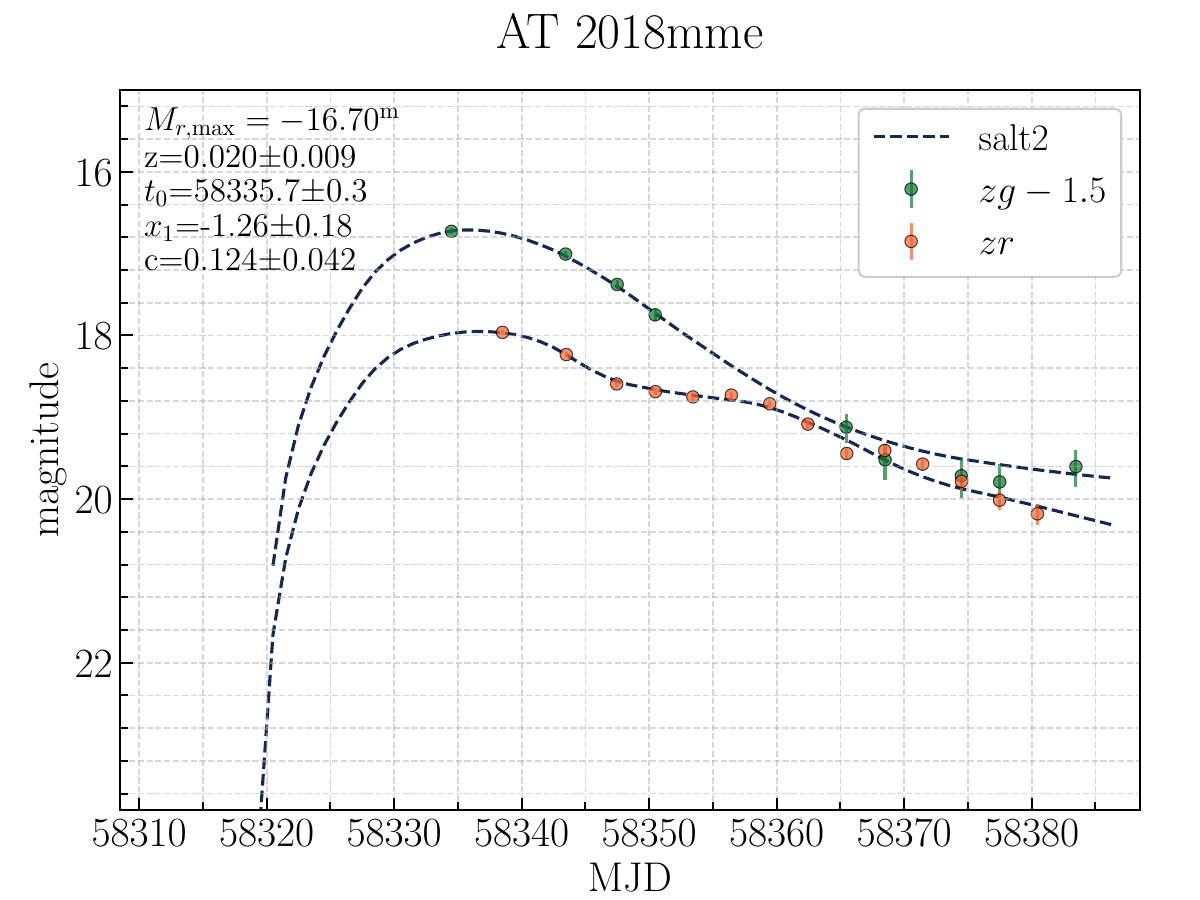}

\includegraphics[width=0.48\linewidth]{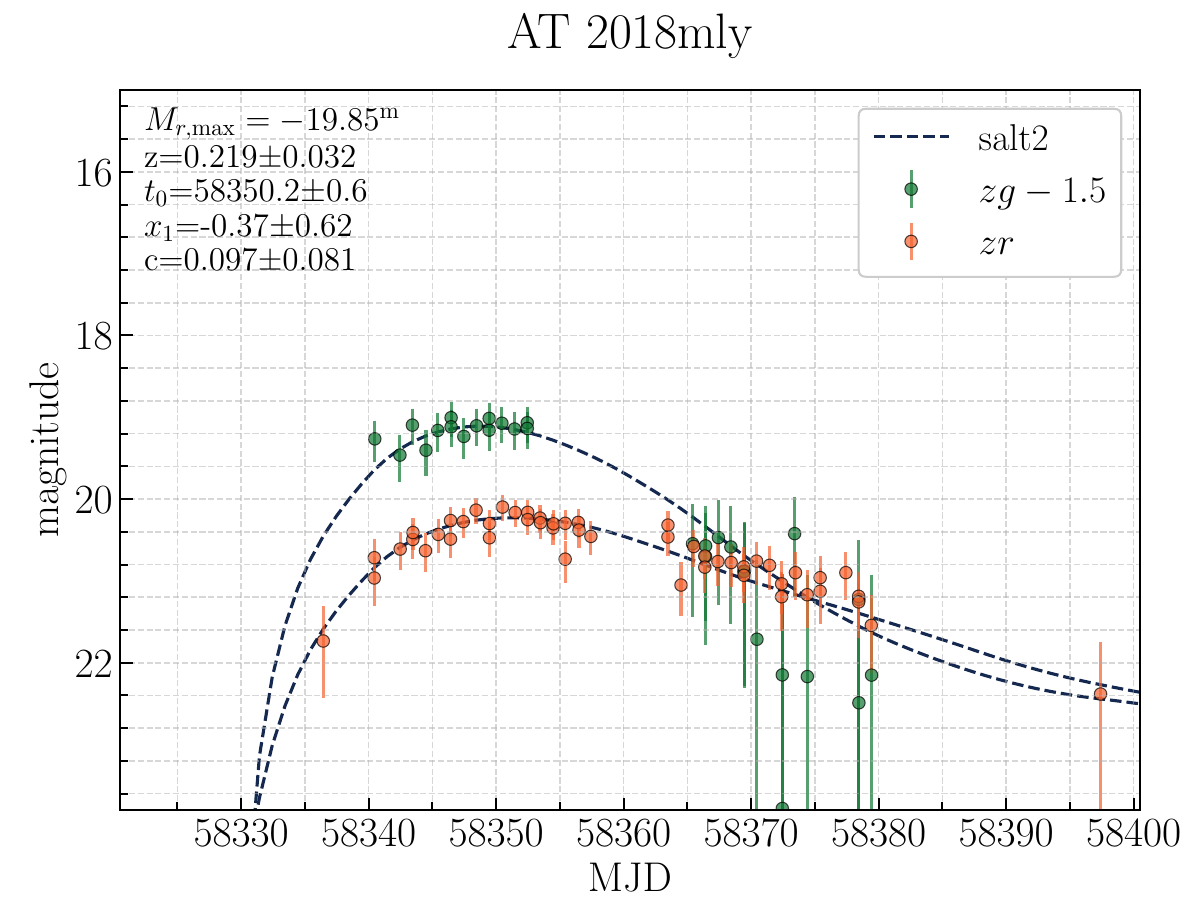}\hfill
\includegraphics[width=0.48\linewidth]{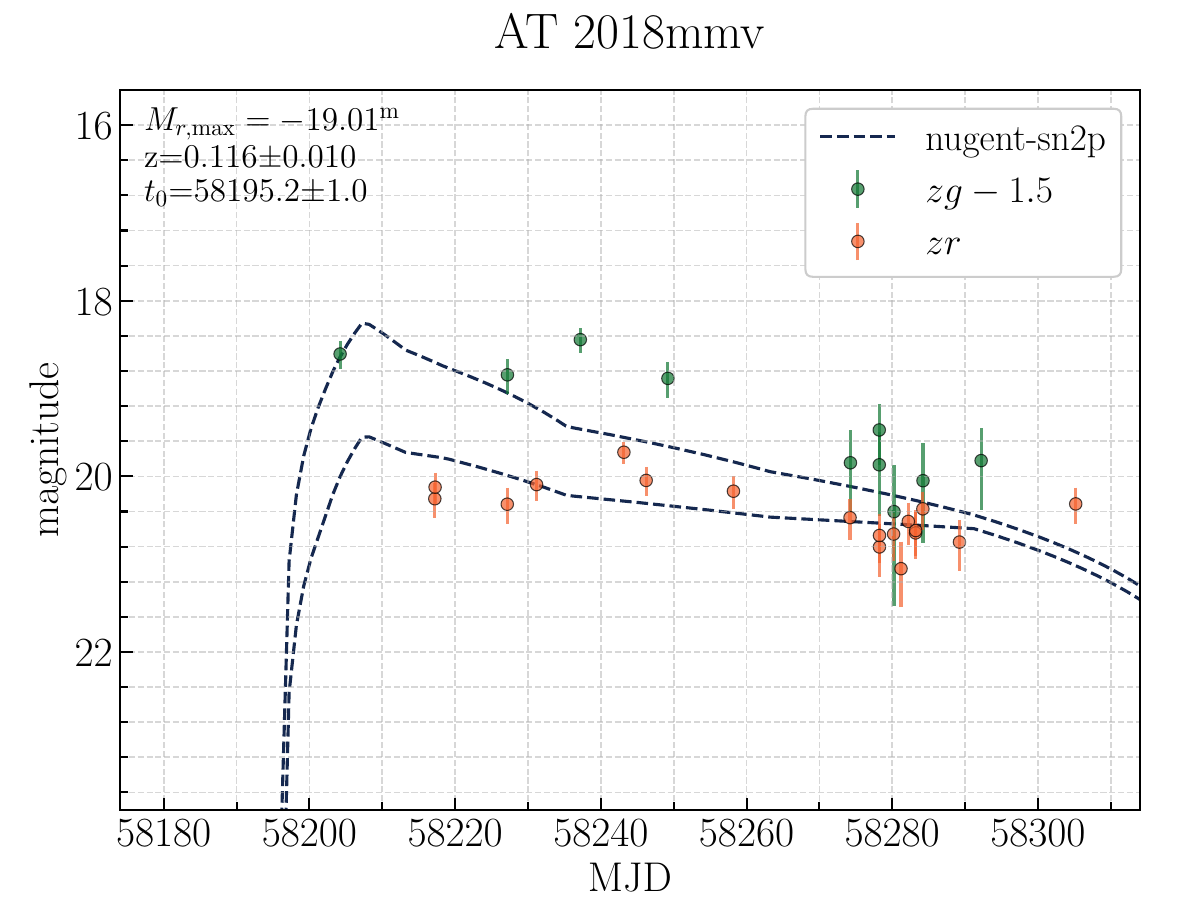}

\includegraphics[width=0.48\linewidth]{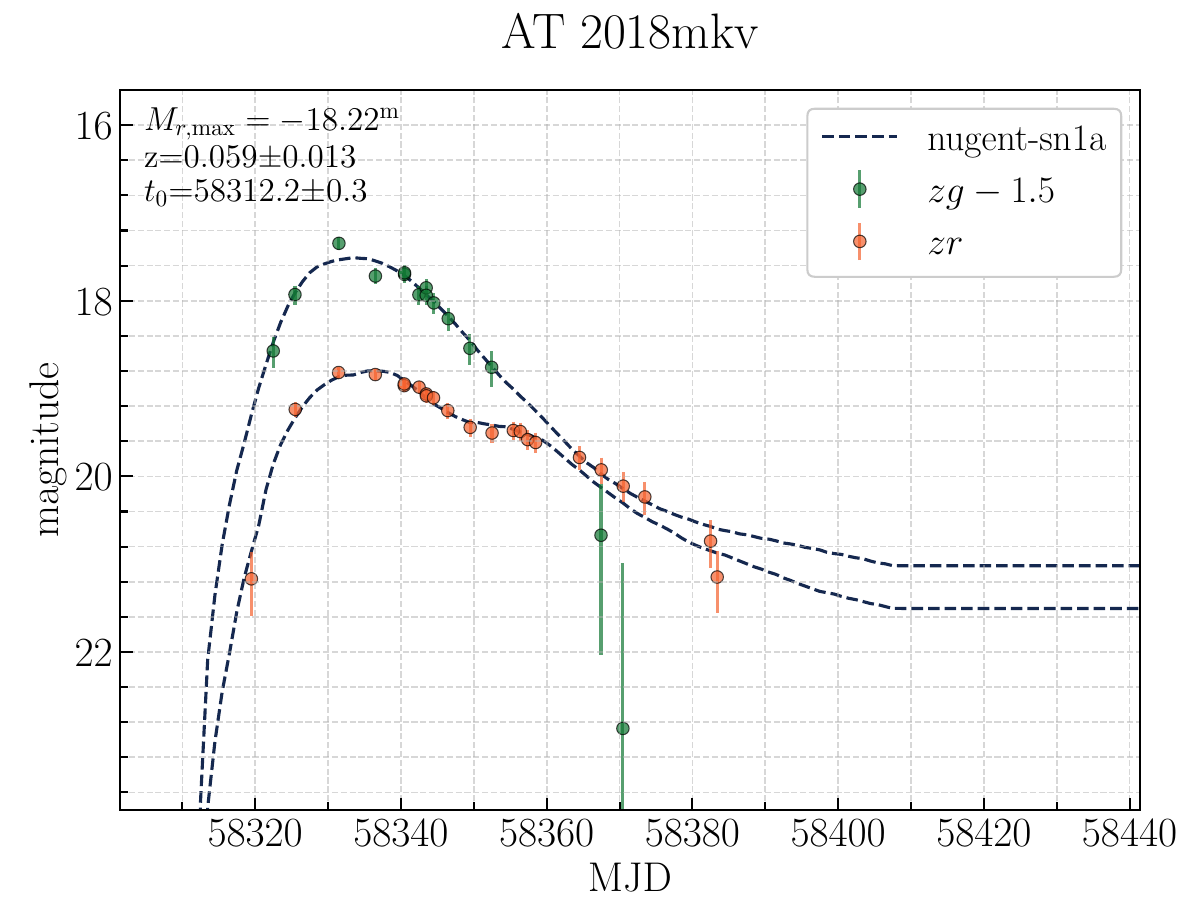}\hfill
\includegraphics[width=0.48\linewidth]{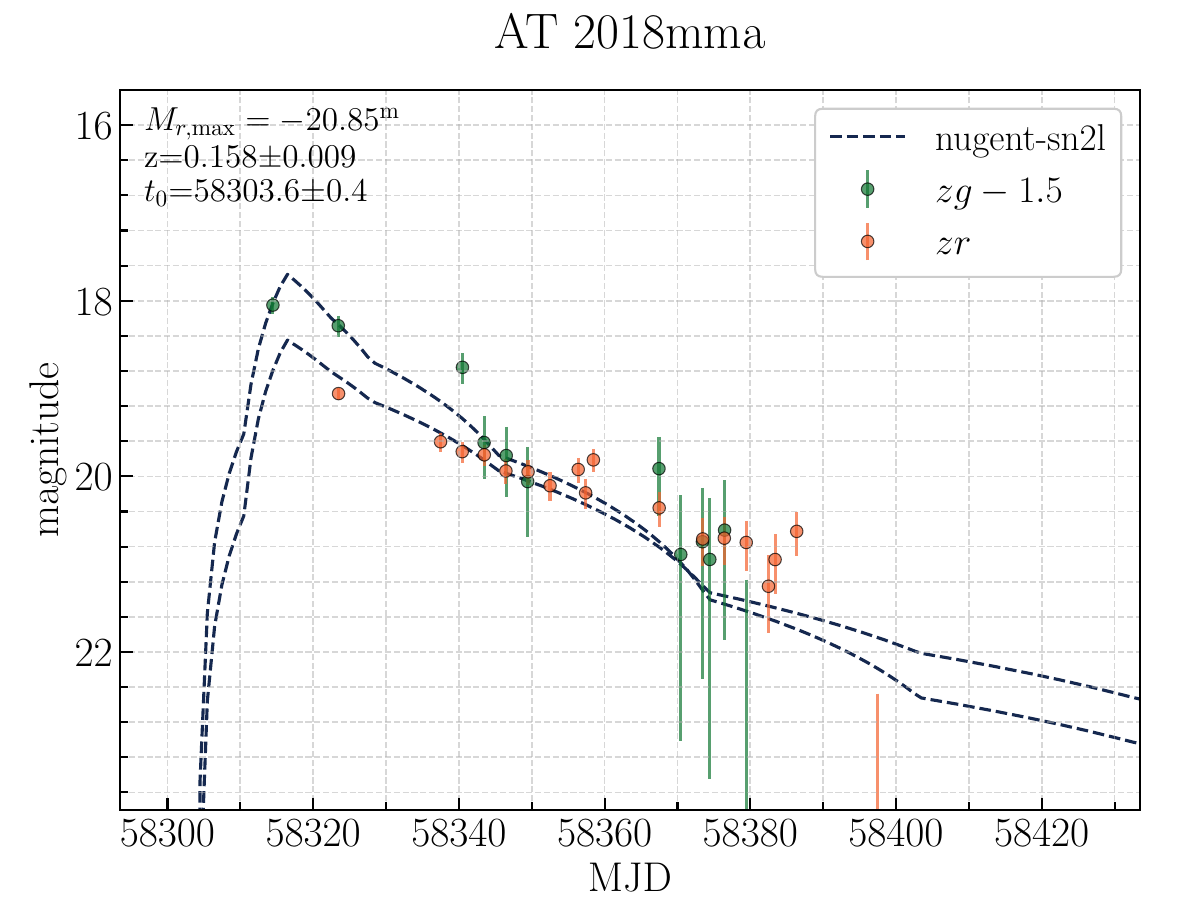}

\caption{ZTF light curves of supernova candidates found by the Fink anomaly detection pipeline, with best-model fits and fitted model parameters (continued).}
\end{figure*}

\begin{figure*}[t]\ContinuedFloat
\centering
\includegraphics[width=0.48\linewidth]{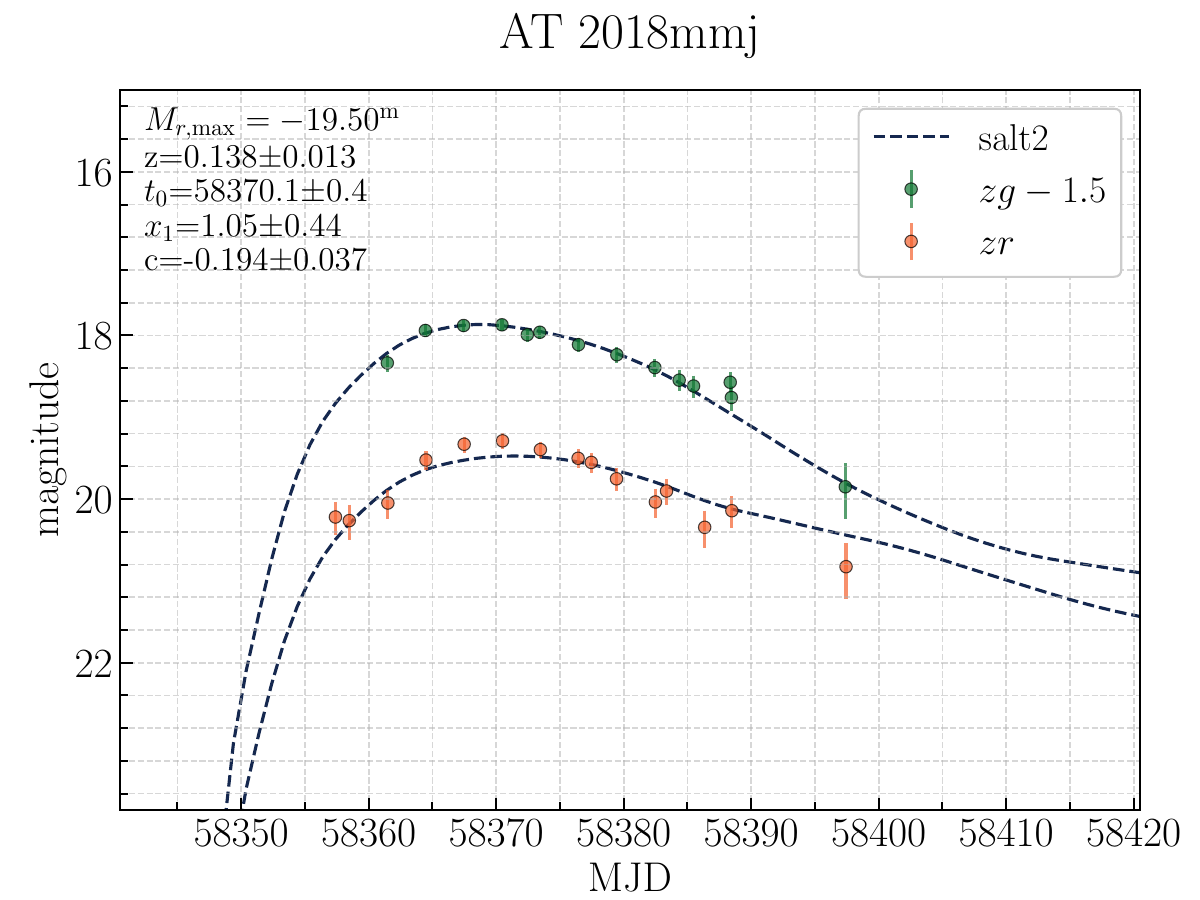}\hfill
\includegraphics[width=0.48\linewidth]{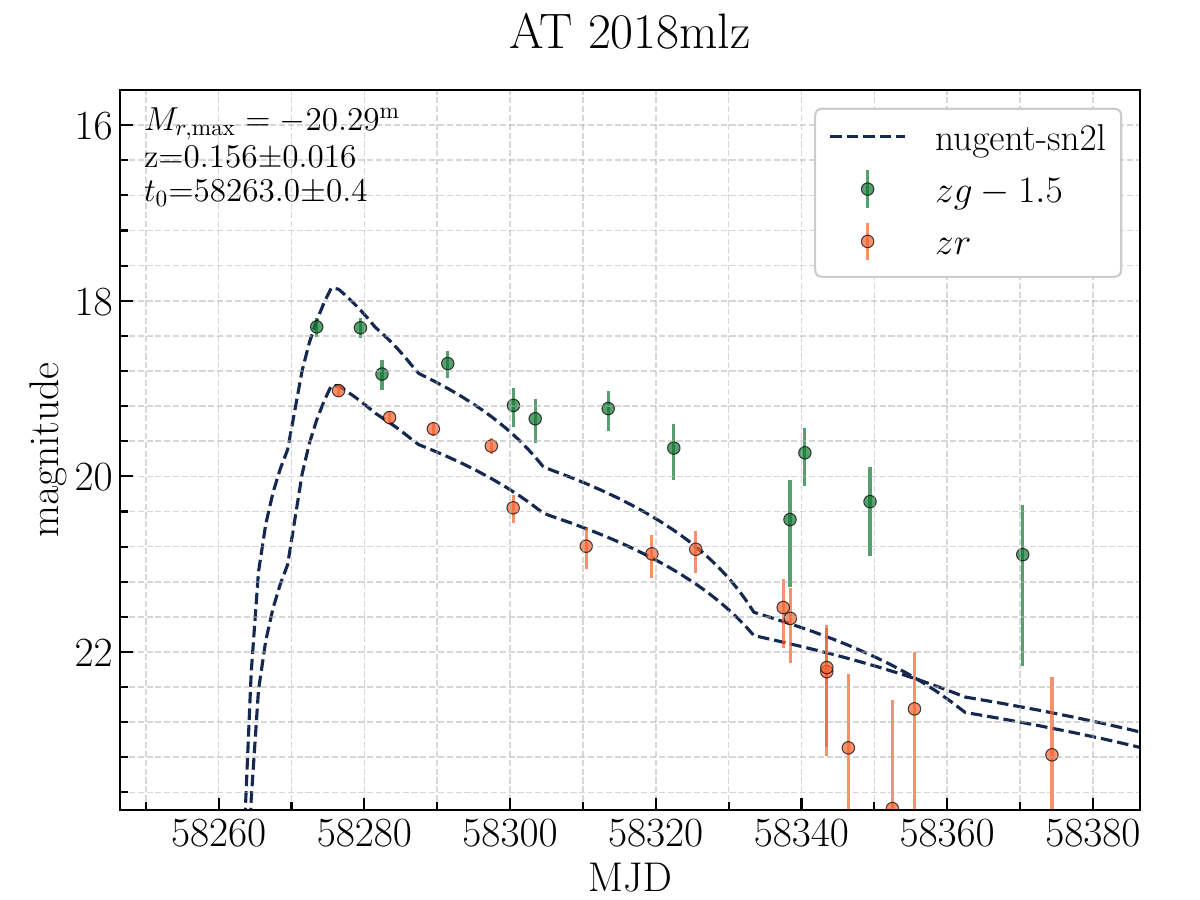}

\includegraphics[width=0.48\linewidth]{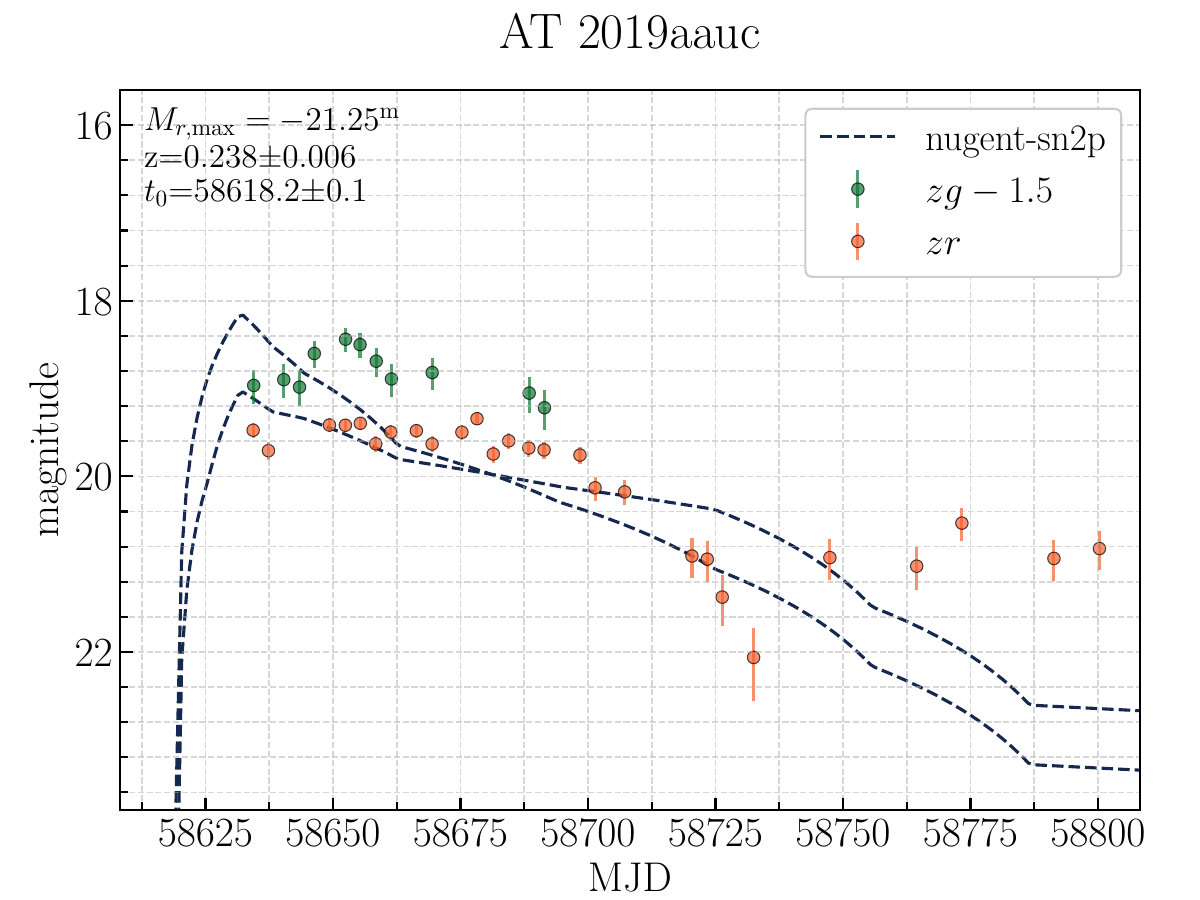}\hfill
\includegraphics[width=0.48\linewidth]{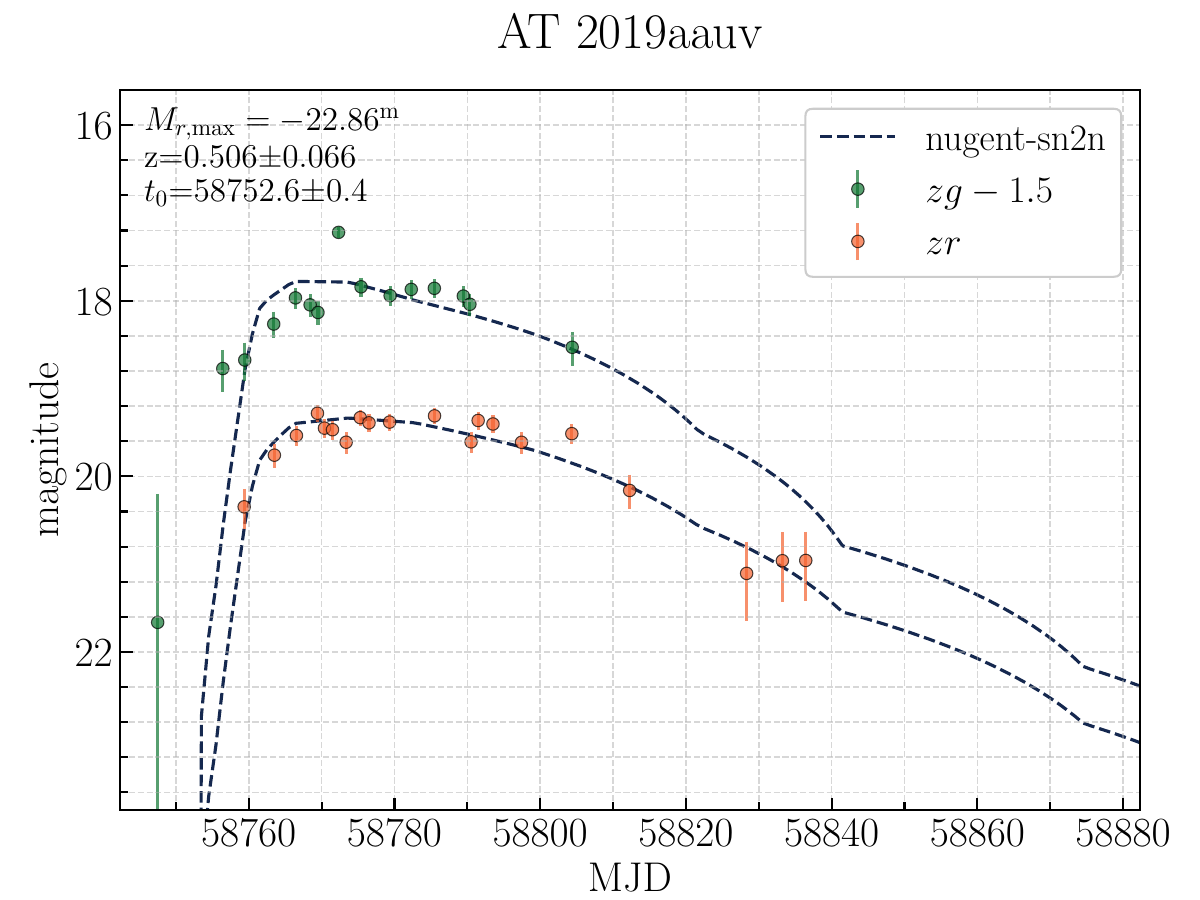}

\includegraphics[width=0.48\linewidth]{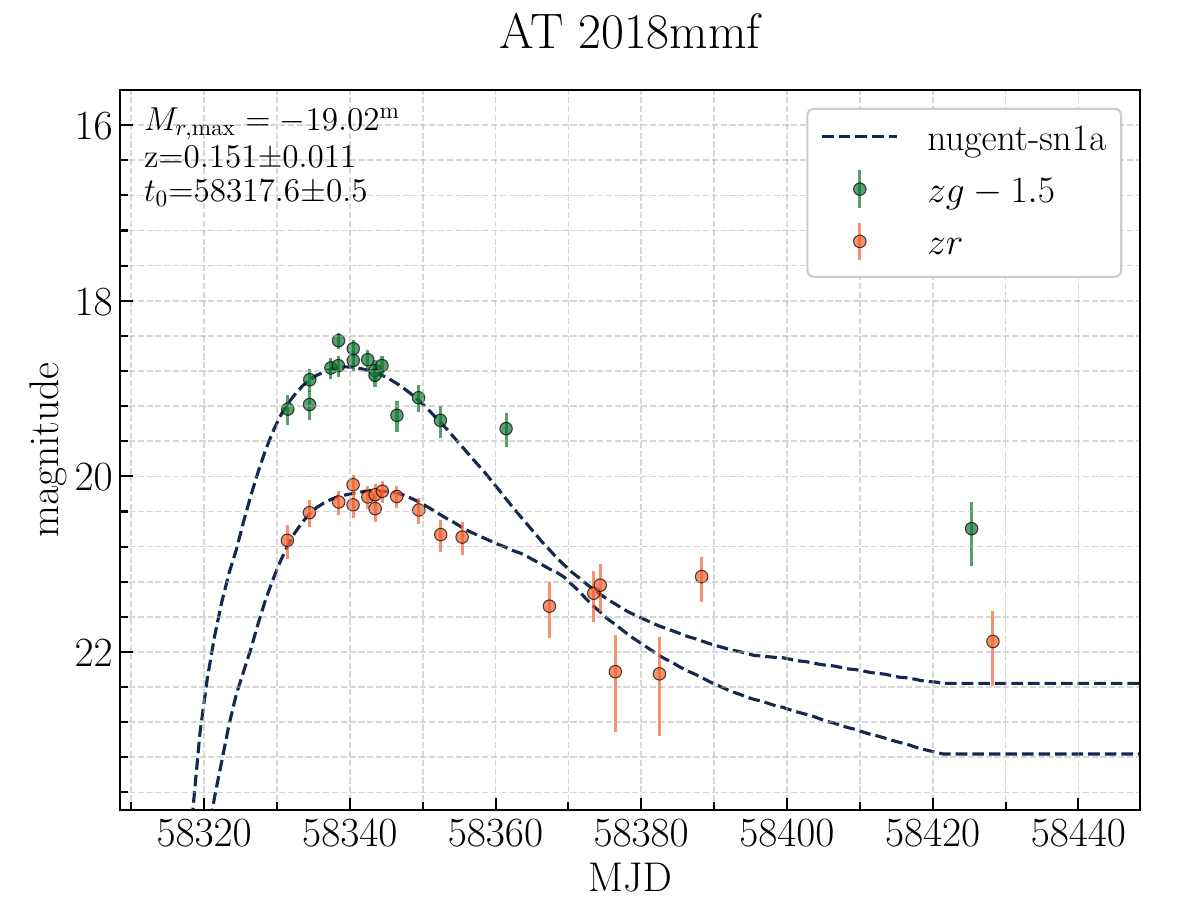}\hfill
\includegraphics[width=0.48\linewidth]{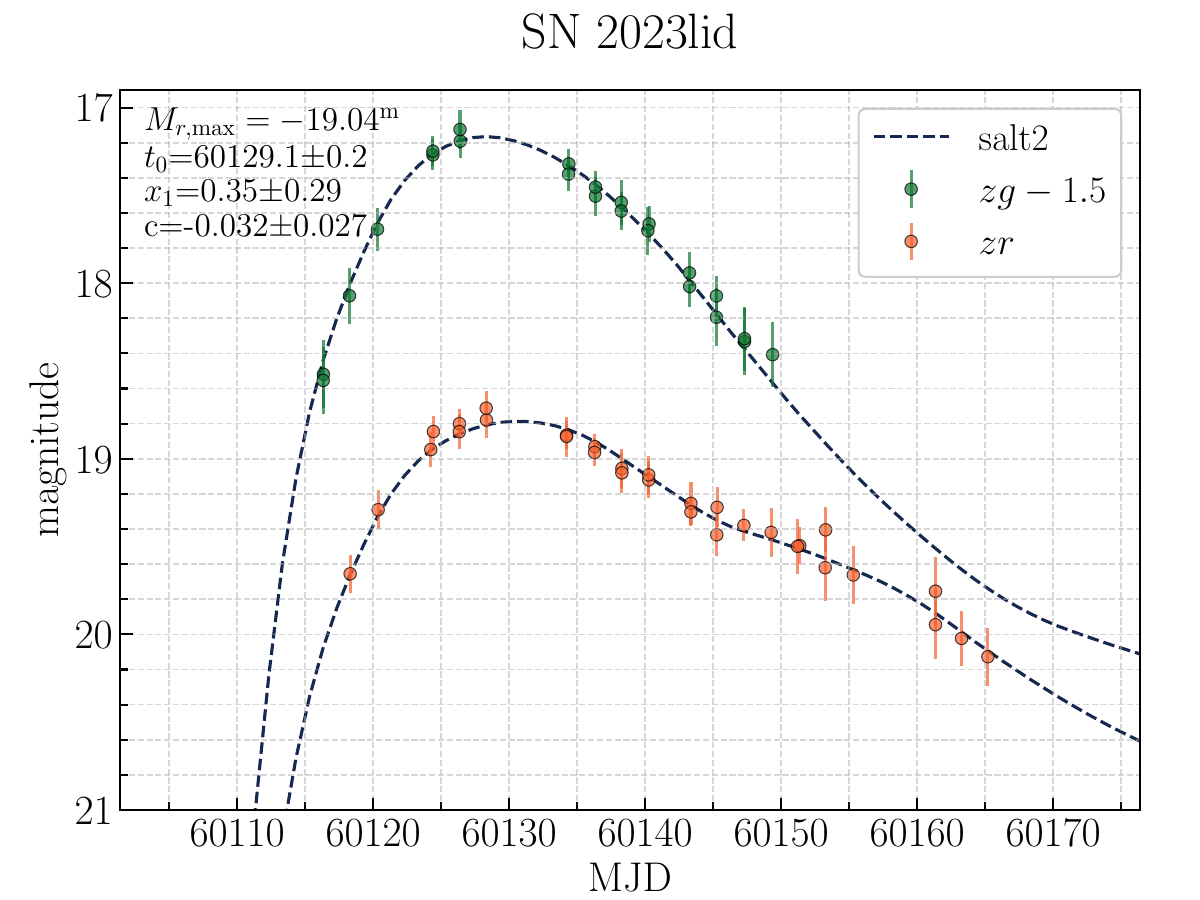}\hfill

\caption{ZTF light curves of supernova candidates found by the Fink anomaly detection pipeline, with best-model fits and fitted model parameters (continued).}
\end{figure*}

\begin{figure*}[t]\ContinuedFloat
\centering
\includegraphics[width=0.48\linewidth]{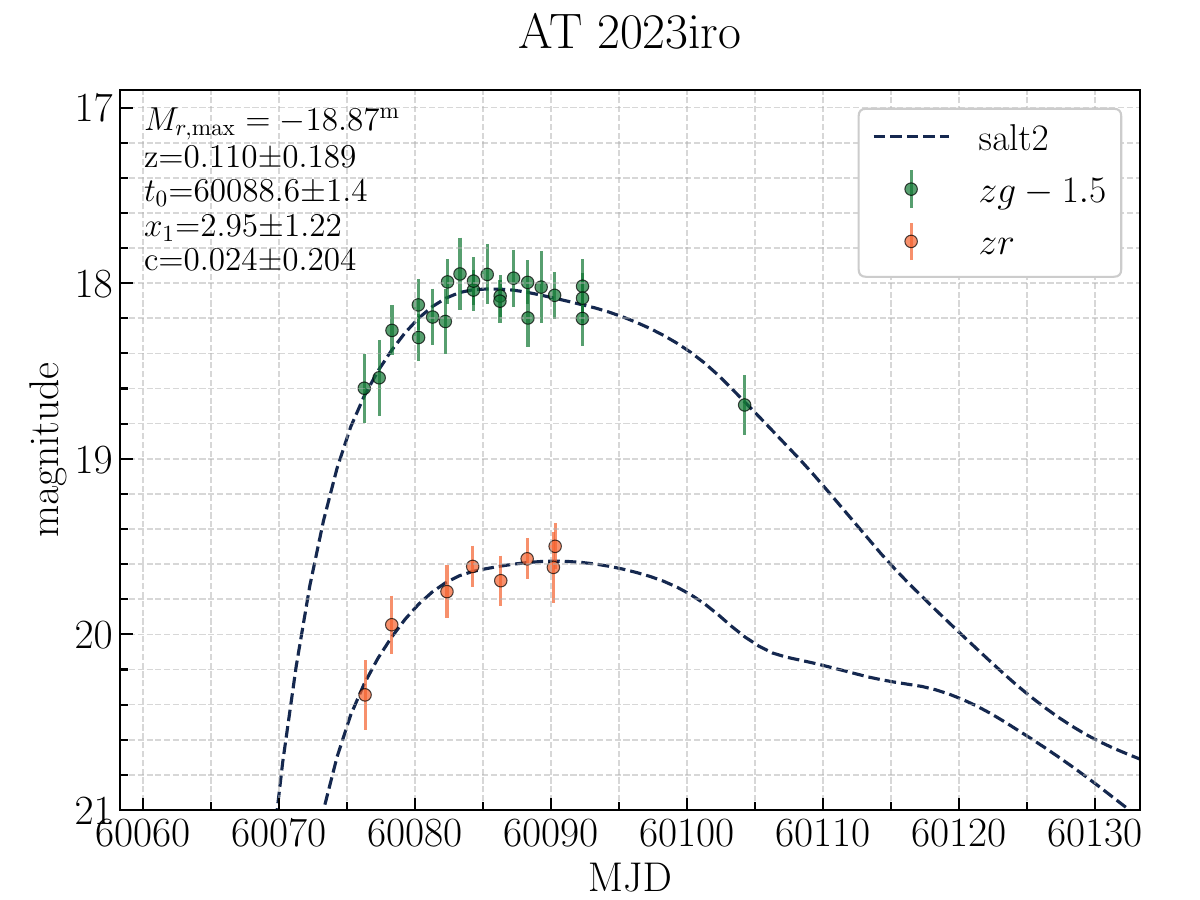}\hfill
\includegraphics[width=0.48\linewidth]{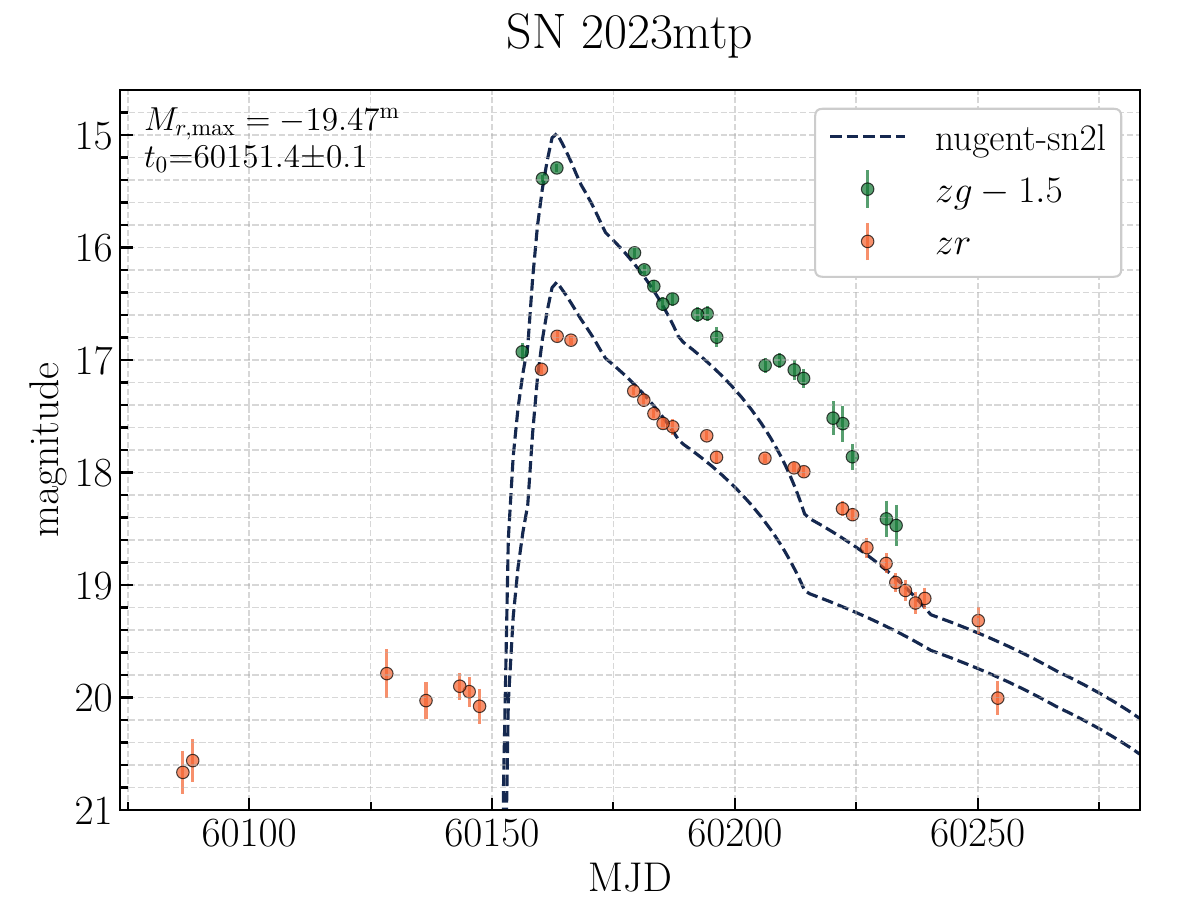}

\includegraphics[width=0.48\linewidth]{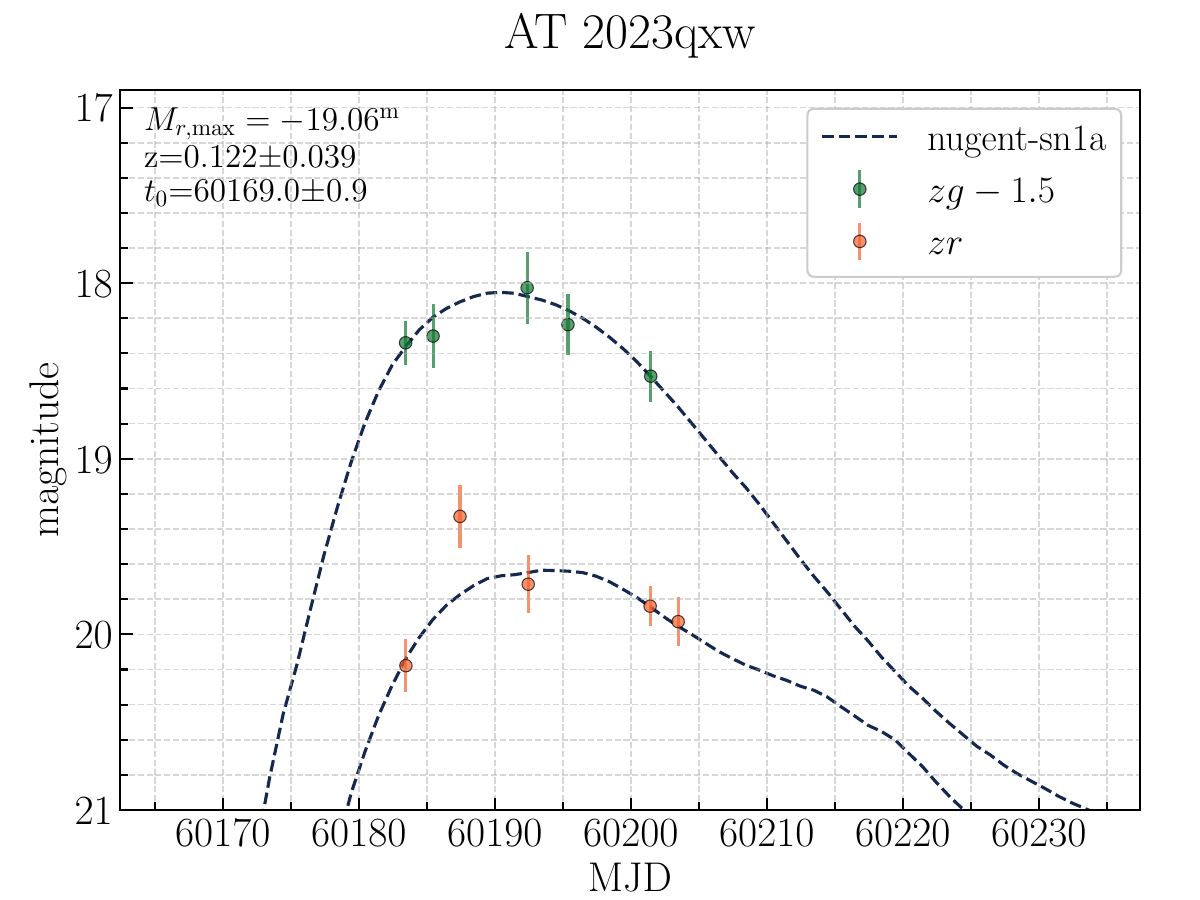}\hfill

\caption{ZTF light curves of supernova candidates found by the Fink anomaly detection pipeline, with best-model fits and fitted model parameters (continued).}
\end{figure*}

\section{Cataclysmic variables light curves}
\label{ap:CV}

We present here the cataclysmic variables observed, identified and registered in the AAVSO VSX database as dwarf novae (see Table~\ref{tab:UG}).

\begin{figure*}[t]\ContinuedFloat
\centering
\includegraphics[width=0.48\linewidth]{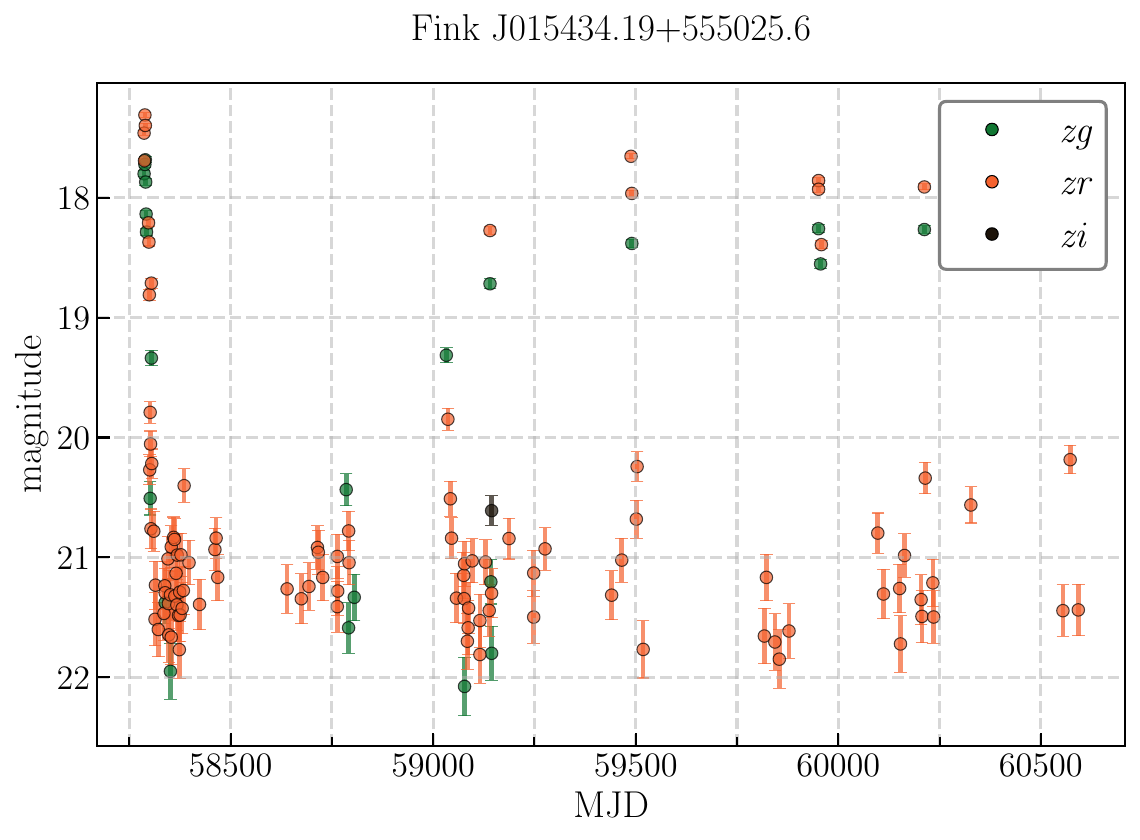}\hfill
\includegraphics[width=0.48\linewidth]{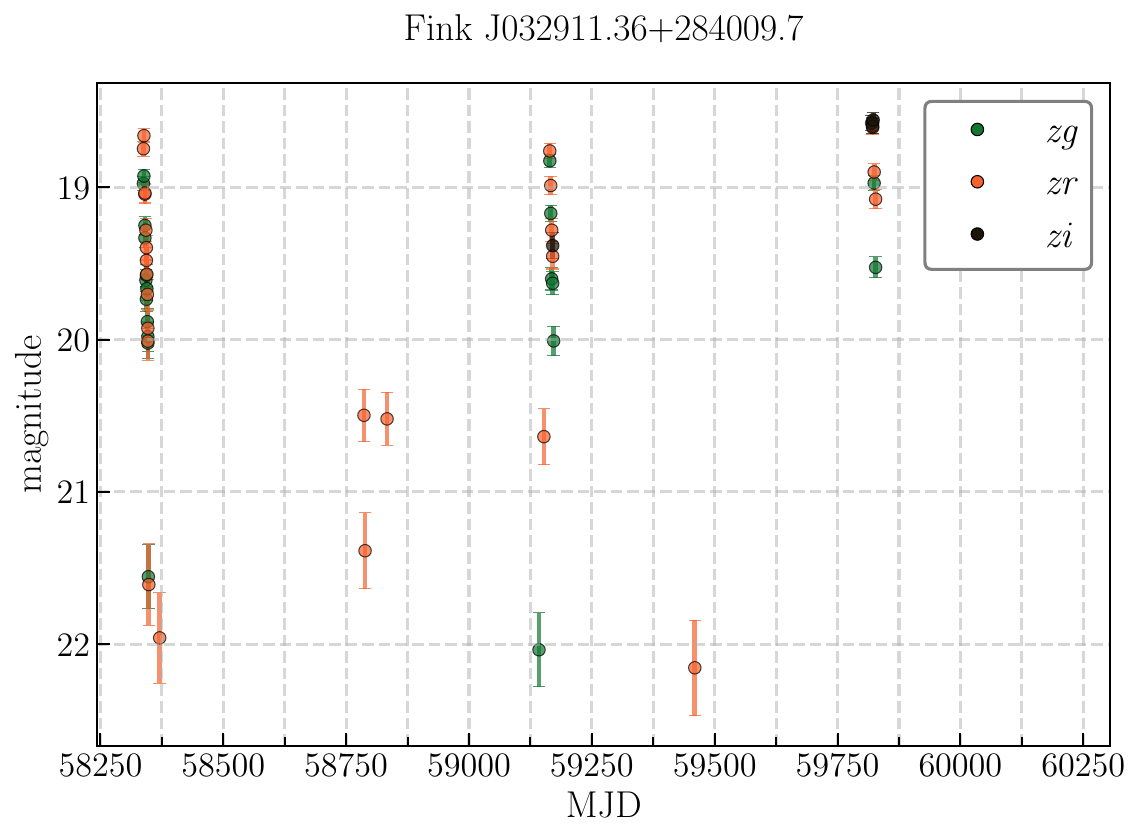}

\includegraphics[width=0.48\linewidth]{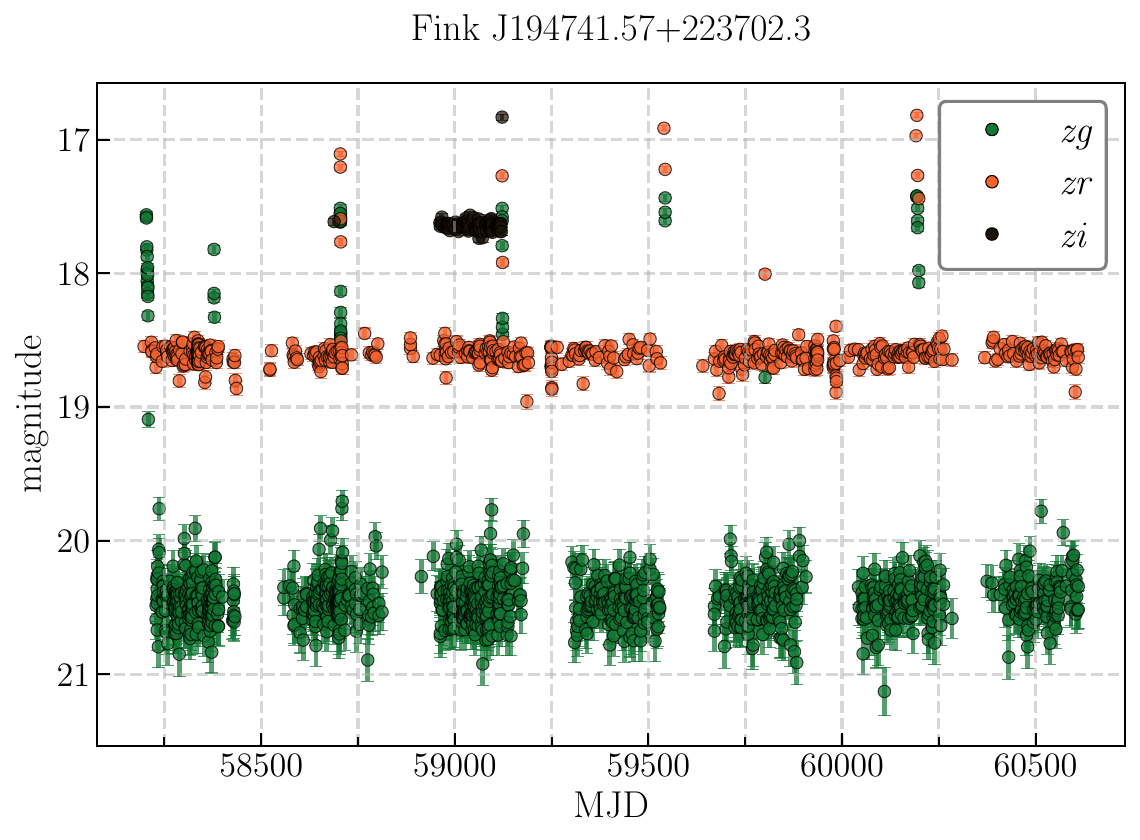}\hfill
\includegraphics[width=0.48\linewidth]{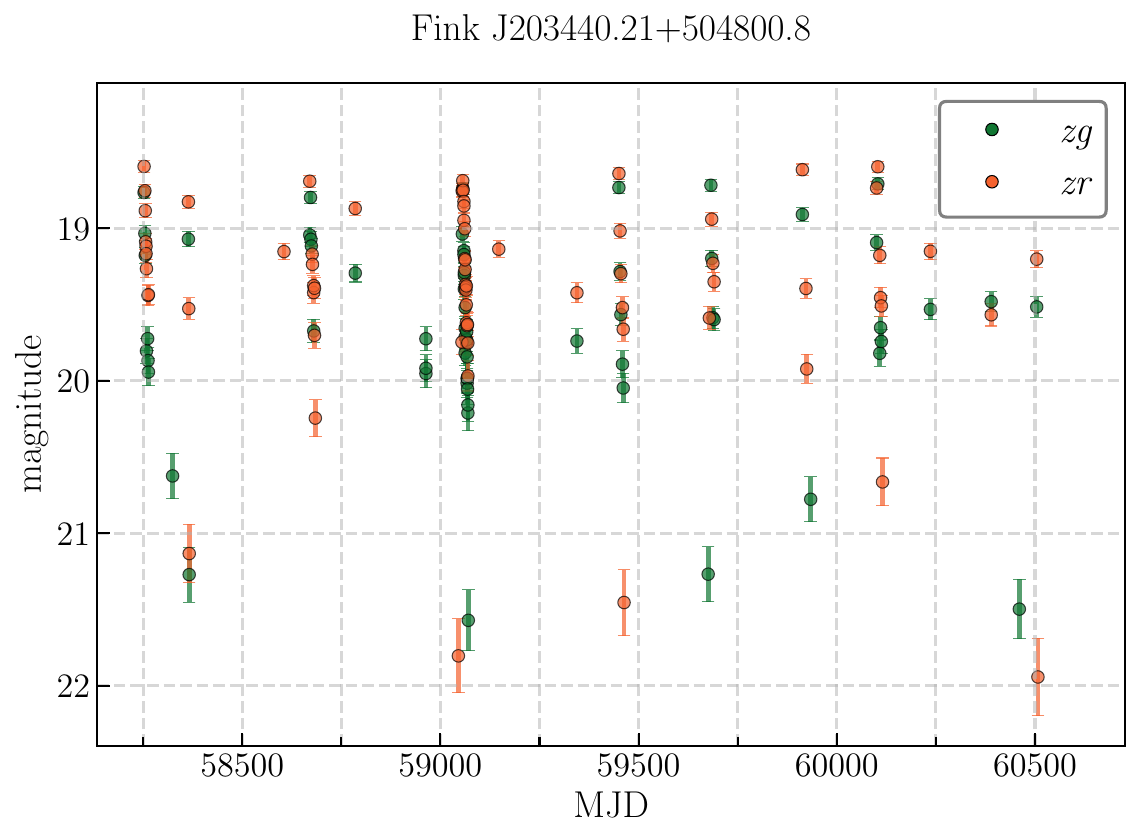}

\includegraphics[width=0.48\linewidth]{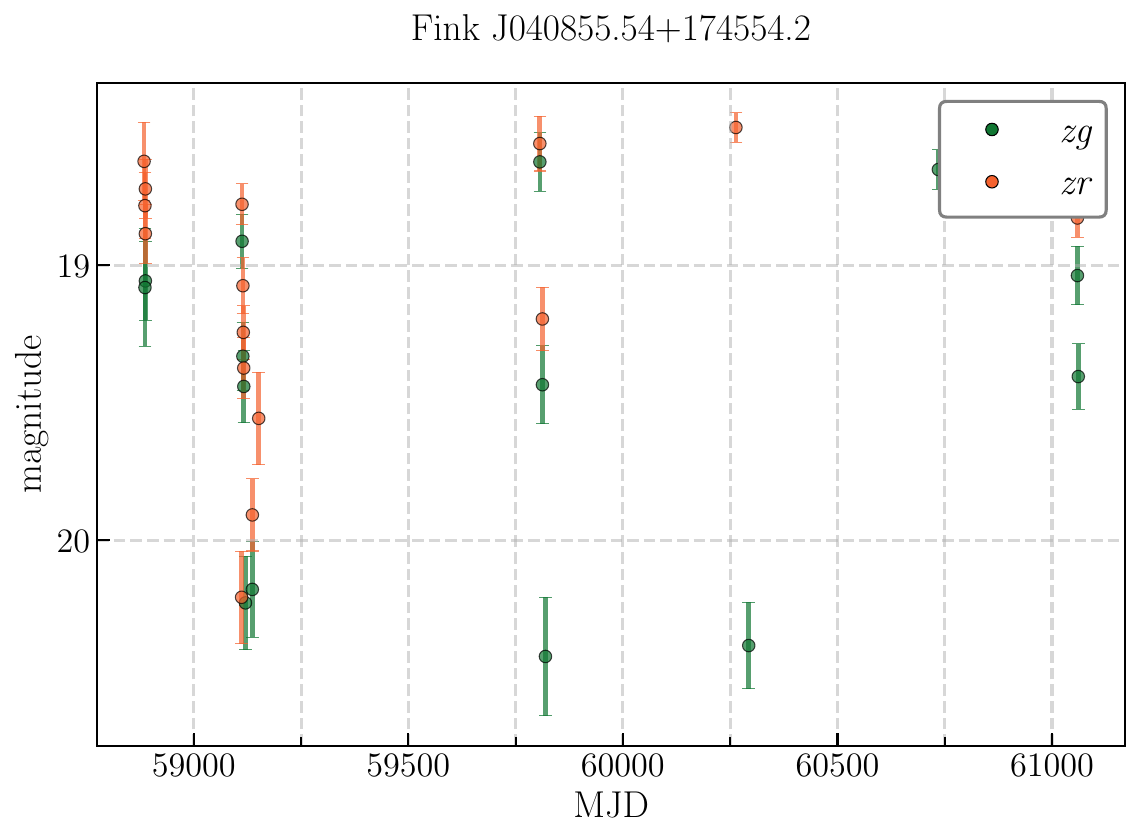}\hfill
\includegraphics[width=0.48\linewidth]{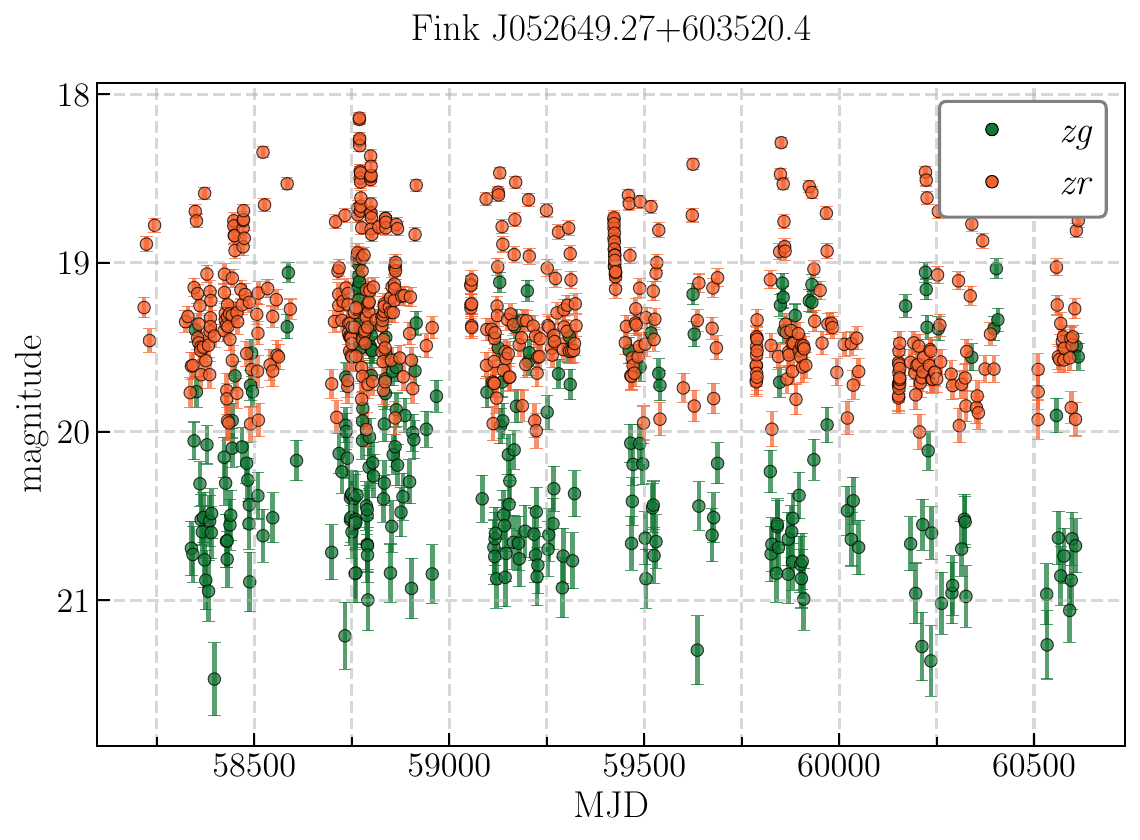}

\caption{ZTF light curves of dwarf novae candidates found with the Fink anomaly detection module.}
\label{fig:UG}
\end{figure*}

\begin{figure*}[t]\ContinuedFloat
\centering
\includegraphics[width=0.48\linewidth]{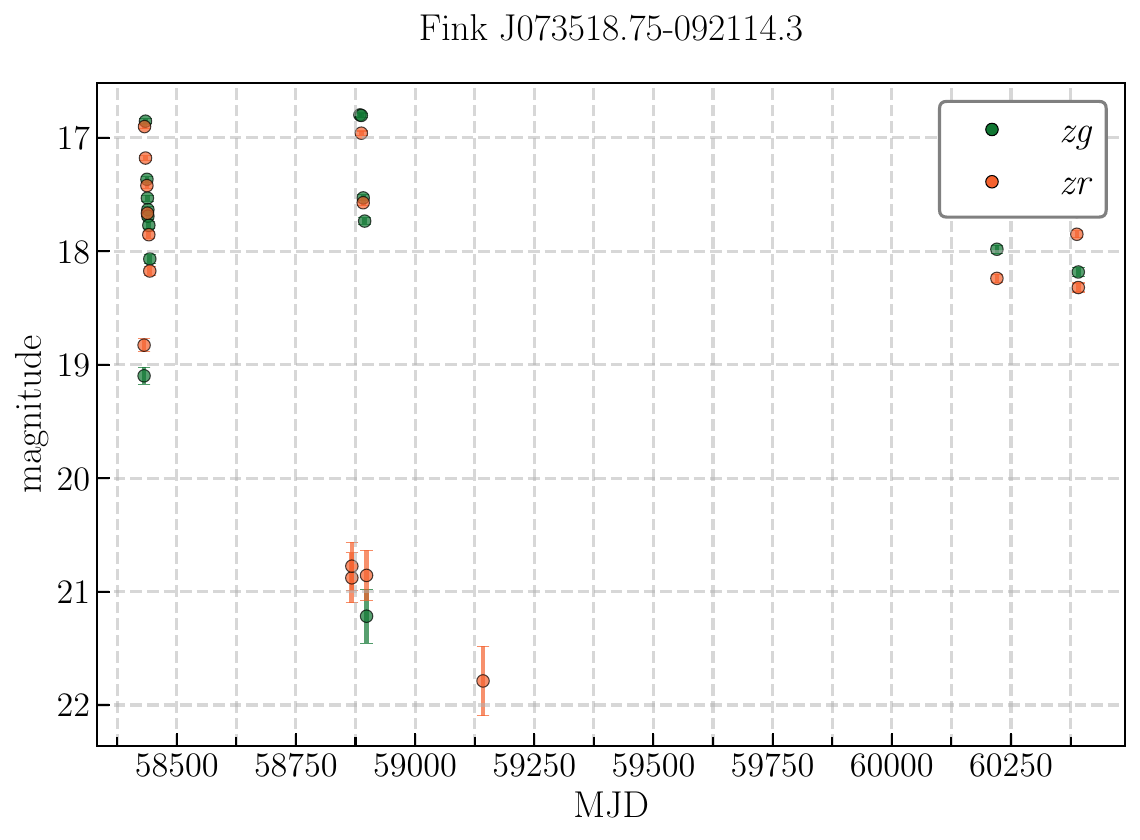}\hfill
\includegraphics[width=0.48\linewidth]{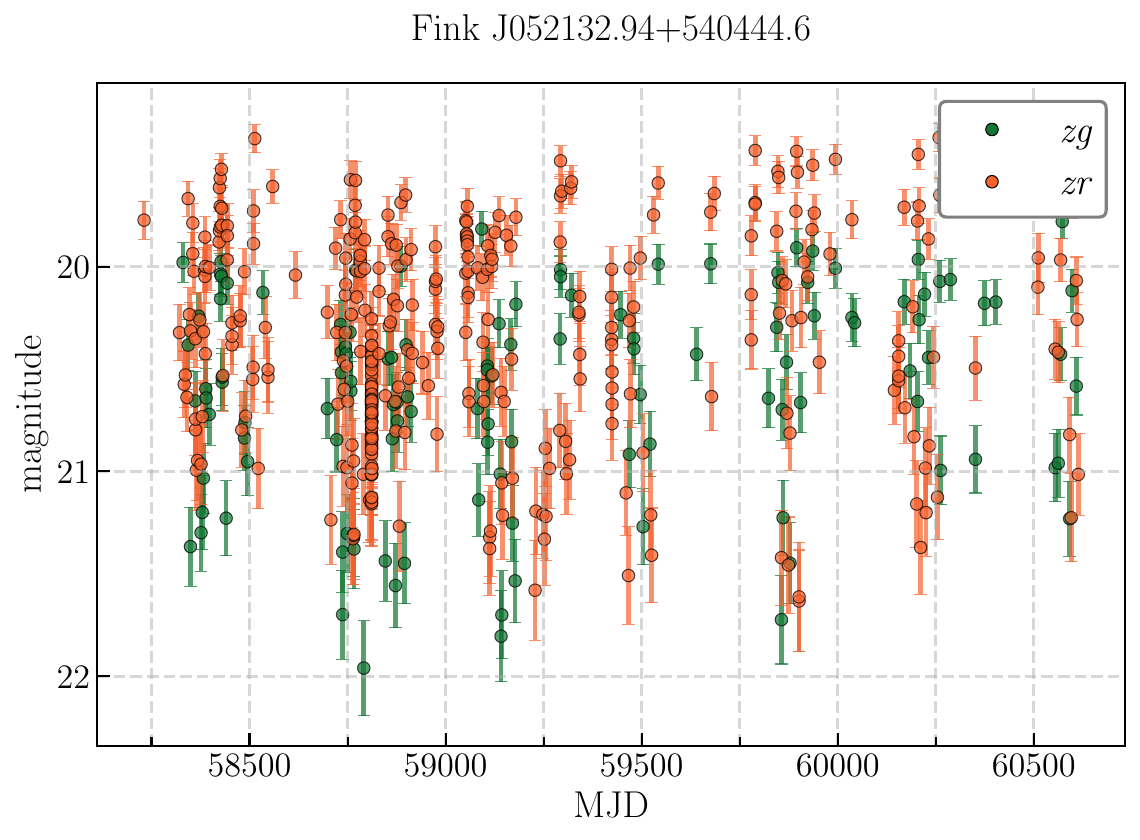}

\includegraphics[width=0.48\linewidth]{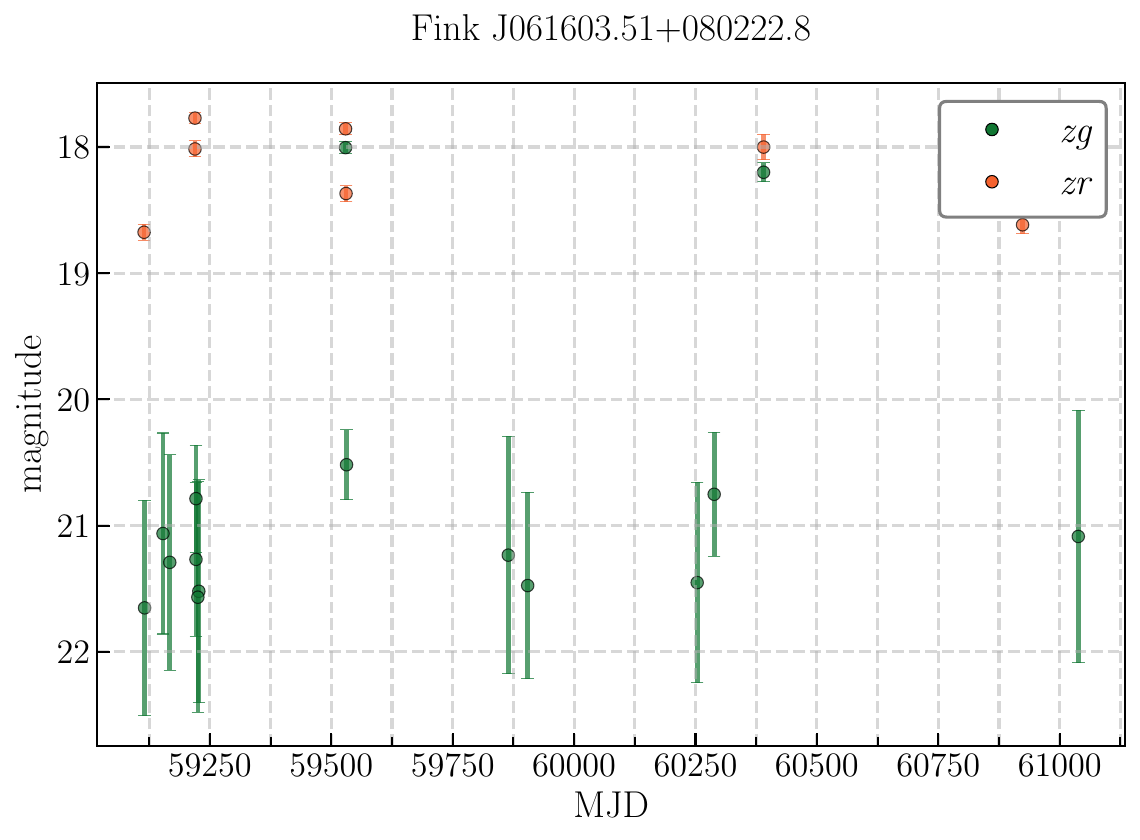}\hfill

\caption{ZTF light curves of dwarf novae candidates found with the Fink anomaly detection module (continued).}
\end{figure*}

\end{document}